\documentclass[seceqno]{style}

\usepackage{graphicx}  
\usepackage{epsfig}
\usepackage{url}

\def\be{\begin{equation}}
\def\ee{\end{equation}}
\def\bea{\begin{eqnarray}}
\def\eea{\end{eqnarray}}
\def\ket#1{\hbox{$\vert #1\rangle$}}   
\def\bra#1{\hbox{$\langle #1\vert$}}   
\def\oneh{{\textstyle {1\over 2}}}

\def\Re{\hbox{\rm Re\,}}
\def\Im{\hbox{\rm Im\,}}

\def          
\simeq{{\ \lower2pt\hbox{$-$}\mkern-13mu \raise2pt \hbox{$\sim$}\ }} 
\def          
\simge{{\ \lower2pt\hbox{$\sim$}\mkern-13mu \raise2pt \hbox{$>$}\ }}

\newcommand{\tvec}[1]{\mbox{\boldmath{$#1$}}}
\newcommand{\svec}[1]{\mbox{\boldmath{$\scriptstyle #1$}}}

\def\dsp{\displaystyle}

\newcommand{\singlesum}[1]{\sum_{\renewcommand{\arraystretch}{0.3}
  \begin{array}{l}\scriptscriptstyle{#1}\end{array}}}
  
\newcommand{\lrpartial}{\raisebox{0.1em}{$
\stackrel{\raisebox{-0.03em}{$\scriptstyle\leftrightarrow$}}{\partial}$}{}}

\renewcommand{\slash}[1]{#1 \hspace{-0.45em} / }

\title{Generalized parton distributions and the structure of the nucleon}

\author{Sigfrido Boffi \atque Barbara Pasquini}

\instlist{\inst{}
Dipartimento di Fisica Nucleare e Teorica, Universit\`a degli Studi di Pavia, Pavia, Italy\\
Istituto Nazionale di Fisica Nucleare, Sezione di Pavia, Pavia}

\PACSes{
\PACSit{12.38.-t} {Quantum chromodynamics}
\PACSit{12.39.-x} {Phenomenological quark models}
\PACSit{13.60.-r}{Photon and charged-lepton interactions with hadrons}
}

\begin{document}

\maketitle

\begin{abstract}
Generalized parton distributions have been introduced in recent years as a suitable theoretical tool to study the structure of the nucleon. Unifying the concepts of parton distributions and  hadronic form factors, they provide a comprehensive framework for describing the quark and gluon structure of the nucleon. In this review their  formal properties and modeling are discussed, summarizing the most recent developments in the phenomenological description of these functions. The status of available data is also presented.

\end{abstract}


\section{Introduction}
\label{sect:intro}

The composite nature of the proton was made manifest by the discovery of its anomalous magnetic moment~\cite{Stern} and confirmed by the observation of the electromagnetic form factors by Hofstadter and coworkers~\cite{Hofstadter}. Nowadays it is firmly believed that the internal dynamics of hadrons such as the proton and neutron (collectively indicated as nucleons), is determined by the strong interactions between quarks exchanging gluons, as governed by quantum chromodynamics (QCD). However, a detailed description of the nucleon structure is still missing because QCD can only be solved in the perturbative regime of short distance phenomena probed in hard collisions, whereas the 
soft part of the interaction corresponding to the long-distance behaviour requires a nonperturbative and/or numerical treatment like, e.g., in lattice simulations.

In order to probe the internal structure of the nucleon large energy ($\nu$) and momentum ($\tvec q$) transfers are necessary as in the so-called deep inelastic scattering (DIS) and ultimately in the Bjorken regime, i.e. when both $\nu$ and $Q^2=-q^2={\tvec q}^2-\nu^2$ become very large with fixed 
$x_B=Q^2/2p\cdot q$ (or $x_B=Q^2/2M_N\nu$ for a nucleon with mass $M_N$ at rest). Under these conditions scaling occurs, i.e. structure functions parametrizing the inclusive DIS cross section become independent of $Q^2$ at fixed $x_B$. In the parton model~\cite{feynman}  this phenomenon is interpreted as the incoherent elastic scattering off the partons with the Bjorken $x_B$ being just the fractional (light-cone, longitudinal) momentum of the struck parton. At the parton level 
one may distinguish three kinds of quark distributions, i.e. the quark density $q(x_B)$, the helicity distribution $\Delta q(x_B)$, and the transversity $\Delta_T q(x_B)$. The first two are well-known quantities: $q(x_B)$ is the probability of finding a quark with a fraction $x_B$ of the longitudinal momentum of the parent (fast-moving) nucleon, regardless of its spin orientation; $\Delta q(x_B)$ gives the net helicity of a quark in a longitudinally polarized nucleon, i.e. it is the number density of quarks with positive helicity minus the number density of quarks with negative helicity, assuming the parent nucleon to have positive helicity; in a transversely polarized nucleon, the transversity $\Delta_T q(x_B)$ is the number density of quarks with polarization parallel to that of the nucleon minus the number density of quarks with antiparallel polarization. Information on the last distribution is missing on the experimental side because $\Delta_T  q(x_B)$ decouples from inclusive DIS and therefore cannot be measured in such a traditional source of information. Other processes, such as polarized Drell-Yan dilepton production, are better suited for accessing transversity. 

Experiments show scaling violation with a $Q^2$ dependence due to the contribution of gluons taking part to the scattering process as active particles or being radiated by the initial and scattered  quarks as described by QCD with the Dokshitzer-Gribov-Lipatov-Altarelli-Parisi (DGLAP) evolution equations~\cite{dglap}. Thus, QCD fits to DIS data also determine the gluon distribution. A large amount of data over the years have provided us with a fair description of quark and gluon distributions (see, e.g., \cite{WG-HERA}). 

More generally, in the Bjorken regime one probes space-time correlations along the light-cone. Whereas inclusive DIS involves diagonal matrix  elements of certain operators, thus allowing a probability interpretation in terms of distributions, a full knowledge of the correlations can only be achieved by considering also the nondiagonal matrix elements of the same operators. This is possible  in exclusive processes under suitable conditions where one can factorize short- and long-distance contributions to the reaction mechanism. These nondiagonal matrix elements can be parametrized in terms of generalized parton distributions (GPDs). GPDs have been introduced in the past in different contexts (see, e.g., \cite{Dittes88,Muller94}), but have raised a large interest in the hadron community only when their importance was stressed in studies of deeply virtual Compton scattering (DVCS)~\cite{Ji97, Radyushkin96a,Ji97a} and hard meson production~\cite{Radyushkin96b} in connection with the 
possibility of factorizing their contribution~\cite{CFS97} and gaining information on the spin structure of the nucleon~\cite{Ji97}. Being related to nondiagonal matrix elements, GPDs do not represent any longer a probability, but rather the interference between amplitudes describing different parton configurations of the nucleon so that they give access to momentum correlations of partons in the nucleon.

The finite momentum transfer to the proton makes a second space-time structure of the process possible. Whereas in inclusive DIS the partonic subprocess is the scattering of a photon on a quark or antiquark, in the case of GPDs the virtual photon can also annihilate on a quark-antiquark pair with 
transverse separation of order $1/Q$ in the proton target.  

The momentum transfer can also have a transverse component. This provides a powerful tool to study hadron structure in three dimensions, because in addition to the information on the (longitudinal) behaviour in momentum space along the direction in which the nucleon is moving (as in the case of ordinary parton distributions), they also give insights on how partons are spatially distributed in the transverse plane~\cite{Burkardt00a} (as in the case of elastic form factors).

The GPDs can also be viewed as the generating functions for the form factors of the twist-two operators  governing the interaction mechanisms of hard processes in the deep inelastic regime. These generalized form factors do not couple directly to any known fundamental interactions, but can be studied indirectly looking at moments of the GPDs. The most peculiar example are the form factors of the energy momentum tensor, which allow to access the total and orbital angular momentum of the nucleon carried by quarks and gluons. 

In the last ten years GPDs have been investigated in great detail from the theoretical point of view, and many review papers already have summarized the progressing status of their understanding and modeling~\cite{Ji98a,Radyushkin, GPVdH,Diehlrep,Jiarnps,BR05}. 

Experimentally, measuring GPDs in exclusive processes is a big challenge requiring high luminosity and resolution. Therefore only very recently first dedicated experiments have been planned and performed.

This review is an attempt to provide a general overview on the most recent developments in the phenomenological description of the nucleon GPDs. After a summary of properties of the GPDs in sect.~\ref{sect:def} and their physical content in sect.~\ref{sect:phys}, the main lines of research in modeling GPDs for the nucleon are presented in sect.~\ref{sect:model}. Results obtained for quantities describing the nucleon structure are discussed in sect.~\ref{sect:appl}, and the status of experimental investigation is presented in sect.~\ref{sect:exp}. Conclusions are drawn in the final section.


\section{Definition and properties of generalized parton distributions}
\label{sect:def}


\subsection{Generalized form factors}
\label{subsec:ff}

In quantum field theory one can construct currents, i.e. Dirac tensors as bilinear combinations of $\Gamma$ matrices ($\Gamma= {\tvec 1}, \gamma_5,\gamma^\mu,\gamma^\mu\gamma_5, \sigma^{\mu\nu}$) and Dirac fields, $\bar\psi\,\Gamma\,\psi$.  Depending on the selected $\Gamma$ matrix and its Lorentz properties one can define scalar, pseudoscalar, vector, axial and tensor currents.

Matrix elements of the above currents on nucleon states with initial (final) momentum $p^\mu$ (${p'}^\mu$) and covariantly normalized as $\bra{p'}p\rangle = 2p^0(2\pi)^3\delta(p'-p)$, are usually expressed in terms of form factors. For example, with $\psi\to \psi_q(z)$ for quarks of flavour $q$, the matrix elements of quark (electroweak) vector and axial currents are decomposed as
\bea
\label{eq:emff}
\bra{p'}\bar \psi_q(0)\gamma^\mu \psi_q(0)\ket{p} &=&  \bar u(p')\left[ F^q_1(t)\gamma^\mu + i\frac{1}{2M_N} F^q_2(t) \sigma^{\mu\nu}\Delta_\nu\right] u(p),
\\
\bra{p'}\bar \psi_q(0)\gamma^\mu\gamma_5 \psi_q(0)\ket{p} &=&  \bar u(p')\left[ g_A^q(t)\gamma^\mu\gamma_5 + \frac{1}{2M_N} g_P^q(t) \Delta^\mu\gamma_5\right] u(p),
\label{eq:weakff}
\eea
where $u(p)$ is the Dirac spinor normalized as $\bar u(p)u(p)= 2M_N$, and $\Delta^\mu={p'}^\mu-p^\mu$ and $t=\Delta^2$. For each separate flavour $q$ the vector current involves the contributions $F_1^q$ and $F_2^q$ respectively to the Dirac and Pauli electromagnetic form factors of proton and neutron, $F^{p,n}_{1}$  and $F^{p,n}_{2}$, while the weak axial current involves the contributions $g_A^q$ and $g_P^q$ respectively to the axial and induced pseudoscalar form factors. 

Restricting oneself to up and down quarks only and referring to their contribution in the proton, the form factors $F^q_{1,2}$ are related to the physical $F^{p,n}_{1,2}$ as
\be
F^u_{1,2} = 2 F_{1,2}^p + F_{1,2}^n, \quad F^d_{1,2} = F_{1,2}^p + 2 F_{1,2}^n .
\ee
For the axial vector form factor one uses the isospin decomposition
\be
g_A^u= \oneh g_A + \oneh g_A^0, \quad g_A^d= -\oneh g_A + \oneh g_A^0,
\label{eq:ga0}
\ee
where $g_A $ ($g_A^0$)  is the isovector (isoscalar) axial form factor with $g_A(0)=1.267$ and, from quark models, $g_A^0(t)= {\textstyle {3\over 5}}\,g_A(t)$. Relations similar to (\ref{eq:ga0}) hold for the induced pseudoscalar form factor containing an important pion pole contribution through the partial conservation of the axial current.

The above currents can be considered as particular cases of more general operators in QCD. In the deep inelastic regime of hard processes, where an operator product expansion is performed in order to overcome the problem of light-cone singularities arising as a consequence of the explored short distances, the major contribution comes from the so-called twist-two tensor operators (see, e.g., \cite{Muta}). Formally, the twist $\tau$ is defined as the dimension $d$ in mass units minus the Lorentz spin $s$ of the operator, $\tau = d-s$. A spin-$s$ tensor transforms as an irreducible representation of the Lorentz group. The maximal spin for a given number of Lorentz indices is achieved when they are all symmetrized. The irreducibility implies that the reduction to lower-spin tensors is not possible: as a consequence, the contraction of any pair
of indices with the metric tensor gives zero. Thus, the Lorentz structure has to be traceless.

In QCD, there are six towers of twist-two operators forming totally symmetric representations of the Lorentz group~\cite{Ji98a,Jiarnps,BR05}:
\begin{eqnletter}
\label{eq:twist2}
    {\mathcal O}^{\mu\mu_1\cdots\mu_{n-1}}_q &=&
       \bar \psi_q  \gamma^{(\mu}\, i\stackrel{\leftrightarrow}{\cal D}{}^{\mu_1}
        \cdots  i\stackrel{\leftrightarrow}{\cal D}{}^{\mu_{n-1})}
        \psi_q\ , \label{eq:twist2.r1} \\
    \tilde {\mathcal O}^{\mu\mu_1\cdots\mu_{n-1}}_q &=& 
       \bar \psi_q\gamma^{(\mu}\gamma_5\, i\stackrel{\leftrightarrow}{\cal D}{}^{\mu_1}
        \cdots  i\stackrel{\leftrightarrow}{\cal D}{}^{\mu_{n-1})}
         \psi_q\ , \label{eq:twist2.r2}  \\
    {\mathcal O}^{\mu\nu\mu_1\cdots\mu_{n-1}}_{qT} &=& 
       \bar \psi_q  \sigma^{\mu(\nu}\,i\stackrel{\leftrightarrow}{\cal D}{}^{\mu_1}
        \cdots  i\stackrel{\leftrightarrow}{\cal D}{}^{\mu_{n-1})}
        \psi_q\ , \label{eq:twist2.r3}  \\
   {\mathcal O}^{\mu\mu_1\cdots\mu_{n-1}\nu}_g &=& 
       F^{(\mu\alpha} i\stackrel{\leftrightarrow}{\cal D}{}^{\mu_1}
        \cdots  i\stackrel{\leftrightarrow}{\cal D}{}^{\mu_{n-1}}
        F_\alpha^{~\nu)} \ , \label{eq:twist2.r4}  \\
  \tilde {\mathcal O}^{\mu\mu_1\cdots\mu_{n-1}\nu}_g &=& 
       -i\,F^{(\mu\alpha} i\stackrel{\leftrightarrow}{\cal D}{}^{\mu_1}
        \cdots  i\stackrel{\leftrightarrow}{\cal D}{}^{\mu_{n-1}}
        \tilde F_\alpha^{~\nu)} \ , \label{eq:twist2.r5}  \\
    {\mathcal O}^{\mu\mu_1\cdots\mu_{n-1}\nu\alpha\beta}_{gT} &=& 
       F^{(\mu\alpha} i\stackrel{\leftrightarrow}{\cal D}{}^{\mu_1}
        \cdots  i\stackrel{\leftrightarrow}{\cal D}{}^{\mu_{n-1}}
        F^{\nu)\beta}\ , \label{eq:twist2.r6} 
\end{eqnletter}
where $\tilde{F}^{\alpha\beta} = \frac{1}{2}  \epsilon^{\alpha\beta\gamma\delta} F_{\gamma\delta}$ is the dual field strength tensor with $\epsilon_{0123}=1$. In Eqs.~(\ref{eq:twist2}) all indices within $(\cdots)$ are symmetrized and traceless, and
\be
\stackrel{\leftrightarrow}{\cal D}_\mu \, \equiv\,
\oneh\left(\stackrel{\rightarrow}{\cal D}_\mu - \stackrel{\leftarrow}{\cal D}_\mu\right),
\qquad
\stackrel{\rightarrow}{\cal D}_\mu \, =\, \stackrel{\rightarrow}{\partial}_\mu - \, i g \, t_a A_\mu^a (z) , 
\qquad
\stackrel{\leftarrow}{\cal D}_\mu \,=\, \stackrel{\leftarrow}{\partial}_\mu + \, i g \, t_a A_\mu^a (z)
\ee
are covariant derivatives expressed in terms of the gluon vector potential $A_\mu^a (z)$. The label $a=1,\dots,8$ is the octet colour label, and the $t_a$  are (one half of) the Gell-Mann matrices for the triplet representation of SU(3), with
\be
[t_a,t_b] = i\,f_{abc} t_c.
\ee
The gluon field tensor is
\be
F^a_{\mu\nu} = \partial_\mu A^a_\nu - \partial_\nu A^a_\mu + g\,f_{abc} A^b_\mu A^c_\nu,
\ee
where $g$ is a constant representing the coupling strength between $\psi_q$ and $A^a_\mu$. In the following when not necessary the colour index will be omitted.

Like the form factors of the electromagnetic current, additional information about nucleon structure 
can be found in the (generalized) form factors of the twist-two operators when the matrix elements are taken between states of unequal momenta. Using Lorentz symmetry and parity and time reversal invariance, one can write down all possible form factors of the spin-$n$ operator ${\mathcal O}^{\mu\mu_1\cdots\mu_{n-1}}_q$ in Eq.~(\ref{eq:twist2.r1})~\cite{Ji98a,Jiarnps}
\begin{eqnarray}
\label{eq:form}
& & \langle p'| \bar \psi_q  \gamma^{(\mu}\, i\stackrel{\leftrightarrow}{\cal D}{}^{\mu_1}
        \cdots  i\stackrel{\leftrightarrow}{\cal D}{}^{\mu_{n-1})}
        \psi_q |p\rangle \\
& &\quad{}={\bar u}(p')\left[   \sum_{i=0 \atop \scriptstyle{\rm even}}^{n-1} 
        \gamma^{(\mu} \Delta^{\mu_1}\cdots \Delta^{\mu_i} 
     {P}{}^{\mu_{i+1}}\cdots
     {P}{}^{\mu_{n-1})} \,A^q_{n,i}(t) \right.\nonumber \\ 
&&\qquad\qquad \quad{}  -   \sum_{i=0 \atop \scriptstyle{\rm even}}^{n-1} 
     \frac{\Delta_\alpha}{2M_N} i\,\sigma^{\alpha(\mu} \,\Delta^{\mu_1}
\cdots \Delta^{\mu_i} 
     {P}{}^{\mu_{i+1}}\cdots{P}{}^{\mu_{n-1})}\, B^q_{n,i}(t) \nonumber \\
&&\qquad \qquad \quad{} \left.+ {1\over M_N} \, \Delta^{(\mu}\Delta^{\mu_1} \cdots \Delta^{\mu_{n-1})} \,
     C_{n}^q(t)\,  {\rm Mod}(n+1,2) \right]u(p), \nonumber
\end{eqnarray}
where $P=\oneh(p+p')$ is the average nucleon momentum, and ${\rm Mod}(n+1,2)$ is 1 for even $n$ and 0 for odd $n$. Thus $C_{n}^q(t)$ is present only when $n$ is even, and in general there are $n+1$ form factors~\cite{JiLebed01,Hagler04}. 

Similar decompositions in terms of generalized form factors are possible also for nucleon matrix elements of the other twist-two operators in Eqs.~(\ref{eq:twist2}). The problem of counting the number of allowed form factors has been addressed in Ref.~\cite{JiLebed01} making use of a method based on partial wave formalism and crossing symmetry, i.e. by considering the number of independent amplitudes in the crossed channel $\bra{p\bar p}{\cal O}\ket{0}$ corresponding to proton-antiproton creation from the twist-two source. In the case of the vector operator one finds exactly $n+1$ form factors, as quoted  above. The situation is more complicated in the case of the axial vector and tensor operators~\cite{Hagler04,Chen05}.

As has been observed in Ref. \cite{Diehlrep}, there is no $C^q_{n}(t)$-like generalized form factor present for the axial vector, and the parametrization reads~\cite{Hagler04}
\begin{eqnarray}
\label{paraav}
& & \left\langle p^{\prime }\right| \bar{\psi}_q(0)\gamma ^{(\mu}\gamma_5 \, 
i{\cal D}^{\mu _{1}}\cdots i{\cal D}^{\mu _{n-1})}\psi_q (0)
\left| p\right\rangle\\
&& \quad{} = %
\bar{u}(p^{\prime })\sum_{\scriptstyle i=0 
\atop\scriptstyle{\rm even}}^{n-1}\left\{
\gamma ^{(\mu }\gamma_5\,\Delta ^{\mu _{1}}\cdots \Delta ^{\mu _{i}}%
{P}{}^{\mu _{i+1}}\cdots{P}{}^{\mu _{n-1})}\tilde{A}^q_{n,i}(t)\right.   \nonumber \\
&& \qquad\qquad \qquad\left. {}+\gamma _{5}\frac{1}{2M_N}\Delta ^{(\mu }\Delta ^{\mu
_{1}}\cdots \Delta ^{\mu _{i}}{P}{}^{\mu _{i+1}}\cdots{P}%
{}^{\mu _{n-1})}\tilde{B}^q_{n,i}(t)\right\} u(p), \nonumber
\end{eqnarray}
for a total of $2[\frac{n-1}{2}]+2$ independent form factors.

\newcommand{\Asym}{\mathop{\mbox{\tvec A}}}
\newcommand{\Sym}{\mathop{\mbox{\tvec S}}}

There is a total of $2[\frac{n-1}{2}]+n +2$ independent form factors for the parametrization of the tensor operator 
${\mathcal O}^{\mu\nu\mu_1\cdots\mu_{n-1}}_{qT}$ in Eq.~(\ref{eq:twist2.r3}) 
~\cite{Hagler04}:
\begin{eqnarray}
\label{parat}
&&
\left\langle p^{\prime }\right| \bar{\psi}_q(0)i\sigma ^{\mu (\nu }
i{\cal D}^{\mu_{1}}\cdots i{\cal D}^{\mu _{n-1})}\psi_q (0)
\left| p\right\rangle   \\
&& \ {} =  \Asym_{[\mu \nu ]} \Sym_{(\nu \mu _{1}\ldots )}%
\bar{u}(p^{\prime })\left[ 
\sum_{\scriptstyle i=0 \atop \scriptstyle{\rm even}}^{n-1}
i\sigma ^{\mu \nu }\Delta ^{\mu _{1}}\cdots \Delta ^{\mu _{i}}%
{P}{}^{\mu _{i+1}}\cdots{P}{}^{\mu _{n-1}}A_{Tn,i}(t)\right.   \nonumber \\
&&\qquad\qquad \quad \qquad \qquad{}+ 
\sum_{\scriptstyle i=0 \atop \scriptstyle{\rm even}}^{n-1}
\frac{{P}{}^{[\mu }\Delta ^{\nu ]}}{M_N^{2}}\Delta ^{\mu _{1}}\cdots
\Delta ^{\mu _{i}}{P}{}^{\mu _{i+1}}\cdots{P}{}^{\mu _{n-1}}%
\tilde{A}_{Tn,i}(t)  \nonumber \\
&&\qquad\qquad \quad \qquad \qquad{}+ 
\sum_{\scriptstyle i=0 \atop \scriptstyle{\rm even}}^{n-1}
 \frac{\gamma ^{[\mu }\Delta ^{\nu ]}}{2M_N}\Delta ^{\mu _{1}}\cdots
\Delta ^{\mu _{i}}{P}{}^{\mu _{i+1}}\cdots{P}{}^{\mu
_{n-1}}B_{Tn,i}(t)  \nonumber \\
&&\qquad\qquad \quad \qquad \qquad{}+\left.
 \sum_{\scriptstyle i=0 \atop \scriptstyle{\rm odd}} ^{n-1}\frac{\gamma ^{[\mu }%
{P}{}^{\nu ]}}{M_N}\Delta ^{\mu _{1}}\cdots \Delta ^{\mu _{i}}{%
P}{}^{\mu _{i+1}}\cdots{P}{}^{\mu _{n-1}}\tilde{B}_{Tn,i}(t)\right] u(p), \nonumber
\end{eqnarray}
where one has first to symmetrize and then to antisymmetrize as indicated~\cite{Geyer99}.

Analogous form factor decompositions are possible for the gluon operators~\cite{Diehlrep}. 


\subsection{Generalized parton distributions}

In hard scattering processes, where hadrons and partons move fast  in the $\hat z$-direction, it is natural to work with light-cone coordinates in terms of two light-like four-vectors $n_+=(1,0,0,1)/\sqrt{2}$ and $n_-=(1,0,0,-1)/\sqrt{2}$, i.e. 
\be
v^\mu\equiv v^+n_+^\mu + v^-n_-^\mu +v_\perp^\mu,
\ee
where $v^+=v\cdot n_-=(v^0+v^3)/\sqrt{2}$, $v^-=v\cdot n_+=(v^0-v^3)/\sqrt{2}$ and $v_\perp=(0,\tvec{v}_\perp,0)$. The relevant momenta are then the light-cone plus-momenta $p^+$ and the relevant $\Gamma$ structures become, e.g., $\slash n_-\equiv\gamma\cdot n_-=\gamma^+$.

Light-cone bilocal operators arising in hard scattering processes can be expanded in terms of the above currents making use of the relation
\bea
 \label{eq:higher-moments}
 \lefteqn{
(P^+)^{n} \int dx\, x^{n-1}  \int \frac{d z^-}{2\pi}\, e^{ix P^+ z^-}
  \Big[ \bar\psi_{q}(-\oneh z)\, \gamma^+ \psi_q(\oneh z) \Big]_{z^+=0,\,
  \svec{z}_\perp=0} 
}\\
 &=& \left. \Big(i \frac{d}{dz^-}\Big)^{n-1}
        \Big[ \bar\psi_{q}(-\oneh z)\, \gamma^+ \psi_q(\oneh z) \Big]
     \right|_{z=0}
  =  \bar\psi_{q}(0)\, \gamma^+ (i \lrpartial^+)^{n-1}\, \psi_q(0)
\nonumber
\eea
and its analogs for operators involving other $\Gamma$ matrices than $\gamma^+$.  In general, with the covariant derivative $\cal D$ instead of $\partial$ on the right-hand side, a Wilson link operator $W[-\oneh z,\oneh z]$ appears between the operators at positions $-\oneh z$ and $\oneh z$, where
\be
W[a,b] = {\cal P}\exp\left(- ig\int _a^b dz^- A^+(z^-n_-)\right),
\ee
and $\cal P$ denotes path ordering between $a$ and $b$. In the light-cone gauge, $A^+=0$, the Wilson link reduces to unity, as it will be assumed in the following. However, the condition $A^+=0$ does not remove all gauge freedom because $z^-$-independent gauge transformations are still possible with consequences on the correlation functions in hard processes sensitive to transverse momenta of partons~\cite{Hoyer02,Collins02,JiYuan02,Collins03,BJY03}. Technical details needed in the calculation of the gauge link corresponding to a given partonic subprocess and a calculational scheme are provided in Refs.~\cite{BMP03,BMP0405}.

Parton distributions are just defined in terms of matrix elements of light-cone bilocal operators between proton states of equal momenta~\cite{Roberts}. In general,  with initial (final) momentum $p$ ($p'$)  and helicity $\lambda_N$ ($\lambda_N'$) one defines a set of generalized quark distributions for a hadron with spin $\frac{1}{2}$ that have been classified in Refs.~\cite{Hoodbhoy98a,Diehl01}:
\begin{eqnarray}
  \label{no-flip-quark}
\lefteqn{ \int \frac{d z^-}{4\pi}\, e^{ix P^+ z^-}
  \langle p',\lambda_N'|\, 
     \bar{\psi}_q(-{\textstyle\frac{1}{2}}z)\, 
     \gamma^+ \psi_q({\textstyle\frac{1}{2}}z)\, 
  \,|p,\lambda_N \rangle \Big|_{z^+=0,\, \mathbf{z}_\perp=0} } 
\\
&&\hspace{2em}{}= \frac{1}{2P^+} \bar{u}(p',\lambda_N') \left[
  H^q\,  \gamma^+  +
  E^q\, \frac{i \sigma^{+\alpha} \Delta_\alpha}{2M_N}
  \right] u(p,\lambda_N) ,
\nonumber\\
\lefteqn{ \int \frac{d z^-}{4\pi}\, e^{ix P^+ z^-}
  \langle p',\lambda_N'|\, 
     \bar{\psi}_q(-{\textstyle\frac{1}{2}}z)\, 
     \gamma^+ \gamma_5\, \psi_q({\textstyle\frac{1}{2}}z)\, 
  \,|p,\lambda_N \rangle \Big|_{z^+=0,\, \mathbf{z}_\perp=0} } 
\\
&&\hspace{2em} {}= \frac{1}{2P^+} \bar{u}(p',\lambda_N') \left[
  \tilde{H}^q\, \gamma^+ \gamma_5 +
  \tilde{E}^q\, \frac{\gamma_5 \Delta^+}{2M_N}
  \right] u(p,\lambda_N) ,
 \nonumber \\
   \label{flip-quark}
\lefteqn{ \int \frac{d z^-}{4\pi}\, e^{ix P^+ z^-}
  \langle p',\lambda_N'|\, 
     \bar{\psi}_q(-{\textstyle\frac{1}{2}}z)\, i \sigma^{+i}\, 
     \psi_q({\textstyle\frac{1}{2}}z)\, 
  \,|p,\lambda_N \rangle \Big|_{z^+=0,\, \mathbf{z}_\perp=0} } 
\\
&&\hspace{2em} {}= \frac{1}{2P^+} \bar{u}(p',\lambda_N') \left[
  H_T^q\, i \sigma^{+i} +
  \tilde{H}_T^q\, \frac{P^+ \Delta^i - \Delta^+ P^i}{M_N^2} \right.
\nonumber \\
&& \left. \hspace{5em} {}+
  E_T^q\, \frac{\gamma^+ \Delta^i - \Delta^+ \gamma^i}{2M_N} +
  \tilde{E}_T^q\, \frac{\gamma^+ P^i - P^+ \gamma^i}{M_N}
  \right] u(p,\lambda_N). \nonumber
\end{eqnarray}

\begin{figure}
\begin{center}
    \epsfig{file=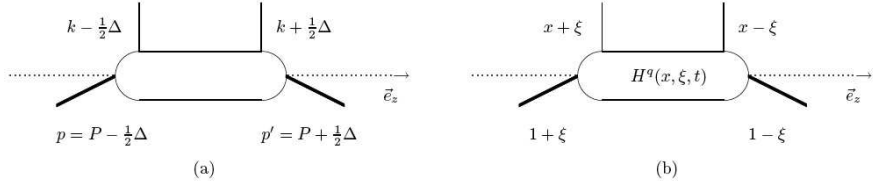, width=12cm}
\end{center}
\caption{\small (a) Kinematic variables in the symmetric frame of reference; (b) parametrization of the GPD $H^q(x,\xi,t)$ in terms of momentum fractions.}
\label{fig:kinematics}
\end{figure}

Because of Lorentz invariance the eight GPDs $H^q$, $E^q$, $\tilde{H}^q$, $\tilde{E}^q$,  $H_T^q$, $ \tilde{H}_T^q$, $E_T^q$, $\tilde{E}_T^q$ can only depend on three kinematical variables (Fig.~\ref{fig:kinematics}), i.e. the (average) quark longitudinal momentum fraction $x=k^+/P^+$, the invariant momentum square $t$ and the skewness parameter $\xi$ given by
\be
\xi = \frac{p^+-{p'}^+}{p^+ +{p'}^+}= -\frac{\Delta^+}{2P^+}.
\ee
In addition, as in the case of parton distributions, there is an implicit scale dependence in the definition of GPDs corresponding to the renormalization scale $\mu^2$, i.e. the scale at which the QCD operators in Eqs.~(\ref{no-flip-quark})-(\ref{flip-quark}) are understood to be renormalized.

The eight GPDs are all required to be real valued as a consequence of time reversal invariance, with support in the interval $x,\xi\in [-1,1]$. 

Similarly for gluons one has~\cite{Diehl01}
\begin{eqnarray}
  \label{no-flip-gluon}
\quad{}&&\lefteqn{ \frac{1}{P^+} \int \frac{d z^-}{2\pi}\, e^{ix P^+ z^-}
  \langle p',\lambda_N'|\, 
     F^{+i}(-{\textstyle\frac{1}{2}}z)\, 
     F_{i}{}^{+}({\textstyle\frac{1}{2}}z)\, 
  \,|p,\lambda_N \rangle \Big|_{z^+=0,\, \mathbf{z}_\perp=0} } 
\\
&&\quad {}=  \frac{1}{2P^+} \bar{u}(p',\lambda_N') \left[
  H^g\, \gamma^+ +
  E^g\, \frac{i \sigma^{+\alpha} \Delta_\alpha}{2M_N} 
  \right] u(p,\lambda_N) ,
\nonumber \\
\quad{}&& \lefteqn{ - \frac{i}{P^+} \int \frac{d z^-}{2\pi}\, e^{ix P^+ z^-}
  \langle p',\lambda_N'|\, 
     F^{+i}(-{\textstyle\frac{1}{2}}z)\, 
          \tilde{F}_{i}{}^{+}({\textstyle\frac{1}{2}}z)\, 
  \,|p,\lambda_N \rangle \Big|_{z^+=0,\, \mathbf{z}_\perp=0} }
\\
&&\quad {}= \frac{1}{2P^+} \bar{u}(p',\lambda_N') \left[
  \tilde{H}^g\, \gamma^+ \gamma_5 +
  \tilde{E}^g\, \frac{\gamma_5 \Delta^+}{2M_N}\, \, 
  \right] u(p,\lambda_N) ,
\nonumber\\
    \label{flip-gluon}
\quad{}&&\lefteqn{ - \frac{1}{P^+} \int \frac{d z^-}{2\pi}\, e^{ix P^+ z^-}
  \langle p',\lambda_N'|\, {\mathbf S}
     F^{+i}(-{\textstyle\frac{1}{2}}z)\,
     F^{+j}({\textstyle\frac{1}{2}}z)
  \,|p,\lambda_N \rangle \Big|_{z^+=0,\,\mathbf{z}_\perp=0} }
\\
&&\quad {}=  {\mathbf S}\,
\frac{1}{2 P^+}\, \frac{P^+ \Delta^j - \Delta^+ P^j}{2 M_N P^+}\,
\bar{u}(p',\lambda_N') \left[
  H_T^g\, i \sigma^{+i} +
  \tilde{H}_T^g\, \frac{P^+ \Delta^i - \Delta^+ P^i}{M_N^2} \right. 
\nonumber \\
&& \left. \hspace{10em} {}+
 E_T^g\, \frac{\gamma^+ \Delta^i - \Delta^+ \gamma^i}{2M_N} +
  \tilde{E}_T^g\, \frac{\gamma^+ P^i - P^+ \gamma^i}{M_N}\, 
   \right] u(p,\lambda_N) , \nonumber
\end{eqnarray}
where a summation over $i=1,2$ is implied and  ${\mathbf S}$ denotes symmetrization in $i$ and $j$ and subtraction of the trace. 

\begin{table}[b]
\caption{Quark and gluon operators, $\mathcal{O}^{q}_{\mu'\mu}$ and $\mathcal{O}^{g}_{\mu'\mu}$ respectively, for leading-twist GPDs and the parton helicity transitions they describe.  It is implied that the first field is taken at $-\oneh z$ carrying helicity $\mu'$ and the second at $\oneh z$ carrying helicity $\mu$.}
\label{tab:helicities}
\begin{center}
{\renewcommand{\arraystretch}{1.3}
\begin{tabular}{ccccccc} 
\hline
 & \multicolumn{2}{c}{$x\le -\xi$}
 & \multicolumn{2}{c}{$x<|\xi|$}
 & \multicolumn{2}{c}{$x\ge \xi$}
  \\
 $\mathcal{O}^{q}_{\mu'\mu}$ &  
 $\mu'$ & $\mu$ &  $\mu'$ & $\mu$  & $\mu'$ & $\mu$ 
\\ \hline
 $\frac{1}{4}\bar\psi_{q} \gamma^+ (1+\gamma_5) \psi_q$  
    & $-$ & $-$ & $-$ & $+$ & $+$ & $+$ \\
 $\frac{1}{4}\bar\psi_{q} \gamma^+ (1-\gamma_5) \psi_q$ 
   & $+$ & $+$ & $+$ & $-$ & $-$ & $-$\\
 $-\frac{1}{4}\,i\,\bar\psi_{q} (\sigma^{+1} - i\sigma^{+2}) \psi_q$ 
    & $+$ & $-$ & $+$ & $+$ & $-$ & $+$\\
 $\phantom{-}\frac{1}{4}\,i\,\bar\psi_{q} (\sigma^{+1} + i\sigma^{+2}) \psi_q$ 
  & $-$ & $+$ & $-$ & $-$ & $+$ & $-$\\
\hline
$\mathcal{O}^{g}_{\mu'\mu}$ & & & & & & 
\\ \hline
 $\oneh [ F^{+i}\, F_{i}{}^{+}
         -i F^{+i}\, \tilde{F}_{i}{}^{+} ]$ & $-$ & $-$ & $-$ & $+$ & $+$ & $+$
\\
 $\oneh [ F^{+i}\, F_{i}{}^{+}
         +i F^{+i}\, \tilde{F}_{i}{}^{+} ]$ & $+$ & $+$ & $+$ & $-$ & $-$ & $-$
\\
$\oneh [ F^{+1} F^{1+} - F^{+2} F^{2+}       
         -i F^{+1} F^{2+} -i F^{+2} F^{1+} ]$ & $+$ & $-$ & $+$ & $+$ & $-$ & $+$
\\
$\oneh [ F^{+1} F^{1+} - F^{+2}\, F^{2+}     
         +i F^{+1} F^{2+} +i F^{+2} F^{1+} ]$ & $-$ & $+$ & $-$ & $-$ & $+$ & $-$     
\\ \hline
\end{tabular}
}
\end{center}
\end{table}

Quark and gluon operators can be usefully rearranged in order to make explicit their action on the parton helicity, as indicated in Tab.~\ref{tab:helicities}. The corresponding matrix elements, 
\begin{eqnletter}  
\label{eq:matrices}
A^q_{\lambda_N'\mu', \lambda_N\mu} &=&
\int \frac{d z^-}{2\pi}\, e^{ix P^+ z^-}
  \bra{p',\lambda_N'}\, {\mathcal O}^{q}_{\mu',\mu}(z)
  \,\ket{p,\lambda_N} \Big|_{z^+=0,\, \mathbf{z}_\perp=0} ,
  \label{eq:matrices.r1}\\
A^g_{\lambda_N'\mu', \lambda_N\mu} &=& \frac{1}{P^+}
\int \frac{d z^-}{2\pi}\, e^{ix P^+ z^-}
  \bra{p',\lambda_N'}\, {\mathcal O}^{g}_{\mu',\mu}(z)
  \,\ket{p,\lambda_N} \Big|_{z^+=0,\, \mathbf{z}_\perp=0} ,
 \label{eq:matrices.r2}
\end{eqnletter}  
are similar to helicity amplitudes~\cite{Diehlrep}. In particular they share the same property from parity invariance,
\be
  \label{parity}
A_{-\lambda_N'-\mu', -\lambda_N-\mu} = (-1)^{\lambda_N'-\mu'-\lambda_N+\mu}\,
                                   \left[A_{\lambda_N'\mu', \lambda_N\mu}\right]^* .
\ee
Using light-cone helicity spinors explicit calculation~\cite{Diehlrep} gives the matrix elements
\begin{eqnletter}  
\label{no-flip-amplitudes}
A_{++,++} &=& \sqrt{1-\xi^2} \left( \frac{H+\tilde{H}}{2} - 
            \frac{\xi^2}{1-\xi^2}\, \frac{E+\tilde{E}}{2} \right) , 
\label{no-flip-amplitudes.r1}\\
A_{-+,-+} &=& \sqrt{1-\xi^2} \left( \frac{H-\tilde{H}}{2} - 
            \frac{\xi^2}{1-\xi^2}\, \frac{E-\tilde{E}}{2} \right) , 
\label{no-flip-amplitudes.r2} \\
A_{++,-+} &=& - e^{i \varphi}\,
              \frac{\sqrt{t_0-t}}{2M_N}\, \frac{E-\xi\tilde{E}}{2} , 
\label{no-flip-amplitudes.r3} \\
A_{-+,++} &=& e^{i \varphi}\,
              \frac{\sqrt{t_0-t}}{2M_N}\, \frac{E+\xi\tilde{E}}{2} ,
\label{no-flip-amplitudes.r4}
\end{eqnletter}
for both quarks and gluons, where $\varphi$ is the azimuthal angle of the vector  $D= p'/(1-\xi) - p/(1+\xi)$, i.e. $e^{i \varphi}= (D^1+iD^2)/\vert\tvec D\vert$. Therefore, the four GPDs $H$, $E$, $\tilde{H}$, $\tilde{E}$ are parton helicity conserving and chiral even. The distributions $H$ and $E$ are
sometimes referred to as unpolarized and $\tilde{H}$ and $\tilde{E}$ as polarized because $H$ and $E$
correspond to the sum over parton helicities, and $\tilde{H}$ and $\tilde{E}$ to the difference.

In the case of parton-helicity flip, for quarks one obtains
\begin{eqnletter}
\label{flip-amplitudes}
A^q_{++,+-} &=&   e^{i \varphi}\, \frac{\sqrt{t_0-t}}{2M_N} \left( \tilde{H}_T^q
              + (1-\xi)\, \frac{E_T^q + \tilde{E}_T^q}{2} \right) , 
\label{flip-amplitudes.r1} \\
A^q_{-+,--} &=&   e^{i \varphi}\, \frac{\sqrt{t_0-t}}{2M_N} \left( \tilde{H}_T^q
              + (1+\xi)\, \frac{E_T^q - \tilde{E}_T^q}{2} \right) , 
\label{flip-amplitudes.r2} \\
A^q_{++,--} &=& \sqrt{1-\xi^2} \left(H_T^q + \
              \frac{t_0-t}{4 M_N^2}\, \tilde{H}_T^q -
              \frac{\xi^2}{1-\xi^2}\, E_T^q +
              \frac{\xi}{1-\xi^2}\, \tilde{E}_T^q \right) , 
\label{flip-amplitudes.r3} \\
A^q_{-+,+-} &=& -  e^{2i \varphi}\, \sqrt{1-\xi^2}\; \frac{t_0-t}{4 M_N^2}\, \tilde{H}_T^q,
\label{flip-amplitudes.r4}
\end{eqnletter}
where
\be
\label{eq:minimalt}
- t_0 = \frac{4{\xi}^2M_N^2}{1-\xi^2}
\label{eq:tmin}
\ee
is the minimal value for $-t$ at given $\xi$. Analogous expressions hold for gluons with an additional global factor $e^{i\varphi} \, \sqrt{1-\xi^2}\, \sqrt{t_0-t} /(2M_N)$ on the right-hand side. Other helicity combinations are given by parity invariance. Therefore, the four GPDs $H_T$, $ \tilde{H}_T$, $E_T$, $\tilde{E}_T$ are parton helicity flipping and chiral odd.

By inverting the set of Eqs.~(\ref{no-flip-amplitudes}), (\ref{flip-amplitudes}) the different GPDs can be separately extracted from the amplitudes $A_{\lambda_N'\mu', \lambda_N\mu}$.


\subsection{Forward limit}

In the forward case, $p=p'$, both $\Delta$ and $\xi$ are zero. As $\xi\to 0$, also $x\to x_B$, where the Bjorken variable $x_B$ is the fraction of the longitudinal momentum carried by the active parton. In this case the  functions $H^q$, $\tilde H^q$ and $H_T^q$ reduce to the usual DIS parton distribution functions, i.e.
\bea
\quad\ H^q(x_B,0,0) &=& \left\{
\begin{array}{rr}
q(x_B), & x_B>0,\\
-\bar q(-x_B), & x_B<0,\\
\end{array}\right. \\ 
\tilde H^q(x_B,0,0) &=& \left\{
\begin{array}{rr}
\Delta q(x_B), & x_B>0, \\
\Delta\bar q(-x_B), & x_B<0,\\
\end{array}\right.\\
\quad\ H_T^q(x_B,0,0) &=& \left\{
\begin{array}{rr}
\Delta_T q(x_B), & x_B>0,\\
-\Delta_T\bar q(-x_B), & x_B<0,\\
\end{array}\right.
\eea
where $q(x_B)$, ($\bar q(x_B)$), $\Delta q(x_B)$ ($\Delta\bar q(x_B)$)  and $\Delta_T q(x_B)$ ($\Delta_T\bar q(x_B)$) are quark (antiquark) density, helicity and transversity distributions, respectively: 
\bea
\label{eq:qdix}
q(x_B) &=& \left. \int  \frac{dz^-}{4\pi}\, e^{ix_B p^+z^-}
\bra{p}\bar\psi(0)\,\gamma^+\,\psi(z)\ket{p}
\right\vert_{z^+={\svec z}_\perp=0},
\\
\label{eq:deltaqdix}
\Delta q(x_B) &=& \left. \int  \frac{dz^-}{4\pi}\, e^{ix_B p^+z^-}
\bra{p S_\parallel}\bar\psi(0)\,\gamma^+\gamma_5\,\psi(z)\ket{pS_\parallel} 
\right\vert_{z^+={\svec z}_\perp=0},
\\
\label{eq:deltatqdix}
\Delta_T q(x_B) &=& \left. \int  \frac{dz^-}{4\pi}\, e^{ix_B p^+z^-}
\bra{p S_\perp}\bar\psi(0)\,\gamma^+\gamma^1\gamma_5\,\psi(z)\ket{pS_ \perp} 
\right\vert_{z^+={\svec z}_\perp=0},
\eea
with $S_\parallel$ ($S_\perp$) being the longitudinal (transverse) nucleon-spin projection. 

No corresponding relations exist for the functions $E^q$, $\tilde E^q$, $E_T^q$ and $\tilde H_T^q$, because in the forward limit they decouple in their defining equations. However, they do not vanish. In particular, $E^q(x,0,0)$ carries important information about the quark orbital angular momentum (see. Eq.~(\ref{eq:aom})). In contrast, $\tilde E_T^q$ vanishes identically being an odd function of $\xi$ by time reversal symmetry~\cite{Diehl01}.

For gluons one has
\be
H^g(x_B,0,0) = x_B\,g(x_B), \qquad \tilde H^g(x_B,0,0) = x_B\,\Delta g(x_B), \quad x_B>0 .
\ee
All the gluon helicity-flip matrix elements go to zero in the forward case corresponding to the collinear limit ($t=t_0$ in Eqs.~(\ref{flip-amplitudes})) and therefore for a spin-$\oneh$ target decouple from any observable for collinear scattering.


\subsection{Polynomiality}

Introducing a light-like vector $n^\mu$, which is conjugate to $P^\mu$ in the sense that $P\cdot n=1$, and contracting both sides of Eq.~(\ref{eq:form}) with $n_\mu n_{\mu_1}\cdots n_{\mu_{n-1}}$, one obtains
\bea
\label{eq:oenne}
&& n_{\mu}n_{\mu_1}\cdots n_{\mu_{n-1}} 
\langle p'| {\mathcal O}^{\mu\mu_1\cdots \mu_{n-1}}_q |p\rangle
\\
&&\quad{}  =   {\bar u}(p') \,\slash{n}\, u(p) \, H^q_{n}(\xi, t)
  + {\bar u}(p')\,\frac{\sigma^{\mu\alpha} n_\mu i\Delta_\alpha} {2M_N}\,u(p)\, E^q_{n}(\xi, t) , 
  \nonumber
\eea
where 
\begin{eqnarray}
   H^q_{n} (\xi,t) &= & \sum_{i=0 \atop \scriptstyle{\rm even}}^{n-1}
     A^q_{n,i}(t) \,(2\xi)^{i} + {\rm Mod}(n+1,2)~ C^q_{n}(t) (2\xi)^{n} , 
    \label{eq:polyh} \\
    E^q_{n} (\xi,t) &= & \sum_{i=0 \atop \scriptstyle{\rm even}}^{n-1}
     B^q_{n,i}(t) \,(2\xi)^{i} - {\rm Mod}(n+1,2)~ C^q_{n}(t) (2\xi)^{n}  .  
     \label{eq:polye}
\end{eqnarray}
Recalling the relation (\ref{eq:higher-moments}) the functions $H^q_{n} (\xi,t)$ and $E^q_{n} (\xi,t)$ are easily recognized to be the $n$-th Mellin moment of the GPDs $H^q(x, \xi, t)$ and $E^q(x, \xi,t)$, respectively,  i.e.
\begin{eqnarray}
\label{eq:accaqn}
     \int^1_{-1} dx \,x^{n-1} H^q(x,\xi,t) &=&H^q_{n}(\xi, t) , \\
     \int^1_{-1} dx \,x^{n-1} E^q(x,\xi,t) &=&E^q_{n}(\xi, t)  . 
\label{eq:eqn}
\end{eqnarray}   
Eqs.~(\ref{eq:polyh}) and (\ref{eq:polye}) show that the $n$-th Mellin moments of the GPDs $H^q(x, \xi, t)$ and $E^q(x, \xi,t)$ are polynomial functions of $\xi$ with highest power $n$. This is called the  polynomiality condition. In QCD the polynomiality property of GPDs follows from their hermiticity and Lorentz, parity and time-reversal invariance~\cite{Ji98a}. 

An alternative notation is often used instead of (\ref{eq:polyh}) and (\ref{eq:polye})~\cite{GPVdH}, i.e.
\bea
\qquad{}&&
    \int_{-1}^1\! d x\:x^{n-1}\,H^q(x,\xi,t)
    = h^{(n)}_0(t) + h^{(n)}_2(t)\, \xi^2 + \dots +
       \cases{h^{(n)}_n(t)\, \xi^n    \!\!\!\!\! & for $n$ even\cr
           h^{(n)}_{n-1}(t)\, \xi^{n-1}\!\!\!\!\! & for $n$ odd,} 
\label{eq:polynom-Hq}\\
&&   
    \int_{-1}^1\! d x\:x^{n-1}\,E^q(x,\xi,t)
    = e^{(n)}_0(t) + e^{(n)}_2(t)\, \xi^2 + \dots +
       \cases{e^{(n)}_n(t)\, \xi^n    \!\!\! & for $n$ even\cr
           e^{(n)}_{n-1}(t)\, \xi^{n-1}\!\!\! & for $n$ odd,}
\label{eq:polynom-Eq} 
\eea
where flavour indices are suppressed for brevity. For a spin-$\oneh$ particle the coefficients in front of the highest power in $\xi$ for even $n$ are related to each other and arise from the so-called $D$-term $D^q(z,t)$ with $z=x/\xi$ \cite{Polyakov99,Teryaev01}, which has finite support only for $|x|<|\xi|$, according to
\be
    h^{q\,(n)}_N(t)=-\,e^{q\,(n)}_N(t)=\int_{-1}^1 dz\; z^{n-1}D^q(z,t).
\label{Eq:relation-h-e}
\ee

Similar polynomiality conditions can be derived for the other twist-two operators. For example, in Eq.~(\ref{paraav}) time reversal invariance constrains $\bar{u}(p') \gamma_5 u(p)$
to come with odd powers of $\Delta^\alpha$.  One then has the following sum rules~\cite{Diehlrep}
\begin{eqnarray}
\label{eq:tilde-h}
\int_{-1}^1 dx\, x^{n-1} \tilde{H}^q(x,\xi,t) &=&
  \sum_{i=0 \atop \scriptstyle{\rm even}}^{n-1} (2\xi)^i 
        \tilde{A}^q_{n,i}(t) ,
\\
\label{eq:tilde-e}
\int_{-1}^1 dx\, x^{n-1} \tilde{E}^q(x,\xi,t) &=&
  \sum_{i=0 \atop \scriptstyle{\rm even}}^{n-1} (2\xi)^i 
        \tilde{B}^q_{n,i}(t) ,
\end{eqnarray}
where the highest power in $\xi$ is $n-1$ instead of $n$.  

For chiral-odd GPDs one has
\begin{eqnarray}
\label{eq:chiral-h}
\int_{-1}^1 dx\, x^{n-1} H^q_T(x,\xi,t) &=&
  \sum_{i=0 \atop \scriptstyle{\rm even}}^{n-1} (2\xi)^i 
        A^q_{Tn,i}(t) ,
\\
\label{eq:chiral-e}
\int_{-1}^1 dx\, x^{n-1}{E}^q_T(x,\xi,t) &=&
  \sum_{i=0 \atop \scriptstyle{\rm even}}^{n-1} (2\xi)^i 
        B^q_{Tn,i}(t) ,\\
\label{eq:chiral-htilde}
\int_{-1}^1 dx\, x^{n-1} \tilde{H}^q_T(x,\xi,t) &=&
  \sum_{i=0 \atop \scriptstyle{\rm even}}^{n-1} (2\xi)^i 
        \tilde{A}^q_{Tn,i}(t) ,
\\
\label{eq:chiral-etilde}
\int_{-1}^1 dx\, x^{n-1} \tilde{E}^q_T(x,\xi,t) &=&-
  \sum_{i=0 \atop \scriptstyle{\rm odd}}^{n-1} (2\xi)^i 
        \tilde{B}^q_{Tn,i}(t) .
\end{eqnarray}

The polynomiality property has been shown to lead to integral relations of the form
\bea
& & \int_{-1}^1 dx\, F(x,\xi,t) \left[ \frac{1}{\omega\xi -x} -\sigma\frac{1}{\omega\xi +x} \right] \\
& & \quad{}=\int_{-1}^1 dx\, F(x,\frac{\xi}{\omega},t) \left[ \frac{1}{\omega\xi -x} -\sigma\frac{1}{\omega\xi +x} \right] + {\cal I}(\omega,t) \nonumber
\eea
for $\sigma=\pm 1$ and any $\omega\ge 1$, where $F$ is one of the GPDs~\cite{DI07}. The only cases with nonzero ${\cal I}(\omega,t)$ occur for unpolarized distributions ($F=H^q,E^q$) and $\sigma=+ 1$, where ${\cal I}(\omega,t)$ takes contribution from the $D$-term, i.e.
\be
{}\pm {\cal I}(\omega,t) = 2\int_{-1}^1 dx\frac{D^q(x,t)}{\omega-x} ,
\ee
with the sign $+$ ($-$) in the l.h.s. to be taken for $H^q$ ($E^q$).
Thus polynomiality ensures the possibility of writing dispersion relations for the GPDs encoding causality and analyticity. In order to account for divergent contributions of both valence and sea quarks fixed-$t$ subtracted dispersion relations have been proposed~\cite{AT07} with  the subtraction constant defined by the $D$-term (see also~\cite{Polyakov07}).


\subsection{Sum rules}

Relevant sum rules are obtained by forming the first Mellin moment of quark GPDs. For $n=1$ in Eq.~(\ref{eq:oenne}), from (\ref{eq:accaqn}) and (\ref{eq:eqn}) one identifies $H^q_{1}$ and $E^q_{1}$ (or $h_0^{(1)}$ and $e_0^{(1)}$ in the polynomial expansions (\ref{eq:polynom-Hq}) and (\ref{eq:polynom-Eq})) with the Dirac and Pauli form factors appearing in Eq.~(\ref{eq:emff}), i.e.
\be
\label{eq:emformfactors}
\int_{-1}^1 dx \, H^q(x,\xi,t) = F_1^q(t), \qquad \int_{-1}^1 dx \, E^q(x,\xi,t) = F_2^q(t).
\ee
As a consequence of Lorentz invariance, correctly the integrals only depend on $t$ because integrating over $x$ removes all reference to the particular light-cone direction with respect to which $\xi$ is defined, so that the result must be independent of $\xi$. At $t=0$ the quark form factors $F_1^u$ and $F_1^d$ are normalized so as to yield the normalization 2 for the up quark distribution in the proton and 1 for the down quark distribution, i.e. $F_1^u(0)=2$, $F_1^d(0)=1$. The quark form factors $F_2^u$ and $F_2^d$ are normalized through the proton and neutron anomalous magnetic moments $\kappa^p=1.793$ and $\kappa^n=-1.913$, i.e. $F_2^u(0)=\kappa^u=2\kappa^p+\kappa^n=1.673$, $F_2^d(0)=\kappa^d=\kappa^p+2\kappa^n=-2.033$.

Similarly, one has
\be
\int_{-1}^1 dx \, \tilde H^q(x,\xi,t) = g_A^q(t), \qquad \int_{-1}^1 dx \, \tilde E^q(x,\xi,t) = g_P^q(t),
\ee
where $g_A^q$ and $g_P^q$ are the axial and induced pseudoscalar form factors of Eq.~(\ref{eq:weakff}). The normalization of $g_A^q$ and $g_P^q$ at $t=0$ can be derived using the isospin decomposition~(\ref{eq:ga0}) and the corresponding one for $g_P^q$.

There are also tensor form factors associated with the first moments of $H_T$, $\tilde H_T$ and $E_T$, whereas
\be
\int_{-1}^1 dx \, \tilde E_T^q(x,\xi,t) = 0.
\ee
In particular
\be
\label{eq:kappaT}
\int_{-1}^1 dx \left[ 2\tilde H^q_T(x,0,0) + E^q_T(x,0,0)\right] = \kappa^q_T,
\ee
where $\kappa^q_T$ describes how far and in which direction the average position of quarks with spin in the $\hat x$-direction is shifted in the $\hat y$-direction in an unpolarized nucleon~\cite{Burkardt05b}. Thus $\kappa^q_T$ governs the spin-flavour dipole moment in an unpolarized nucleon and plays a role similar to the anomalous magnetic moment $\kappa^q$ for unpolarized quarks in a transversely polarized nucleon.

The second Mellin moment of GPDs is relevant for the spin structure of the nucleon. The angular momentum operator in QCD contains quark and gluon spin and orbital angular momentum~\cite{JM90} with an important role of the orbital angular momentum in the formation of the total spin of the nucleon~\cite{Sehgal74,Ratcliffe87}. It can be written in gauge-invariant form as $\tvec J = \sum_q\tvec J_q + \tvec J_g$, with
\be
J^i_{q,g} = \oneh \epsilon^{ijk} \int d^3x\, \left[ {\mit\Theta}_{q,g}^{0k}x^j  - {\mit\Theta}_{q,g}^{0j}x^k\right] ,
\ee
where
\be
{\mit\Theta}_q^{\mu\nu} = \bar\psi_q\, \gamma^{(\mu} i\stackrel{\leftrightarrow}{\cal D}{}^{\nu)}\psi_q ,
\quad{\mit\Theta}^{\mu\nu}_g = F^{\rho(\mu}  F^{\ \nu)}_{\rho}
\ee
are the gauge-invariant quark and gluon parts of the (Belinfante improved) symmetric energy-momentum (stress) tensor. They are twist-two operators of the type ${\mathcal O}_q^{\mu\nu}$ and ${\mathcal O}_g^{\mu\nu}$ discussed in section 2.1, and their matrix elements can be expanded in terms of generalized form factors:
\bea
  \label{eq:emt}
 \quad{}    \bra{p'} {\mit\Theta}_{q,g}^{\mu\nu} (0) \ket{p}
    &=& \bar u(p')\biggl[
    A^{q,g}_{2,0}(t)\,\frac{\gamma^\mu P^\nu+\gamma^\nu P^\mu}{2}+
    B^{q,g}_{2,0}(t)\,\frac{i(P^{\mu}\sigma^{\nu\rho}+P^{\nu}\sigma^{\mu\rho})\Delta_\rho}{4M_N}
    \\
     &&\qquad{}+ C^{q,g}_{2}(t)\,\frac{\Delta^\mu\Delta^\nu-g^{\mu\nu}\Delta^2}{M_N}
    \pm \bar c(t) g^{\mu\nu} \biggr]u(p)\, ,
    \nonumber
\eea
where the form factor $\bar c(t)$ enters the quark and gluon parts with opposite signs in order to account for conservation of the total energy-momentum tensor. Alternatively, by means of the Gordon identity
\be
2M_N \bar u(p')\gamma^\alpha u(p)= \bar u(p')(i\sigma^{\alpha\kappa}\Delta_\kappa+2P^\alpha)u(p),
\ee
Eq.~(\ref{eq:emt}) can be rewritten as~\cite{GPVdH,GGOPSSU07}
\bea
 \label{eq:emtvar} 
     \bra{p'} {\mit\Theta}^{\mu\nu}_{q,g}(0) \ket{p}
    &=& \bar u(p')\left[M_2^{q,g}(t)\,\frac{P^\mu P^\nu}{M_N}+
    J^{q,g}(t)\ \frac{i(P^{\mu}\sigma^{\nu\rho}+P^{\nu}\sigma^{\mu\rho})
    \Delta_\rho}{2M_N}\right.
    \\
    &  &\qquad\quad  {} + d^{q,g}_1(t)\,\left.
    \frac{\Delta^\mu\Delta^\nu-g^{\mu\nu}\Delta^2}{5M_N}
    \pm \bar c(t)\,g^{\mu\nu} \right]u(p)\, ,
\nonumber
\eea
with
\begin{eqnletter}
\label{eq:emtalt}
    A^{q,g}_{2,0}(t)   &=& M_2^{q,g}(t), 
    \label{eq:emtalt.r1} \\
    A^{q,g}_{2,0}(t)+B^{q,g}_{2,0}(t) &=& 2\,J^{q,g}(t), 
    \label{eq:emtalt.r2} \\
    C_{q,g}(t)   &=& {\textstyle {1\over 5}}\,d_1^{q,g}(t)\,.
    \label{eq:emtalt.r3}
\end{eqnletter}

From Eq.~(\ref{eq:emt}) one immediately has the following sum rules for the GPDs:
\bea
\int dx\, x\, \left[H^{q}(x,\xi,t) + E^{q}(x,\xi,t)\right] =A^{q}_{2,0}(t) + B^{q}_{2,0}(t) , \\
\int dx\, \left[H^{g}(x,\xi,t) + E^{g}(x,\xi,t)\right] =A^{g}_{2,0}(t) + B^{g}_{2,0}(t) , 
\eea
where the $\xi$ dependence drops out eliminating the contribution of $C_{q,g}(t)$, and the different power of $x$ in the integrals reflects the different forward limit of quark and  gluon distributions. Extrapolating the sum rules to $t=0$, the angular momentum carried by each parton species is found to be~\cite{Ji97,Ji97a}
\be
\label{eq:jisumrule}
\langle J^3_{q}\rangle = \oneh \left[A^{q}_{2,0}(0) + B^{q}_{2,0}(0)\right], \quad
\langle J^3_{g}\rangle = \oneh \left[A^{g}_{2,0}(0) + B^{g}_{2,0}(0)\right].
\ee 
Since $A^{q,g}_{2,0}(0)$ give the momentum fractions of the nucleon carried by quarks and gluons, 
\be
\int dx\, \left[ x\,\sum_q H^q(x,0,0) + H^g(x,0,0) \right ] = \sum_q A^{q}_{2,0}(0) + A^{g}_{2,0}(0) =1,
\ee 
the total $B(0)=\sum_q B^{q}_{2,0}(0)+ B^{g}_{2,0}(0)$ vanishes for any composite system~\cite{Ji98,BHMS01}. Then, for the total angular momentum of a proton moving in the $\hat z$-direction and polarized in the helicity eigenstate $\lambda_N=+1$ one recovers the value $\langle J^3\rangle=\sum_q\langle  J^3_{q}\rangle+ \langle J^3_{g}\rangle=\oneh$. In addition, according to an extension of the equivalence principle of general relativity to describe the interaction of the nucleon with the external gravitational field one arrives to the interpretation of $B(0)$ as an anomalous gravitomagnetic moment being the analog of the anomalous magnetic moment~\cite{teryaev}. There is also evidence supporting the conjecture that the equivalence principle is valid separately for quarks and gluons resulting in exact equipartition of momenta and angular momenta in the nucleon. The most precise numerical support comes from lattice calculations~\cite{Hagler07}.

The sum rules (\ref{eq:jisumrule}), known as Ji's sum rules, are renormalization scale dependent. By solving the combined evolution equations for quark and gluon helicity and orbital angular momentum one finds in the asymptotic limit~\cite{JiTH96}
\be
\langle J^3_{q}\rangle = \frac{1}{2} \frac{3n_f}{16 + 3n_f}, \quad 
\langle J^3_{g}\rangle = \frac{1}{2} \frac{16}{16 + 3n_f},
\ee
where $n_f$ is the number of flavours. Thus the partition of the nucleon spin between quarks and gluons follows the well-known partition of the nucleon momentum~\cite{GW74} when it is probed at infinitely small distance scale.

It would be desirable to further split the Ji's sum rules (\ref{eq:jisumrule}) into spin and orbital angular momentum parts as~\cite{JM90}
\be
\label{eq:jaffe}
\oneh = \sum_q\left[\oneh\Sigma^q + L_q\right] + \Delta g + L_g,
\ee
where 
\be
\Sigma^q=\int_{-1}^1 dx\,\Delta q(x), \quad L_q = \int_{-1}^1 dx\, L_q(x)
\ee
are the quark spin and orbital angular momentum expressed in terms of the helicity distribution $\Delta q(x)=\tilde H^q(x,0,0)$ and the orbital angular momentum density $L_q(x) $. Analogous definitions hold for the gluon spin and orbital angular momentum, $\Delta g$ and $L_g$, respectively. However, the matrix elements of the gluon spin are not invariant under general gauge transformations~\cite{Hoodbhoy99b}. One can only achieve a gauge-invariant separation of the quark orbital angular momentum~\cite{Hoodbhoy99}:
\be
\label{eq:aom}
L_q(x) = \oneh x\,\left[ H^q(x,0,0) + E^q(x,0,0)\right] - \oneh\tilde H^q(x,0,0).
\ee

When dealing with transverse total angular momentum, since there is no gluon spin contribution one can establish the transverse spin sum rule as follows~\cite{Bakker04}:
\be
\oneh = \oneh \sum_{q,\bar q} \int_0^1 dx\,\Delta_T q(x) + \sum_{q,\bar q,g} \langle L_\perp\rangle ,
\ee
where $\Delta_T q(x)=H_T^q(x,0,0)$ is the transversity distribution and $L_\perp$ is the component of the angular momentum along the proton spin, polarized perpendicular to the proton direction of motion.

The quark angular momentum can be further decomposed with respect to quarks of definite transversity making use of the second Mellin moments $A^q_{T2,0}$, $\tilde A^q_{T2,0}$ and $B^q_{T2,0}$ of the chiral-odd GPDs $H^q_T$, $\tilde H^q_T$ and $E^q_T$, respectively. If $J^x_{q,+\hat x}$ is the angular momentum carried by quarks with transverse polarization in the $+\hat x$-direction in a proton polarized along $\hat x$, the expectation value of the transverse asymmetry $\delta^x J^x_q = J^x_{q,+\hat x} - J^x_{q,-\hat x}$ is given by~\cite{Burkardt05b}
\be
\label{eq:deltajx}
\langle \delta^x J^x_q\rangle = \oneh \left[ A^q_{T2,0}(0) + 2 \tilde A^q_{T2,0}(0) + B^q_{T2,0}(0)\right].
\ee


\subsection{Positivity}

Positivity bounds have extensively been studied in~\cite{Pire99,Pobylitsa1,Pobylitsa2}, where one can also  find reference to previous work. They are all based on the positivity of the norm in the Hilbert space of states. There is a hierarchy of inequalities relating GPDs to the three quark distributions, i.e. the unpolarized, polarized and transversity distributions: $i$) strong inequalities where GPDs are bounded by combinations of all three distributions; $ii$) weaker inequalities without the transversity distribution; $iii$) still weaker inequalities where GPDs are bounded only by the unpolarized distribution.

In particular, for the region $\vert\xi\vert<\vert x\vert$ one has~\cite{Pobylitsa1}
\be
\label{eq:ineq1}
\quad{}\left[ H^q(x,\xi,t) - \frac{\xi^2}{1-\xi^2} E^q(x,\xi,t) \right]^2 +
\left[ \frac{\sqrt{t_0-t}}{2M_N\sqrt{1-\xi^2}} E^q(x,\xi,t) \right]^2 \le \frac{q(x_1)q(x_2)}{1-\xi^2}.
\ee
The function $q(x)$ is the usual (forward) distribution function for unpolarized quarks of flavour $q$ taken at values
\be
x_1 = \frac{x+\xi}{1+\xi}, \quad x_2 = \frac{x-\xi}{1-\xi}.
\ee
The bound (\ref{eq:ineq1}) is stronger than the following inequality derived in~\cite{DFKJ01}:
\be
\left\vert H^q(x,\xi,t) - \frac{\xi^2}{1-\xi^2} E^q(x,\xi,t) \right\vert \le \sqrt{\frac{q(x_1)q(x_2)}{1-\xi^2}}.
\ee
From (\ref{eq:ineq1}) one also finds the bound for $E^q$ derived earlier in~\cite{DFKJ01} up to a typo there:
\be
\left\vert E^q(x,\xi,t)\right\vert \le \frac{2M_N}{\sqrt{t_0-t}} \sqrt{q(x_1)q(x_2)}.
\ee
Similarly,
\be
\left\vert \tilde E^q(x,\xi,t)\right\vert \le \frac{2M_N}{\xi\sqrt{t_0-t}}\sqrt{q(x_1)q(x_2)}
\ee
and
\be
\left\vert \tilde H^q(x,\xi,t)\right\vert \le \sqrt{ \frac{-t}{{t_0-t}}\frac{q(x_1)q(x_2)}{1-\xi^2} } .
\ee


\section{Physical content of generalized parton distributions}
\label{sect:phys}


\subsection{Parton interpretation}

The physical content of the bilocal operators entering the definition of quark GPDs becomes transparent in light-cone coordinates decomposing the Dirac field $\psi_q$ into the sum of the  `good' and `bad' components corresponding to the independent and dependent degrees of freedom, i.e. $\psi_{q,+}=P_+\psi_q$ and $\psi_{q,-}=P_-\psi_q$, with projection operators $P_\pm =\oneh\gamma^\mp\gamma^\pm$.

\begin{figure}
\begin{center}
\epsfig{file=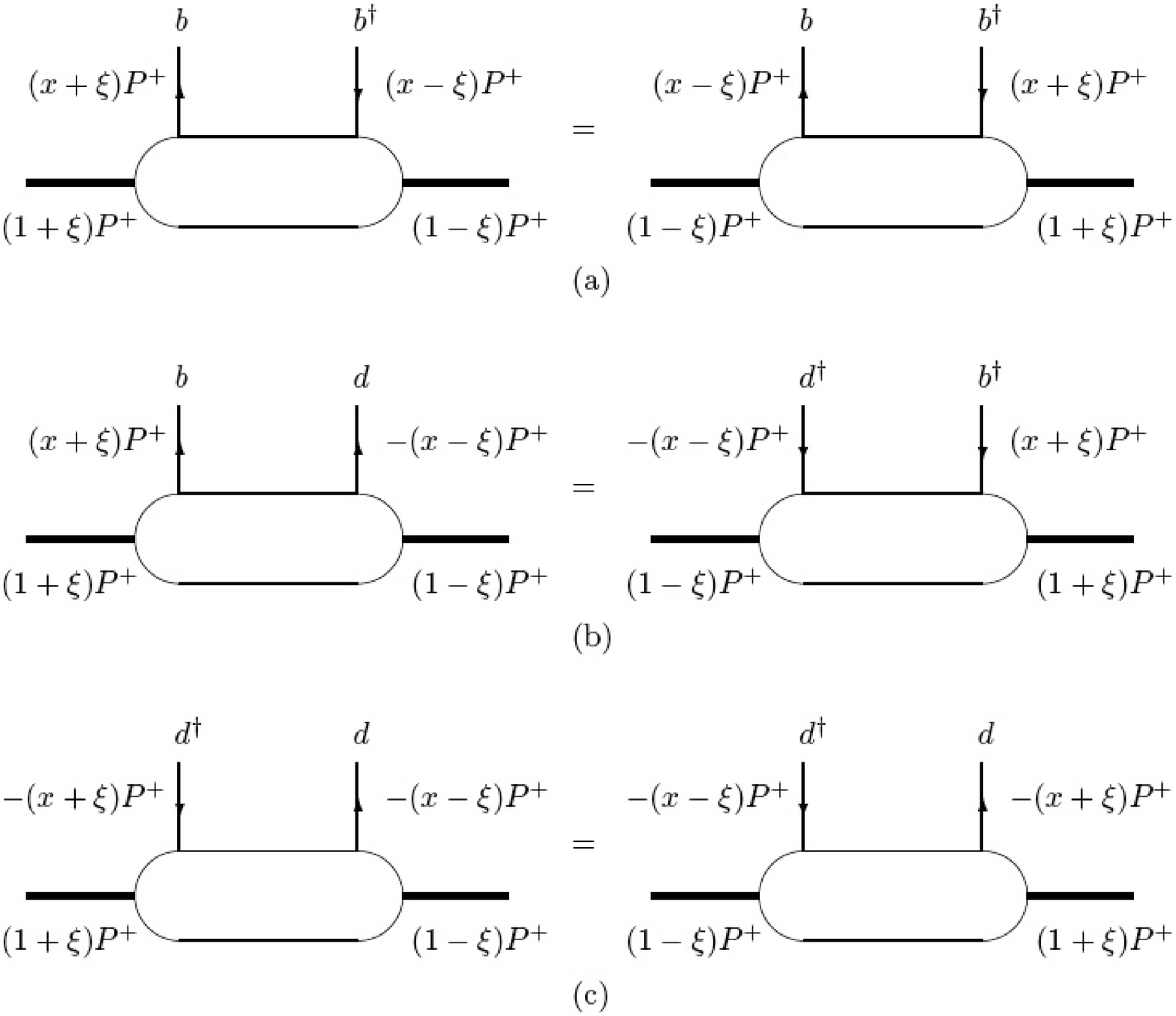, width=11cm}
\end{center}
\caption{\small Diagrams showing $\xi$ symmetry of GPDs: a) $x>\xi$, b) $-\xi< x<\xi$, c) $x<-\xi$.}
\label{fig:qqbar}
\end{figure}

The `good' fields have the following Fourier expansion in momentum space~\cite{BL89}
\bea
\label{eq:momexp}
\quad\psi_{q,+}(z^-,{\tvec z}_\perp) =
\int \frac{dk^+d{\tvec k}_\perp}{2k^+ (2\pi)^3} \Theta(k^+)
&& \sum_\mu\left\{ b_{q\mu}(k^+, {\tvec k}_\perp) u_+(k,\mu)
\,e^{-ik^+z^- + i {\svec k}_\perp\cdot{\svec z}_\perp}
\right.\\
&&  \quad\left. + d^\dagger_{q\mu}(k^+, {\tvec k}_\perp) v_+(k,\mu)
\,e^{+ik^+z^- - i {\svec k}_\perp\cdot{\svec z}_\perp}\right\},
 \nonumber
\eea
and similarly for $\bar\psi_{q,+}$. The spinors $u_+(k,\mu) = P_+u(k,\mu)$ and $v_+(k,\mu) = P_+v(k,\mu)$ are the projections of the usual quark and antiquark spinors with helicity $\mu$. The quark (antiquark) creation and annihilation operators $b^\dagger$ ($d^\dagger$) and $b$ ($d$) obey the anticommutation relations
\bea
\label{eq:anticommute}
\{b_{q'\mu'}({k'}^+, {\tvec k}'_\perp),b^\dagger_{q\mu}(k^+, {\tvec k}_\perp)\} &=& \{d_{q'\mu'}({k'}^+, {\tvec k}'_\perp),d^\dagger_{q\mu}(k^+, {\tvec k}_\perp)\} \\
&=& 2k^+(2\pi)^3\,\delta({k'}^+-k^+)\,\delta({\tvec k}'_\perp -{\tvec k}_\perp)\,
\delta_{q'q}\,\delta_{\mu'\mu}. \nonumber
\eea
Then, for example, the bilocal quark field operator $\bar\psi_q(-\oneh z)\gamma^+\psi_q(\oneh z)$ can be written as a density operator in terms of `good' light-cone components, i.e. 
$\sqrt{2}\psi^\dagger_{q,+}(-\oneh z)\psi_{q,+}(\oneh z),$ and the operator on the left-hand side of Eq.~(\ref{no-flip-quark}) becomes~\cite{DFKJ01}
\bea
\label{eq:ottosei}
&&\int \frac{dz^-}{4\pi}\, e^{ix P^+z^-}
\bar\psi_q(-\oneh z)\,\gamma^+\,\psi(\oneh z)
\\
&&\qquad{} =
\sqrt{2}\int \frac{d{k'}^+d{\tvec k}'_\perp}{2{k'}^+ (2\pi)^3} \Theta({k'}^+)
\int\frac{dk^+d{\tvec k}_\perp}{2k^+ (2\pi)^3} \Theta(k^+)
\nonumber \\
&& \qquad\quad\times{}\sum_{\mu,\mu'}\left\{
\delta(2xP^+ - {k'}^+ - k^+)\,b^\dagger_{q\mu'}({k'}^+, {\tvec k}'_\perp)b_{q\mu}({k}^+, {\tvec k}_\perp)\,
 u_+^\dagger(k',\mu')u_+(k,\mu)
\right.\nonumber\\
&& \qquad\qquad{} + 
\delta(2xP^+ + {k'}^+ + k^+)\,d_{q\mu'}({k'}^+, {\tvec k}'_\perp)d^\dagger_{q\mu}({k}^+, {\tvec k}_\perp)\,
 v_+^\dagger(k',\mu')v_+(k,\mu)
\nonumber\\
&& \qquad\qquad{} + 
\delta(2xP^+ + {k'}^+ - k^+)\,d_{q\mu'}({k'}^+, {\tvec k}'_\perp)b_{q\mu}({k}^+, {\tvec k}_\perp)\, v_+^\dagger(k',\mu')u_+(k,\mu)
\nonumber\\
&& \qquad\qquad{} + \left.
\delta(2xP^+ - {k'}^+ + k^+)\,b^\dagger_{q\mu'}({k'}^+, {\tvec k}'_\perp)d^\dagger_{q\mu}({k}^+, {\tvec k}_\perp) \,
u_+^\dagger(k',\mu')v_+(k,\mu)\right\}.
 \nonumber
\eea
Which of the four terms in (\ref{eq:ottosei}) contributes is determined by the positivity conditions $k^+\ge 0$ and ${k'}^+\ge 0$ for the parton momenta, together with momentum conservation, which imposes $k^+-{k'}^+=p^+-{p'}^+=2\xi P^+$. Assuming $\xi>0$, the interval $x\in[-1,1]$ falls into three regions according to whether $\vert x\vert>\xi$ or $\vert x\vert<\xi$:
\begin{enumerate}
\item for $x\in[\xi,1]$ both momentum fractions $x+\xi$ and $x-\xi$ are positive. The corresponding GPD describes emission and reabsorption of a quark;
\item for $x\in[-\xi,\xi]$ one has $x+\xi\ge 0$ and $x-\xi\le 0$. The second momentum fraction can be interpreted as belonging to an antiquark with momentum fraction $\xi- x$ emitted from the initial proton;
\item for $x\in[-1,-\xi]$ both $x+\xi$ and $x-\xi$ are negative. The corresponding GPD describes emission and reabsorption of antiquarks with momentum fractions $\xi-x$ and $-\xi-x$, respectively.
\end{enumerate}
The three situations are illustrated by the diagrams on the left-hand side of Fig.~\ref{fig:qqbar}. By assuming $\xi<0$ a similar analysis gives the diagrams on the right-hand side of the same figure~\cite{Ji98a}.

Diagrams (a) and (c) in Fig.~\ref{fig:qqbar}, arising from the $b^\dagger b$ and $d^\dagger d$ terms in Eq.~(\ref{eq:ottosei}), thus generalize the situation illustrated for the ordinary parton distributions. In these regions GPDs will evolve according to modified DGLAP equations~\cite{dglap}. Therefore these are called DGLAP regions. Diagram (b) in Fig.~\ref{fig:qqbar}, corresponding to the middle region, $-\xi< x<\xi$, does not have a counterpart in parton distributions. This diagram arises from the $db$ term in Eq.~(\ref{eq:ottosei}) and corresponds to the emission of a quark-antiquark pair with momentum $-\Delta$. The middle region is thus similar to that in a meson amplitude and hence will evolve according to modified
Efremov-Radyushkin-Brodsky-Lepage (ERBL) equations~\cite{erbl}. In this ERBL region, GPDs contain completely new information about the nucleon structure, because this region is completely absent in DIS, which corresponds to the limit $\xi\to 0$.


\subsection{Impact parameter space}

GPDs are defined in terms of amplitudes in momentum representation involving proton states with a sharp plus-momentum. One can also use a mixed representation keeping momentum in the light-cone plus-direction and Fourier transforming from transverse momentum to transverse position. In such a representation hadron states with definite plus-momentum can be localized in the transverse plane at a definite position to be interpreted as an impact parameter~\cite{Soper77,Burkardt00a}. This is possible according to the uncertainty principle and as a consequence of the fact that transverse boosts are purely kinematical, i.e. in the light-front framework they form a Galilei subgroup of the Poincar\'e group. The position in the transverse plane coincides with  the `center of momentum' $\tvec R_\perp = \sum_i k^+_i  \tvec{b}_i^{\phantom{+}} /\sum_i k^+_i$ of the partons in the proton, given in terms of their plus-momenta $k^+_i $ and transverse positions (impact parameters) $\tvec{b}_i$. 

In the limit $\xi\to 0$, but $t\ne 0$, i.e. when the momentum transfer is purely transverse, a two-dimensional Fourier transform gives the distribution~\cite{Burkardt02,Burkardt03}
\be
  \label{impact-gpds}
{\mathcal H}^q(x,b^2) = \int \frac{d^2 \tvec{\Delta}}{(2\pi)^2}\, e^{- i\svec{b}\cdot \svec{\Delta}}\, H^q(x,0,-\tvec{\Delta}^2),
\ee
that depends on the impact parameter $\tvec b$ only via its square thanks to rotation invariance. In addition, as a consequence of the Fourier transformation the matrix element is now diagonal in the plus-momentum and the impact parameter of the proton states that can be taken at $\tvec R_\perp =0$. Therefore, the impact-parameter dependent distribution (\ref{impact-gpds}) can be interpreted as the probability density to find a quark with momentum fraction $x$ at transverse distance ${\tvec b}$ from the center of momentum $\tvec R_\perp$.

Similar considerations can be done for the other GPDs. In particular, in order to develop a probabilistic interpretation for 
\be
{\mathcal E}^q(x,b^2)\equiv \int \frac{d^2{\tvec\Delta}}{(2\pi)^2}e^{-i{\svec\Delta}\cdot{\svec b}} E^q(x,0,-{\tvec\Delta}^2),
\ee
it is necessary to consider helicity flip amplitudes because otherwise $E(x,0,t)$ does not contribute. 

When $\Delta^+=0$ for polarized nucleon states one has
\bea
\label{eq:no-flip}
\quad{}\int \frac{dz^-}{4\pi} e^{ixP^+z^- } \left\langle p+\Delta, \uparrow
\left| \bar\psi_{q}\left(0\right)\gamma^+\psi_ q\left({z^-}\right) \right| p,\uparrow\right\rangle
&=& H^q(x,0,-{\tvec \Delta}^2) ,
\\
\label{eq:flip}
\int \frac{dz^-}{4\pi} e^{ixP^+z^- } \left\langle p+\Delta, \uparrow
\left| \bar\psi_{q}\left(0\right)\gamma^+ \psi_q\left({z^-}\right)
\right| p,\downarrow\right\rangle
&=& {}-\frac{\Delta_x-i\Delta_y}{2M_N}E^q(x,0,-{\tvec \Delta}^2).
\eea
Therefore, taking the nucleon polarized in the $\hat x$-direction, one finds 
\bea
\label{eq:qX}
\rho_{\rm x}(x,{\tvec b})
&=&
\int \frac{d^2{\tvec\Delta}}{(2\pi)^2} \, e^{-i{\svec\Delta}\cdot{\svec b}}
\left[ H^q(x,0,-{\tvec\Delta}^2)  + \frac{i\Delta_y}{2M_N} E^q(x,0,-{\tvec\Delta}^2)\right]
\\
&=& {\mathcal H}^q(x,b^2) - \frac{1}{2M_N}\frac{\partial} {\partial b_y} {\mathcal E}^q(x,b^2).
\nonumber
\eea
Due to the contribution of  ${\mathcal E}^q$ in Eq.~(\ref{eq:qX}) that breaks the axial symmetry of ${\mathcal H}^q$, the quark distribution in a transversely polarized proton is distorted sideways in the transverse plane. This transverse distortion, combined with attractive final-state interactions, may give rise to the so-called Sivers effect~\cite{Siversa} and relatively large transverse single-spin asymmetries~\cite{Burkardt02,Burkardt04a,Burkardt04b,BurkHwang04}.

With probes exploring the nucleon on the scale of $1/Q\ll 1$ fm, the impact-parameter representation thus offers the possibility of performing a real femto-photogra\-phy of the interior structure of the nucleon~\cite{RalstonPire}.

This kind of analysis can be extended to the case of nonzero $\xi$~\cite{Diehl02}. In this case GPDs in impact parameter space probe partons at transverse position $\tvec b$ with the initial and final state proton localized around $\tvec R_\perp = 0$ but shifted relative to each other by an amount of order $\xi\tvec b$. In the DGLAP region the impact parameter gives the location where a quark or antiquark is pulled out of and put back into the proton. In the ERBL region the impact parameter describes the transverse location of a quark-antiquark pair in the initial proton.

An extension of this framework to give a quark imaging in the proton has been proposed in Ref.~\cite{BJY04} developing the concept of the quantum phase-space Wigner distribution for quark and gluons in the rest frame of the proton and relating it to the transverse-momentum dependent parton distributions and GPDs.

Otherwise, the Fourier transform of the DVCS amplitude with respect to the skewness parameter $\zeta=2\xi/(1+\xi)$ at fixed momentum transfer $t$ provides an image of the nucleon in a boost-invariant variable $\sigma$, the coordinate conjugate to light-front time~\cite{Brodsky067}. The results obtained in a simple relativistic model for spin-$\oneh$ systems are analogous to the diffractive scattering of a wave in optics where the distribution in $\sigma$ measures the physical size of the scattering center in a one-dimensional system.


\section{Modeling GPDs}
\label{sect:model}

It is quite difficult to calculate GPDs from first principles of QCD. Lattice simulations have received increasing attention in recent years and promising results have appeared. However, we are still far from explicitly calculating GPDs on the lattice. Deeply virtual Compton scattering and hard exclusive electroproduction of mesons give a theoretical possibility of experimentally constrain GPDs. However, this is a quite difficult task since observables involve convolution of the GPDs with hard scattering coefficients and not GPDs themselves. Therefore, there is continuing interest in modeling GPDs. 

There are basically two approaches. One is using ans\"atze to parametrize GPDs to be used in phenomenological analyses. Here, the most popular choice is to parametrize the hadronic matrix elements which define GPDs in terms of double distributions~\cite{Radyushkin97,Radyushkin99a}, modeled by assuming a factorized $t$-dependence determined by some form factors. However, since this assumption is not strictly valid, alternatives have been suggested with the so-called dual representation~\cite{GP06}, with constraints derived from data on the first Mellin moments, i.e. form factors~\cite{Guidal04,Jakob05}, or from simultaneous fits of data and lattice calculations of the higher Mellin moments~\cite{ahmad}. 

Fitting procedures require suitable ans\"atze as well as theoretical developments. A promising approach along these lines has been discussed quite recently combining dispersion relation and operator product expansion techniques and deriving the conformal partial wave decomposition of the DVCS amplitude in terms of complex conformal spin to twist-two accuracy~\cite{Kumericki}. The time-ordered product of the two electromagnetic currents sandwiched between the initial and final hadronic states in the Compton tensor is expanded in the basis of so-called conformal operators~\cite{BKM03}. Based on the analyticity of the Compton tensor the conformal operator product expansion can be combined with the dispersion relation technique to express the Taylor expansion of the Compton form factors with respect to $
\omega=1/\xi$ in terms of the Mellin-Barnes integral representation with coefficients that are independent of the renormalization/factorization scheme. With suitable ans\"atze for the conformal GPD moments in terms of hadronic partial wave amplitudes in the $t$-channel the Mellin-Barnes representation is adequate for building flexible fitting procedures of DVCS and hard meson production~\cite{Kumericki2}.

The other approach is a direct calculation using effective quark models. After the first calculation within the MIT bag model~\cite{JiMS97}, GPDs were calculated in the chiral quark-soliton model~\cite{PPPBGW98,Penttinen00,Ossmann05,Wakamatsu05}, the Nambu-Jona-Lasinio model~\cite{Mineo05}, the light-front Hamiltonian approach~\cite{Asmita}, the nonrelativistic~\cite{Scopetta04} and light-cone~\cite{BPT03,BPT04,BPT05,PPB05} constituent quark model, the meson-cloud model~\cite{PB06}.

After a brief account of the status of lattice simulations, in the following subsections the main lines of research within the two approaches are briefly reviewed. 


\subsection{Lattice simulations}

Lattice QCD offers a unique opportunity to calculate Mellin moments of GDPs from first principles. The first investigations of GPDs including studies of the quark angular momentum contributions to the nucleon spin have been presented by the QCDSF collaboration in quenched QCD~\cite{QCDSF04} and by LHPC/SESAM in $n_f=2$ lattice QCD~\cite{LHPC03}. Lattice results on nucleon GPDs published since then have provided important insights into the transverse structure of unpolarized nucleons~\cite{LHPC04}, the lowest moments of polarized~\cite{LHPC04a} and tensor GPDs~\cite{QCDSF05}, and transverse spin densities of quarks in the nucleon~\cite{QCDiehl05,QCDSF06a}. With the exception of several initial studies~\cite{LHP05,LHPC05a}, all previously published lattice results on GPDs have been obtained from calculations in a two-flavour `heavy pion world' with pion masses in the range of 550 to over 1000 MeV. In Ref.~\cite{Hagler07} previous studies have been improved by presenting a comprehensive analysis of the lowest three moments of unpolarized and polarized GPDs in $n_f=2+1$ lattice QCD with pion masses as low as 350 MeV and volumes as large as (3.5 fm)$^3$ (for further reading, see also~\cite{QCDSF03,QCDSF05a,QCDSF06,LHPC05,AliQCDSF04,AliQCDSF05,LHPC06,LHPC06a}).


\subsection{Double distributions}

The most popular choice when parametrizing GPDs is to use a factorized form with a $t$-dependent part determined by some form factors and a $t$-independent part given in terms of double distributions~\cite{Radyushkin97,Radyushkin99a}. Although this assumption is not strictly valid, it simplifies the QCD evolution considerably because in this way the $t$ dependences of quarks and gluons (which mix under evolution) are not modified during evolution~\cite{Radyushkin99b,Musatov00}. The $t$-independent part $H^q(x, \xi) \equiv H^q( x, \xi, t = 0)$ is parametrized by a two-component form:
\be
H^q(x, \xi) = H^q_{DD}(x, \xi) +
\theta(|\xi|-|x|)\,  D^q\left(\frac{x}{\xi}\right) ,
\label{eq:dd}
\ee
where 
\be
H^q_{DD}(x,\xi)= \int_{-1}^{1}d\beta
\int_{-1+|\beta|}^{1-|\beta|} d\alpha\,
\delta(x-\beta-\alpha\xi)\, F^q(\beta,\alpha) .
\label{eq:dd2}
\ee
A similar decomposition is assumed for the spin-flip quark GPD $E^q$~\cite{GPVdH}, i.e.
\be
E^q(x, \xi) = E^q_{DD}(x, \xi) -
\theta(|\xi|-|x|)\,  D^q\left(\frac{x}{\xi}\right) ,
\label{eq:dde}
\ee
where 
\be
E^q_{DD}(x,\xi)= \int_{-1}^{1}d\beta
\int_{-1+|\beta|}^{1-|\beta|} d\alpha\,
\delta(x-\beta-\alpha\xi)\, K^q(\beta,\alpha) .
\label{eq:dde2}
\ee
The $D$-term contribution $D^q$ in Eqs.~(\ref{eq:dd}) and (\ref{eq:dde}) completes the parametrization of GPDs, restoring the correct polynomiality properties of GPDs~\cite{Ji98a,Polyakov99}. It has a support only for $|x| \leq |\xi|$, so that it is invisible in the forward limit. The $D$-term contributes to the singlet-quark and gluon distributions and not to the non-singlet distribution. Its effect under evolution is at the level of a few percent~\cite{FmcD02b}.

According to Radyushkin's suggestion~\cite{Radyushkin99b}, the double distributions (DDs) entering Eqs.~(\ref{eq:dd2}) and (\ref{eq:dde2}) can be written as
\be
F^q(\beta, \alpha) = h(\beta, \alpha) \, q(\beta) , \quad K^q(\beta, \alpha) = h(\beta, \alpha) \, e(\beta) ,
\label{eq:ddunpol}
\ee
where $q(\beta)$ is the forward quark distribution (for the flavour $q$), while the spin-flip parton distribution $e(\beta)$ cannot be extracted from DIS data and has to be modeled (see, e.g.~\cite{Ellinghaus06}). The profile function $h(\beta, \alpha)$ is parametrized as~\cite{Musatov00}
\begin{eqnarray}
h(\beta , \alpha) = 
 \frac{\Gamma(2b+2)}{2^{2b+1}\Gamma^2(b+1)}\,
\frac{\bigl[(1-|\beta|)^2-\alpha^2\bigr]^{b}}{(1-|\beta|)^{2b+1}} .
\label{eq:profile}
\end{eqnarray}
In Eq.~(\ref{eq:profile}), the parameter $b$ determines the width of the profile function $h(\beta , \alpha)$ and characterizes the strength of the $\xi$ dependence of the GPDs. It is a free parameter for the valence ($b_{val}$) and sea ($b_{sea}$) contributions to GPDs. In such an approach $b_{val}$ and $b_{sea}$ can be used as fit parameters in the extraction of GPDs from hard electroproduction observables. The favoured choice is $b_{val} = b_{sea} = 1.0$, corresponding to maximum skeweness. With a similar assumption adopted for the gluon distribution one defines $b_{gluon}=2$. The limiting case $b\to\infty$ gives $h(\beta, \alpha)\to\delta(\alpha)\,h(\beta)$ and corresponds to the $\xi$-independent ansatz for the GPD, i.e. $H^q(x,\xi)\to H^q(x_B,\xi=0)=q(x_B)$, as used in Refs.~\cite{GuichonVdH98,VGG98}. 

Assuming a phenomenological parton distribution one can therefore construct model GPDs which  include all general constraints and are flexible enough to allow for a fit of the different observables~\cite{GuichonVdH98,VGG98,VGG99,KPVdH01,GPVdH,Guidal04}.

Starting from a similar ansatz in the DGLAP region a parametrization has been proposed~\cite{Freund} that satisfies the requirements of polynomiality and positivity conditions and is also suitable to study twist-3 effects. Other parametrizations were proposed in Ref.~\cite{Mankiewicz99}.

Including the $D$-term the double-distribution representation satisfies polynomiality but does not guarantee positivity. An integral representation obeying both polynomiality and positivity has been worked out in Ref.~\cite{Pobylitsa2} on pure mathematical grounds and in Ref.~\cite{Pobylitsa03} from an analysis of simple perturbative graphs for GPDs. Further proposals to construct double distributions with  dynamical content in connection with the Fock expansion of light-cone wave functions can be found in Refs.~\cite{tiburzi,TDM}.


\subsection{The dual parametrization of GPDs}

An alternative way to parametrize GPDs is the so-called dual parametrization of GPDs~\cite{GP06}. It  is based on the partial wave expansion of the GPDs in the $t$ channel~\cite{Polyakov02}. GPDs are then presented as an infinite series of $t$-channel exchanges, which reminds us of the idea of duality in hadron-hadron scattering. The applicability of such a procedure is justified by recognizing crossing relations between the GPDs of DVCS and the generalized distribution amplitudes of $\gamma^*\gamma\to h\bar h$.  

In fact, the $\gamma^*  h\to\gamma h$ (DVCS) amplitude at large $Q^2$ and small squared momentum transfer between the hadrons is related by crossing symmetry to the crossed channel amplitude of the $\gamma^*\gamma\to h \bar h$ process with a highly virtual photon near threshold, i.e. in the regime where the squared c.m. energy $W^2$ is much smaller than the photon virtuality $Q^2$. At tree level, similar to the DVCS case~\cite{Ji97,Radyushkin96a,Ji97a}, the $\gamma^*\gamma$ amplitude can be written as a convolution of a hard photon-parton scattering amplitude and a generalized distribution amplitude for the soft transition from parton to hadrons~\cite{DGPT98}. As in the case of DVCS, in order to give a nonzero $\gamma^*\gamma\to h \bar h$ amplitude the virtual photon must have the same transverse helicity of the real one as a consequence of chiral invariance in the collinear hard scattering process.

The dual parametrization fulfills the polynomiality condition and allows for flexible modeling of the $t$ dependence of the GPDs designed primarily for small and medium-size values of $x_B$, $x_B\le 0.2$~\cite{GP06}. In addition, crossing is related to the possibility of establishing dispersion relations between the real and imaginary parts of the elementary DVCS amplitude~\cite{DI07,AT07} that are helpful to model GPDs. It must be said, however, that mathematically one can not completely restore the complete image of the nucleon contained in the GPDs from the knowledge of the elementary DVCS amplitude, because one has to deal with an inversion problem. This is a typical problem of tomography which is usually solved with the help of Radon transformations~\cite{Radon}. The dual parametrization combined with the Radon transformation clarifies which part of the image can be analyzed~\cite{Polyakov07}.


\subsection{Constraints from observables}

The connection between GPDs and Dirac and Pauli form factors through the first Mellin moment (\ref{eq:emformfactors}) is suggesting a simple factorized ansatz for the $x$ and $t$ dependences of the GPDs at $\xi=0$, as in the VGG model~\cite{VGG98,GuichonVdH98,VGG99}. Here one assumes
\be
H^u(x,\xi=0,t) = u_v(x)\,F_1^u(t)/2, \quad H^d(x,\xi=0,t) = d_v(x)\,F_1^d(t),
\ee
where $u_v(x)$ and $d_v(x)$ are the (unpolarized) distributions for valence up and down quarks, respectively. A similar expression is assumed for $E^q$. The ansatz for $\tilde H^q$ is
\be
\tilde H^u(x,\xi=0,t) = \Delta u_v\,g_A^u(t)/g_A^u(0), \quad 
\tilde H^d(x,\xi=0,t) = \Delta d_v\,g_A^d(t)/g_A^d(0),
\ee
where $\Delta  u_v(x)$ and $\Delta d_v(x)$ are the polarized (helicity) distributions of valence up and down quarks, respectively. The quark distributions are taken from parametrizations fitting the world DIS data. The $\xi$ dependence is obtained in the VGG model by matching this ansatz with the double-distribution approach. 

However, a complete factorization of the $x$ and $t$ dependences of GPDs seems rather unrealistic. At small $t$ and small $x$ one can expect Regge behaviour of $H^q(x,\xi=0,t)$, so that one can assume an exponential ansatz:
\be
H^q(x,\xi=0,t) = q_v(x)\,\exp[t\,f^q(x)], \quad E^q(x,\xi=0,t) = e^q_v(x)\,\exp[t\,g^q(x)],
\ee
where $e^q_v(x)$ is just the forward limit of $E^q(x,\xi=0,t)$ normalized as
\be
\int_0^1 dx\, e^q_v(x) = \kappa^q,
\ee
and the functions $f^q(x)$ and $g^q(x)$ parametrize how the profile of the quark distribution in the impact parameter plane changes with $x$. Specific forms have been proposed for them in Refs.~\cite{Guidal04,Jakob05} in order to be consistent with dominance of the leading meson trajectories known from Regge phenomenology and to satisfy the Drell-Yan-West relation~\cite{drellyanwest} between the large-$x$ power behaviour of parton distributions at $t=0$ and the large-$t$ power behaviour of the associated elastic form factors. With only four parameters in Ref.~\cite{Guidal04} and with much more flexibility in Ref.~\cite{Jakob05} a rather good fit to elastic nucleon electromagnetic form factors is obtained. Considerable ambiguities have been found in determining $e^q_v(x)$ since the forward limit of $E^q(x,\xi,t)$ is not known. The form factor data disfavour an identical shape in $x$ and $t$ of $E^u(x,\xi=0,t)$ and $E^d(x,\xi=0,t)$~\cite{Jakob05}.

A physically motivated parametrization for the unpolarized quark GPDs $H$ and $E$ has been recently proposed~\cite{ahmad} using a combination of constraints from simultaneous fits of the experimental data on both the nucleon elastic form factors, the DIS structure functions in the non-singlet sector and the lattice calculations of Mellin moments with $n\ge 1$. The method is thus able to produce $\xi$- and $t$-dependent $H^u-H^d$ and $E^u-E^d$ satisfying the polynomiality condition by construction.


\subsection{The chiral quark-soliton model}

The chiral quark-soliton model ($\chi$QSM)~\cite{Diakonov2} is based on the principles of chiral symmetry breaking and the limit of a large number of colours $N_c$. The underlying effective relativistic quantum field theory was derived from the instanton model of the QCD vacuum~\cite{Diakonov1}, which provides a mechanism of dynamical chiral symmetry breaking, and is valid at low energies below a scale of about $\rho_{\rm av}^{-1}\approx 600$ MeV, where $\rho_{\rm av}$ is the average instanton size. Thus the $\chi$QSM can be considered as a realization of the idea that in the large-$N_c$ limit the nucleon can be viewed as a classical soliton of the pion field.

In the leading order of the large-$N_c$ limit the pion field is static, and one can determine the spectrum of the effective one-particle Hamiltonian of the theory:
\be
\hat{H}_{\rm eff}\ket{n} = E_n \ket{n}, \quad
\hat{H}_{\rm eff} = -i\gamma^0\gamma^k\partial_k+\gamma^0MU^{\gamma_5},
\label{eff-Hamiltonian}
\ee
where $U^{\gamma_5}= \exp(i\gamma_5\tau^a\pi^a)$ with $U=\exp(i\tau^a\pi^a)$ denoting the $SU(2)$ chiral pion field. The spectrum consists of an upper and a lower Dirac continuum, which are distorted by the pion field as compared to continua of the free Dirac-Hamiltonian,
\be
\hat{H}_0\ket{n_0} = E_{n_0}\ket{n_0},  \quad
\hat{H}_0 = -i\gamma^0\gamma^k\partial_k+\gamma^0 M ,
\label{free-Hamiltonian}\ee
and of a discrete bound state level of energy $E_{\rm lev}$. By occupying the discrete level and the states of lower continuum each by $N_c$ quarks in an anti-symmetric colour state, one obtains a state with unity baryon number. The soliton energy $E_{\rm sol}$ is a functional of the pion field,
\be
E_{\rm sol}[U] = N_c 
\biggl[E_{\rm lev}+\singlesum{E_n<0}(E_n-E_{n_0})\biggr] .
\label{soliton-energy}
\ee 
Minimization of $E_{\rm sol}[U]$ determines the self consistent solitonic pion field $U_c$. The nucleon mass $M_N$ is given by $E_{\rm sol}[U_c]$. Quantum numbers of the baryon -- like momentum, spin and isospin -- are described by considering zero modes of the soliton. Corrections in $1/N_c$ can be included by considering time-dependent pion field configurations. The results of the $\chi$QSM respect all general counting rules of the large-$N_c$ phenomenology. 

In the $\chi$QSM one can evaluate in a parameter-free way nucleon matrix elements of QCD quark bilinear operators as 
\bea
\label{matrix-elements} 
\qquad{}&& 	
\bra{N',{\tvec p'}} \bar{\psi}_q(z_1)\Gamma\psi_q(z_2) \ket{N,{\tvec p }}\\
&& \ \ = A^{\mbox{\tiny$\Gamma$}}_{\mbox{\tiny$NN'$}}\;
	2M_N N_c \singlesum{n, \rm occ}
	\int d^3{\tvec X}\:e^{i({\svec p'-\svec p})\cdot{\svec X}}\,
	\overline{\Phi}_n({\tvec z}_1-{\tvec X})\Gamma\Phi_n({\tvec z}_2-{\tvec X})\,
	e^{iE_n(z^0_1-z^0_2)} 
	\nonumber\\
& & \quad{} + \dots \nonumber
\eea
where for $z_1\neq z_2$ the insertion of the gauge link is understood on the left hand side. The dots in Eq.~(\ref{matrix-elements}) denote terms subleading in the $1/N_c$ expansion. In Eq.~(\ref{matrix-elements}) $\Gamma$ is some Dirac and flavour matrix, $A^{\mbox{\tiny$\Gamma$}}_{\mbox{\tiny$NN'$}}$ a constant depending on $\Gamma$, the spin and flavour quantum numbers of the nucleon state 
$\ket{N}=\ket{S_3,T_3}$, and $\Phi_n({\tvec x}) = \bra{\tvec x}n\rangle$ are the coordinate-space wave-functions of the single quark states $\ket{n}$ defined in Eq.~(\ref{eff-Hamiltonian}). The sum in Eq.~(\ref{matrix-elements}) goes over occupied levels $n$ (i.e. $n$ with $E_n\le E_{\rm lev}$), and vacuum subtraction is implied for $E_n < E_{\rm lev}$ as in Eq.~(\ref{soliton-energy}).

Many static nucleonic observables, like magnetic moments, electric polarizabilities, axial properties,  have been computed in the $\chi$QSM in the way sketched in Eq.~(\ref{matrix-elements}). 
The results were found in good agreement with data (see Ref.~\cite{Christov96} for a review).
In particular the model describes data on electromagnetic form factors up to $|t| \sim 1\,{\rm GeV}^2$ within (10-30)$\%$~\cite{Christov95,Christov96}. In Ref.~\cite{Diakonov96} it has been demonstrated that the model can be applied to the description of twist-2 quark and anti-quark distribution functions of the nucleon. The consistency of the approach has been shown by giving proofs that  the model expressions satisfy all general requirements of QCD~\cite{Diakonov96}. The distribution functions computed in the $\chi$QSM \cite{Diakonov96,Diakonov97,Pobylitsa98} (and also~\cite{WGR,Weigel98,Wakamatsu99}) refer to a low normalization scale of around $600\,{\rm MeV}$, and agree with available parametrizations performed at comparably low scales~\cite{Gluck94} within (10-30)$\%$.

The $\chi$QSM lacks explicit gluon degrees of freedom. Therefore one deals with quark GPDs only. Different flavour combinations of the GPDs exhibit different behaviour in the large-$N_c$ limit~\cite{GPVdH}
\begin{eqnletter}
\qquad\quad & (H^u+H^d)(x,\xi,t) = N_c^2 f(u,v,t) ,\quad
& (E^u-E^d)(x,\xi,t) = N_c^3 f(u,v,t) , 
\label{eq:ordernc.r1}
\\
& (H^u-H^d)(x,\xi,t) = N_c\, f(u,v,t) , \quad
& (E^u+E^d)(x,\xi,t) = N_c^2 f(u,v,t) ,
\label{eq:ordernc.r2}
 \\
& (\tilde H^u-\tilde H^d)(x,\xi,t) = N_c^2 f(u,v,t) , \quad 
& (\tilde E^u-\tilde E^d)(x,\xi,t) = N_c^4 f(u,v,t) , 
\label{eq:ordernc.r3}
\\
& (\tilde H^u+\tilde H^d)(x,\xi,t) = N_c\, f(u,v,t) , \quad
& (\tilde E^u+\tilde E^d)(x,\xi,t) = N_c^3 f(u,v,t) . 
\label{eq:ordernc.r4}
\end{eqnletter}
The functions $f(u,v,t)$ are stable in the large-$N_c$ limit for fixed values of the  ${\cal O}(N_c^0)$ variables $u=N_cx$, $v=N_c\xi$ and $t$, and of course different for the different GPDs.

\begin{figure}
\begin{center}
\epsfig{file=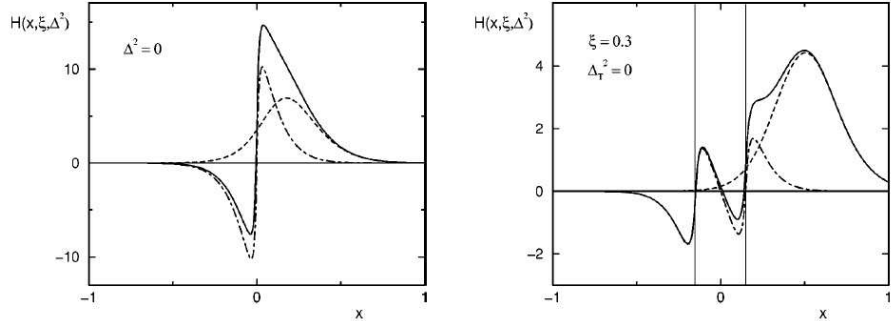,width=12 cm}
\end{center}
\caption{\small The isosinglet distribution $H(x,\xi,\Delta^2) = H^u(x,\xi,\Delta^2) + H^d(x,\xi,\Delta^2)$ in the forward limit, $\Delta^2=0$ and $\xi=0$, (left panel) and for $\Delta^2_\perp = -\Delta^2 -\xi^2M_N^2 = 0$ and $\xi = 0.3$ (right panel). Dashed lines for the contribution from the discrete level; dashed-dotted lines for the contribution from the Dirac continuum; solid lines for the total contribution~\cite{PPPBGW98}.}
\label{fig:petrov}
\end{figure}

The model expressions for the leading GPDs in Eqs.~(\ref{eq:ordernc.r1})  and (\ref{eq:ordernc.r3}) were derived in Refs.~\cite{PPPBGW98,Penttinen00}, respectively, where it was also demonstrated that they are correctly normalized to the corresponding electroweak form factors. In Ref.~\cite{Schweitzer} the expressions for the leading (chiral-even) GPDs have been proved to also satisfy the polynomiality property. Numerical estimates show the relevant role of the Dirac continuum in determining the shape of the GPDs (Fig.~\ref{fig:petrov}). The forward limit of the unpolarized spin-flip isoscalar and isovector  GPDs, $(E^u+E^d)(x,\xi,t)$ and $(E^u-E^d)(x,\xi,t)$, have been computed in Refs.~\cite{Ossmann05} and \cite{Wakamatsu05}, respectively.


\subsection{Constituent-quark models}

The modeling of GPDs in constituent quark models (CQMs) relies on the basic assumption that there exists a scale $Q^2_0$ where the short range (perturbative) part of the interaction is negligible and, therefore, the glue and sea are suppressed, while a long range (confining) part of the interaction produces a proton composed by (three) valence quarks, mainly~\cite{PaPe}. Jaffe and Ross~\cite{JaRo} proposed to ascribe the quark model calculations of matrix elements to that hadronic scale $Q_0^2$, 
typically $Q^2_0=0.2-0.4$ GeV$^2$. In this way, quark models summarizing a great deal of hadronic properties may substitute for low-energy parametrizations, while evolution to larger momentum $Q^2$ is dictated by perturbative QCD~\cite{BPT05}. CQMs were applied in the calculation of unpolarized and polarized GPDs in Ref.~\cite{Scopetta04} in terms of overlap of Schr\"odinger wave functions for three constituent quarks obtained within the instant-form quantization and for nonrelativistic kinematics. A fully covariant formalism has been adopted in Refs.~\cite{BPT03,BPT04,PPB05} for the calculation of the GPDs both in chiral-even and chiral-odd sectors within the framework of light-cone quantization, using the representation of GPDs in terms of overlaps of light-cone wave functions (LCWFs) describing the $N$-parton composition of a hadronic state~\cite{DFKJ01,BDH01}. The valence-quark contribution is obtained by specializing to the case $N=3$. The corresponding valence-quark component of the LCWF can be related to a wave function obtained in the (canonical) instant form through appropriate Melosh rotations. Therefore, one can link LCWFs to wave functions derived in CQMs where they are obtained as eigenfunctions of the nucleon Hamiltonian in the 
instant-form dynamics. Of course, this link is useful in the kinematic range where valence quark degrees of freedom are effective. In this region GPDs exhibit the exact forward limit, reproducing the parton distribution with the correct support and automatically fulfilling the particle number and momentum sum rules. In addition, from the LCWF overlap representation of the PDFs for the valence-quark contribution, it is possible to derive the following relations valid at the hadronic scale $Q^2_0$~\cite{PPB05}:
\be
\label{eq:sofferlike}
2\Delta_T u(x)=\Delta u(x)+{\textstyle\frac{2}{3}} u(x),
\quad
2 \Delta_T d(x)= \Delta d(x)-{\textstyle\frac{1}{3}} d(x).
\ee
These relations are compatible with the Soffer inequality~\cite{Soffer95} and generalize to the case of parton distributions the results obtained in Refs.~\cite{SS97,MaSS98} for the axial ($\Delta q$) and tensor 
($\delta q$) charges:
\be
2\delta q = \Delta q + \Delta q_{\rm NR},
\ee
where $\Delta q_{\rm NR}$ is the axial charge in the nonrelativistic limit,
 i.e.
\be
\Delta q_{\rm{NR}}=\left({\textstyle\frac{4}{3}}\delta_{\tau_q 1/2}
-{\textstyle\frac{1}{3}}\delta_{\tau_q -1/2}\right).
\label{eq:tensor-nr}
\ee

Using LCWFs derived from the relativistic CQM of Ref.~\cite{Schlumpf94a} the resulting tensor charges 
are $\delta u = 1.16$ and $\delta d = -0.29$~\cite{PPB06b}, whereas those predicted by QCD sum rules~\cite{HeJi95} are $\delta u = 1.33\pm 0.53$ and $\delta d = 0.04\pm -0.02$. When evolved in leading-order QCD from the intrinsic scale of the model to $Q^2=10$ GeV$^2$ they become $\delta u = 0.79$ and $\delta d = -0.20$, within the range of values calculated in the different models considered in Ref.~\cite{Barone02} and in fair agreement with lattice QCD calculations~\cite{AliQCDSF04}.

With the same model the sum rule (\ref{eq:deltajx}) for the angular momentum carried by transversely polarized quarks gives quite substantial contributions for up and down quarks, i.e.~\cite{PPB06b}
\be
\langle \delta^x J^x_u\rangle = 0.54 ,\quad \langle \delta^x J^x_d\rangle 
= 0.37,
\ee
and for the quantity $\kappa^q_T$ defined in Eq.~(\ref{eq:kappaT}) one finds $\kappa^u_T=3.98$ and  $\kappa^u_T=2.60$. The same positive sign of $\kappa^q_T$ for both up and down quarks would imply~\cite{Burkardt05b} a negative Boer-Mulders function $h^{\perp q}_1$ describing the asymmetry of the transverse momentum of quarks perpendicular to the quark spin in an unpolarized nucleon~\cite{BM98}.


\subsection{The meson-cloud model}

The pion cloud of the nucleon associated with chiral-symmetry breaking was first discussed in the DIS context by Feynman~\cite{FeyPHI} and Sullivan~\cite{Sull72}. As realized by Thomas~\cite{Thomas83}, it can give an explanation of the flavour-symmetry violation in the sea-quark distributions of the nucleon, 
thus naturally accounting for the excess of $\bar d$ (anti)quarks over $\bar u$ (anti)quarks as observed through the violation of the Gottfried sum rule~\cite{EMC,NA51,E866-98,Hermes98}. The connection between the meson-cloud model and the chiral properties of GPDs was also studied by investigating the nonanalytic behaviour of the $\bar d - \bar u$ distribution in~\cite{TMS00,Chen01}.
Although the nucleon's nonperturbative antiquark sea cannot be ascribed entirely to its virtual meson cloud~\cite{KFS96}, the meson-cloud model for the physical nucleon has a long and successful history in explaining properties such as form factors~\cite{DHSS97,Miller02,PB07a} and parton distributions~\cite{STh98,Londergan98,Kumano98}. It has been revisited and applied for the first time to study quark GPDs in~\cite{PB06}. The meson cloud was also taken into account to study the unpolarized gluon distribution in Ref.~\cite{SW04}.

The basic assumption of the meson-cloud model is that the physical nucleon $\tilde N$ is made of a bare nucleon $N$ dressed by the surrounding meson cloud, so that the state of the physical nucleon can be decomposed according to the meson-baryon Fock-state expansion as a superposition of a bare-nucleon state and states containing virtual mesons associated with recoiling baryons. This state, with four-momentum 
$p_N^\mu=(p^-_N,p^+_N,{\tvec p}_{N\perp})\equiv(p^-_N,\tilde p_N)$ and helicity
 $\lambda_N$, is an eigenstate of the light-cone Hamiltonian
\begin{equation}
H_{LC}= \sum_{B,M} \left[H_0^B(q) + H_0^M(q)+H_I(N,BM)\right].
\label{eq:1}
\end{equation}
In Eq.~(\ref{eq:1}) $H_0^B(q)$ stands for the effective-QCD Hamiltonian which governs the consti\-tuent-quark dynamics, and leads to the confinement of three quarks in a baryon state; analogously, $H_0^M(q)$ describes the quark interaction in a meson state, $H_I(N,BM)$ is the nucleon-baryon-meson interaction, and the sum is over all the possible baryon and meson configurations in which the nucleon can virtually fluctuate.

In a perturbative treatment of the meson effects, the Fock-space expansion is truncated to the one-meson components, and the nucleon wave function is expanded in terms of eigenstates of the bare Hamiltonian $H_0\equiv H_0^B(q) + H_0^M(q)$. The corresponding state of the physical nucleon $\ket{\tilde N}$ can be written as
\begin{eqnarray}
\label{eq:14}
&& |\tilde p_N,\lambda_N;\tilde N\rangle
=
\sqrt{Z}|\tilde p_N,\lambda_N; N\rangle
\\
&& \qquad {}+
\sum_{B,M}
\int \frac{{\rm d}y{\rm d}^2{\tvec k}_{\perp}}{2(2\pi)^3}\,
\frac{1}{\sqrt{y(1-y)}}
\sum_{\lambda',\lambda''}
\phi_{\lambda'\lambda''}^{\lambda_N \,(N,BM)}(y,{\tvec k}_\perp)
\nonumber \\
&&\qquad \qquad{}\times
|yp^+_N,{\tvec k}_{\perp}+y{\tvec p}_{N\perp},\lambda';B\rangle\,
|(1-y)p^+_N,-{\tvec k}_{\perp}+(1-y){\tvec p}_{N\perp},\lambda'';M\rangle,
\nonumber
\end{eqnarray}
where the function $\phi_{\lambda'\lambda''}^{\lambda_N\,(N,BM)}(y,{\tvec k}_\perp)$ defines the probability amplitude for a nucleon with helicity $\lambda_N$ to fluctuate into a virtual $BM$ system with the baryon having helicity $\lambda'$, longitudinal momentum fraction $y$ and transverse momentum ${\tvec k}_\perp$, and the meson having helicity $\lambda''$, longitudinal momentum fraction 
$1-y$ and transverse momentum $-{\tvec k}_\perp$. The normalization constant $Z$ in Eq.~(\ref{eq:14}) gives the probability that the nucleon is a bare core state and ensures the correct normalization of the nucleon wave function:
\begin{equation}
\langle p'^+,{\tvec p}'_\perp, \lambda ';H\vert p^+,{\tvec p}_\perp\lambda;
H\rangle=
2(2\pi)^3p^+\delta(p'^+-p^+)\delta^{(2)}({\tvec p}'_\perp -{\tvec p}_\perp)
\delta_{\lambda\lambda'}.
\label{eq:7}
\end{equation}

In the region $\xi\le x\le 1$, where a quark is taken out of the nucleon with momentum fraction $x+\xi$ and reinserted back with momentum fraction $x-\xi$, the virtual photon can hit either the bare nucleon $N$ or one of the higher Fock states. As a consequence, the matrix elements defining the GPDs in Eq. (\ref{eq:matrices.r1}) can be written as the sum of two contributions, 
\begin{eqnarray}
A^q_{\lambda_N'\mu', \lambda_N\mu} =
Z\,
A^{q, {\rm bare}}_{\lambda'_N\mu',\lambda_N\mu} 
+\delta A^q_{\lambda'_N\mu',\lambda_N\mu}\,,
\label{eq:6}
\end{eqnarray}
where $A^{q, {\rm bare}}$ is the contribution from the bare proton, described in terms of Fock states with three valence quarks. This term can be calculated in the light-front overlap representation derived in Ref.~\cite{DFKJ01}, and applied to the case of three valence quarks in Refs.~\cite{BPT03,BPT04}, where one can also find explicit expressions in terms of bare-nucleon LCWFs derived in a CQM. The $\delta A^q$ term in Eq.~(\ref{eq:6}) is the contribution from the $BM$ Fock components of the nucleon state, corresponding to five-parton configurations. This term can further be split into two contributions, with the active quark belonging either to the baryon ($\delta A^{q/(B'B)M}$) or with the active quark inside the meson ($\delta A^{q/(M'M)B}$), i.e.
\begin{eqnarray}
\delta A^{q}_{\lambda'_N\mu',\lambda_N\mu}=
\sum_{B,B',M}\delta A^{q/(B'B)M}_{\lambda'_N\mu',\lambda_N\mu}
+\sum_{M',M,B}\delta A^{q/(M'M)B}_{\lambda'_N\mu',\lambda_N\mu}.
\label{eq:7a}
\end{eqnarray}
\begin{figure}[ht]
\begin{center}
\epsfig{file=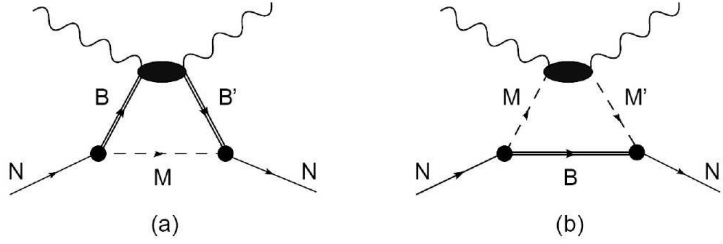,  width=10cm}
\end{center}
\caption{\small Deeply virtual Compton scattering from the virtual (a) baryon 
and (b) meson components of a dressed nucleon.}
\label{fig:fig2}
\end{figure}

The term $\delta A^{q/(B'B)M}$ is schematically represented in Fig.~\ref{fig:fig2}a. It is explicitly given by
\begin{eqnarray}
\label{eq:conv-bar}
\delta A^{q/(B'B)M}_{\lambda'_N\mu',\lambda_N\mu}(x,\xi,{\tvec \Delta}_\perp)&=&
\sum_{\lambda,\lambda',\lambda''}
\int_{x}^1
\frac{{\rm d}\bar y_B}{\bar y_B}
\int
\frac{{\rm d}^2{\bar{\mathbf k}}_{B\perp}}{2(2\pi)^3}
A^{q/(B'B)}_{\lambda'\mu',\lambda\mu}(\frac{x}{\bar y_B},
\frac{\xi}{\bar y_B},\mathbf \Delta_\perp)\\
& &\times
\phi^{\lambda_N\,(N,BM)}_{\lambda''\lambda}
(\tilde y_B,\tilde{{\mathbf k}}_{B\perp})\,
[\phi^{\lambda'_N\,(N,B'M)}_{\lambda''\lambda'}
(\hat y_{B;},\hat{\mathbf k}'_{B'\,\perp})]^*,\nonumber
\end{eqnarray}
where $\bar k_B=(\tilde k_B+\hat k_{B'})/2$ is the average momentum of the baryon in the initial and final state, and $\bar y_B=\bar k_B^+/P^+$ is the average fraction of the plus-momentum of the nucleon carried by the baryon. The term $A^{q/(B'B)}_{\lambda'\mu',\lambda\mu}$ in Eq.~(\ref{eq:conv-bar}) is the matrix element of Eq.~(\ref{eq:matrices.r1}) evaluated between the initial $(B)$ and final $(B')$ baryon state of the baryon-meson component of the nucleon and can 
explicitly be calculated in terms of overlap of bare-baryon LCWFs.

Analogously, the contribution from the meson in the $BM$ fluctuation (Fig.~\ref{fig:fig2}b) is calculated
by interchanging the role of the baryon and meson substates. It reads 
\begin{eqnarray}
\label{eq:conv-mes}
& &
\delta A^{q/(M'M)B}_{\lambda'_N\mu',\lambda_N\mu}(x,\xi,{\tvec \Delta}_\perp)
=\sum_{\lambda,\lambda',
\lambda''}
\int_{x}^1
\frac{{\rm d}\bar y_M}{\bar y_M}
\int
\frac{{\rm d}^2{\bar{\mathbf k}}_{M\perp}}{2(2\pi)^3}
A^{q/(M'M)}_{\lambda'\mu',\lambda\mu}(\frac{x}{\bar y_M},
\frac{\xi}{\bar y_M},\mathbf \Delta_\perp)\\
& &\times
\phi^{\lambda_N\,(N,BM)}_{\lambda''\lambda}
(1-\tilde y_M,-{\tilde{\mathbf k}}_{M\perp})\,
[\phi^{\lambda'_N\,(N,B'M)}_{\lambda''\lambda'}
(1-\hat y_{M'},-\hat{\mathbf k}'_{M'\,\perp})]^*,\nonumber
\end{eqnarray}
with $A^{q/(M'M)}_{\lambda'\mu',\lambda\mu}$ the matrix element between the initial and final meson 
participating to the interaction process. As a result,  the contribution from the $BM$ components in Eqs.~(\ref{eq:conv-bar}) and (\ref{eq:conv-mes}) are obtained by folding the matrix elements of the baryon and meson constituents with the probability amplitudes describing the distributions of these constituents in the dressed initial and final nucleon. We also note that the sum in Eq.~(\ref{eq:7a}) over all the possible $BM$ configurations includes contributions from both the diagonal matrix elements with the same hadrons in the initial and final state ($B'=B$  and $M'=M$ in Eq.~(\ref{eq:conv-bar}) and (\ref{eq:conv-mes}), respectively), and the matrix elements involving the  transition between different hadron states (i.e. the terms with $B'\neq B$ and $M'\neq M$ in Eq.~(\ref{eq:conv-bar}) and (\ref{eq:conv-mes}), respectively).

In the region $-1\le x\le -\xi$ where emission and reabsorption refer to an antiquark with momentum fraction $-(x+\xi)$ and $-(x-\xi)$, respectively, the only nonvanishing contribution comes from the active antiquark in the meson substate of the $BM$ Fock component of the nucleon wave function, i.e.
\begin{eqnarray}
A^q_{\lambda'_N\lambda_N}(x,\xi,{\tvec \Delta}_\perp)
=\sum_{M,M',B}
\delta A^{q/(M'M)B}_{\lambda'_N\lambda_N}(x,\xi,{\tvec \Delta}_\perp),
\end{eqnarray}
where $\delta A^{q/(M'M)B}_{\lambda'_N\lambda_N}$ is given by the convolution formula~(\ref{eq:conv-mes}) taken for negative value of $x$.

In the region $-\xi\le x\le \xi$ the scattering amplitude describes the emission of a quark-antiquark pair from the initial nucleon, so that in the Fock-state decomposition the parton content of the initial nucleon must differ from the final one by such a $q\bar q$ pair. In the meson-cloud model, the final state is described by the three-valence quark configuration of the bare nucleon and the initial state is given by the five-parton component of the $BM$ substate, with the active quark belonging to the baryon and the active antiquark to the meson. The contribution with both the active quark and antiquark belonging to the meson substate of the initial nucleon is vanishing by orthogonality between the final bare-nucleon state and the initial baryon state~\cite{PB06}.

This convolution formalism is applied here to calculate the polarized and unpolarized GPDs, taking into account the meson-cloud contribution corresponding to $\pi,$ $\rho,$ and $\omega$, with the accompanying baryon in the $BM$ component of the dressed nucleon being a nucleon or a $\Delta.$ For the bare-hadron states we use the LCWFs specified in Ref.~\cite{PB07a}, which are able to give an overall good description of the nucleon electroweak form factors.

\begin{figure}[ht]
\begin{center}
\epsfig{file=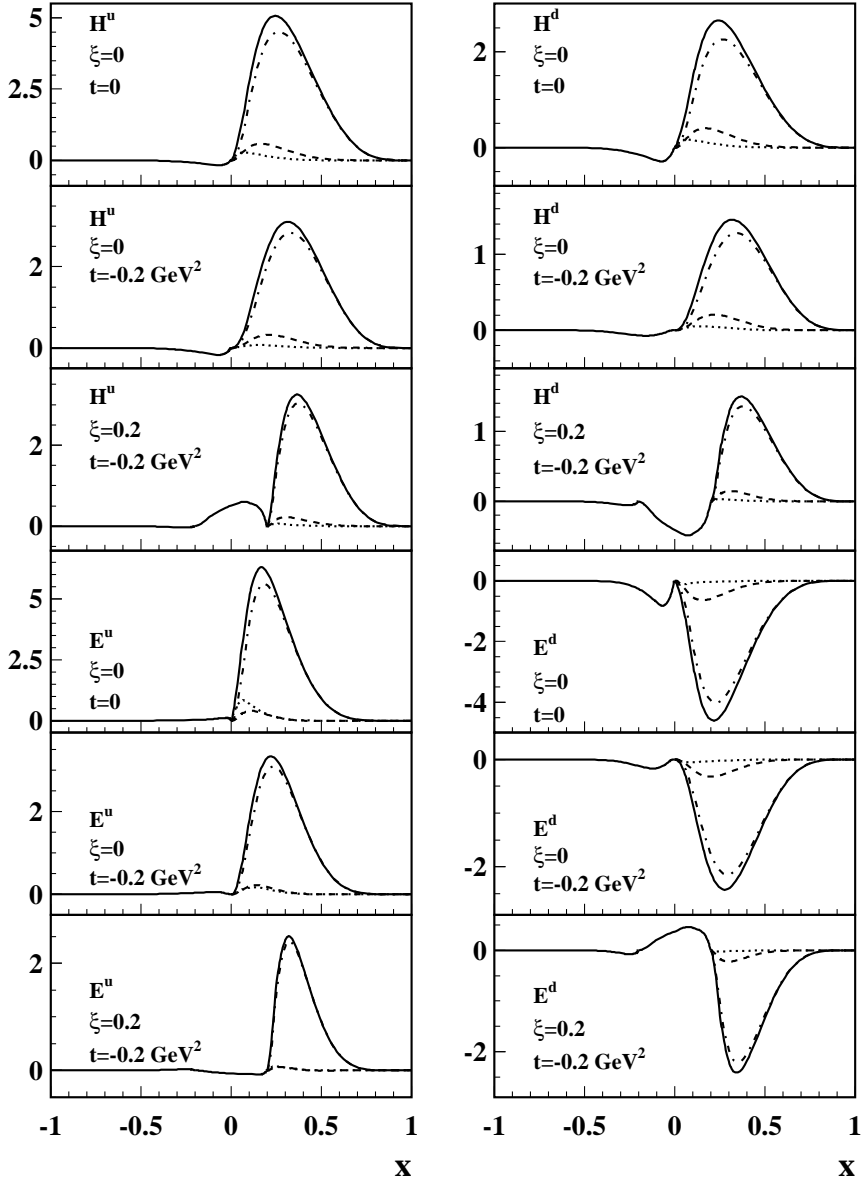, width=16cm}
\end{center}
\caption{\small The spin averaged ($H^q$, three upper panels) and the helicity flip ($E^q$, three lower panels) generalized parton distributions for the $u$ (left panels) and $d$ (right panels) flavours, at fixed
values of $\xi$ and $t$ as indicated. Dashed-dotted curves: bare contribution. Dashed curves:  contribution from the active baryon. Dotted curves:  contribution from the active meson. Solid curves: total result.}
\label{gpd_summary}
\end{figure}

\begin{figure}[ht]
\begin{center}
\epsfig{file=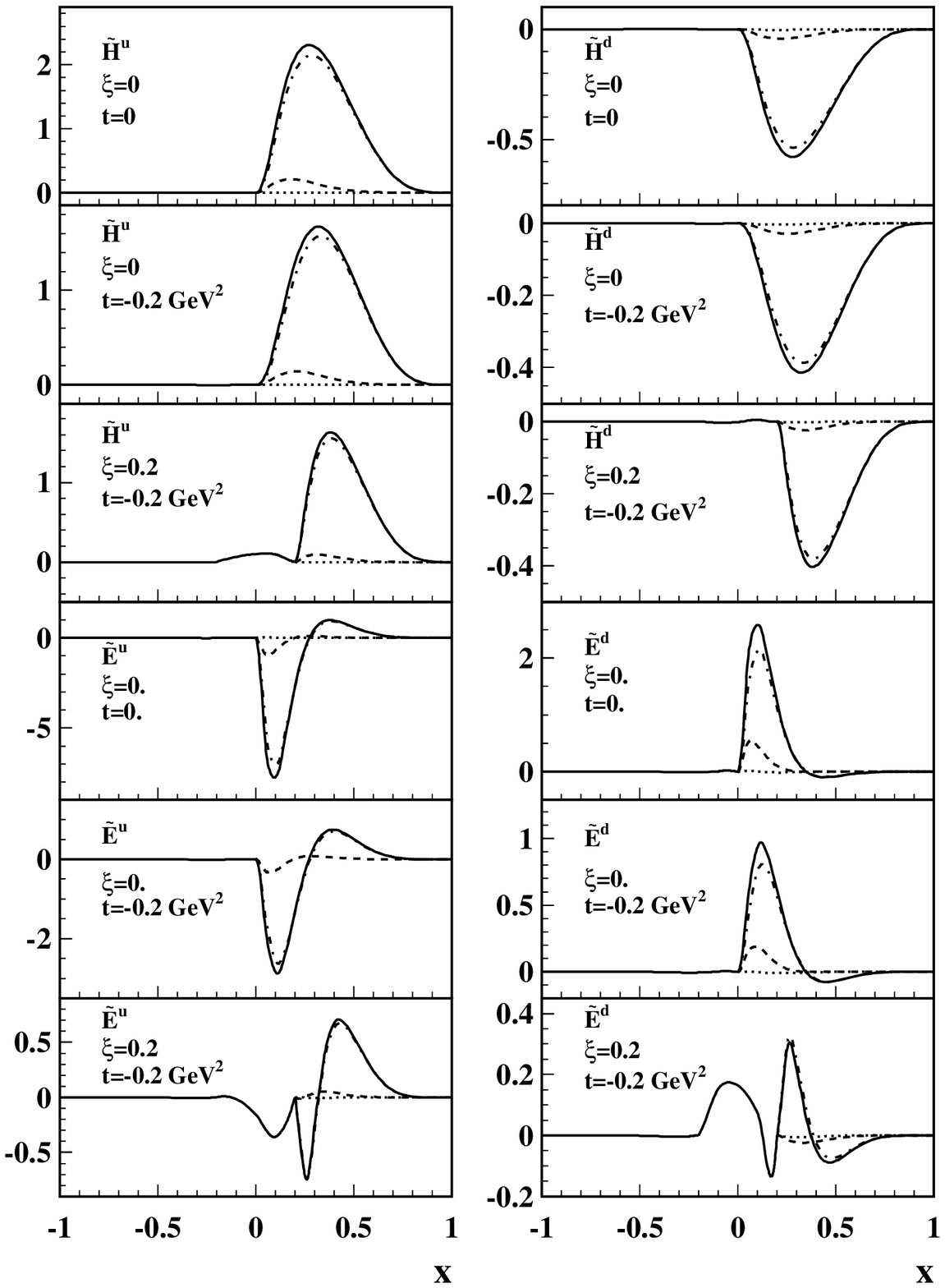, width=16cm}
\end{center}
\caption{\small The helicity dependent spin averaged ($\tilde H^q$, three upper panels) and the  helicity flip ($\tilde E^q$,  three lower panels) generalized parton distributions for the $u$ (left panels) and $d$ (right panels) flavours, at fixed values of $\xi$ and $t$ as indicated. Dashed-dotted curves: bare contribution. Dashed curves:  contribution from the active baryon. Dotted curves:  contribution from the active meson. Solid curves: total result.}
\label{gpd_tilde_summary}
\end{figure}

In Fig.~\ref{gpd_summary} the spin-averaged $H^q$ and the helicity-flip $E^q$ GPDs are plotted as function of $x$ at different fixed values
 of $\xi$ and $t$ together with the separated contributions from the bare proton (dashed-dotted lines), the baryon (dashed lines) and the meson (dotted lines) in the baryon-meson fluctuation. All these contributions add up incoherently to give the total result (full curves). The bare proton is always positive within its support ($0\le x\le 1$) with the exception of $E^d$ for which it is negative. The same behaviour characterizes the baryon contribution from the baryon-meson fluctuation that is also limited to the range $0\le x\le 1$, consistently with the assumption that the only active degrees of freedom for such a baryon are the valence quarks. Both contributions vanish at the end points of their support. The contribution from the mesons in the higher Fock state components extend all over the full range $-1\le  x\le 1$, and it is determined at positive (negative) $x$ by the quark (antiquark) residing in the mesons. The resulting effect of the meson cloud is thus to add a contribution for negative $ x$ and to increase the magnitude of the GPDs for positive $ x$ with respect to the case of the bare proton. In particular, for positive and small $x$ the meson-cloud contribution as a whole is comparable to that of the bare proton, confirming the important role of the sea at small $x$ found within the chiral quark-soliton model~\cite{PPPBGW98,Ossmann05,Wakamatsu05}. Although we adopted instant-form hadron wave function which are SU(6) symmetric, the helicity flip GPDs are different from zero because of the orbital angular momentum generated by the Melosh rotations in the transformation from the canonical spin to the light-front spin. Therefore the faster  fall off of $E^q$ with respect to $H^q$ for $x\to 1$ is a consequence of the decreasing role of the Melosh transform to generate angular momentum in $E^q$ with increasing quark momentum. From the first moment of $E^q$ at $t=0$ and $\xi=0$ we obtain the following values for the quark anomalous magnetic moment: $\kappa^u=1.94$ and $\kappa^d=-1.73$, corresponding to $
\kappa^p=1.87$ and $\kappa^n=-1.80$. These values include a contribution from the higher Fock states of the order of $15\%,$ which is essential to bring the values of $\kappa ^N$ pretty close to the experimental values $\kappa^p=1.793$ and $\kappa^n=-1.913$.

Although we use a covariant formalism based on the LCWF overlap representation of GPDs, one notices from Fig.~\ref{gpd_summary} that the first moment of $H^q$ and $E^q$ is not $\xi$-independent as required by Lorentz invariance. This is a consequence of the model calculation where we introduced phenomenological LCWFs which violate Lorentz symmetry. As outlined in Ref.~\cite{HM07}, the functional form of LCWFs is dictated by the underlying Lorentz symmetry, i.e. the longitudinal and transverse variables are tied to each other in a certain but non apparent manner. Hence, phenomenological LCWFs may result in a violation of the GPDs polynomiality property.

With a nonvanishing $\xi$ one can explore the ERBL region with 
$\vert x\vert\le \xi$ where  in our model only transition amplitudes between
 the bare-proton and baryon-meson components are contributing. We see that the
 GPDs in the ERBL region are rather regular functions over the whole range, 
with zeros at the endpoints $x=\pm\xi$. This result is quite different from
 the oscillatory behaviour predicted by the chiral quark-soliton 
model~\cite{PPPBGW98} where the valence contribution of the discrete level is 
a smooth function extending into the ERBL region and adding to the sea 
contribution. Here this is forbidden because the support of the valence
 contribution is limited to the DGLAP region.

The results for the bare-nucleon and $BM$-component contributions to the 
helicity-dependent GPDs
are shown in Fig.~\ref{gpd_tilde_summary}.
Comparing the results at $\xi=0$ and different $t$ values, we see that
 both $\tilde H^u$ and
$\tilde H^d$ exhibit a small $t$-dependence. At constant $t=-0.2$ GeV$^2$, 
the spin-averaged GPDs $\tilde H$ have a rather weak  
$\xi$-dependence, at variance with the behaviour of the spin-flip GPDs $\tilde E$.  Due to
the opposite sign of $\tilde H^u$ and $\tilde H^d$ in all kinematic conditions,
their difference is positive and peaked at a value of $x$ comparable
to the results obtained in the chiral quark-soliton model in the leading order of
the $1/N_c$ expansion~\cite{Penttinen00}. 
Also $\tilde E^u$ and $\tilde E^d$ have
opposite sign as functions of $x$, and at small $x>0$ exhibit a negative and 
positive peak, respectively, with a sign change at intermediate value of $x$.
The contribution in the ERBL region from the interference between
$BM$ higher-Fock states and the bare-nucleon state is more pronounced
in the case of the spin-flip GPDs. 
The isovector combination $\tilde E^u-\tilde E^d$ in the ERBL region has the 
same  negative sign as in  quark-soliton model, but
it is much smaller in absolute value~\cite{Penttinen00}.
We also note that in our calculation the pion-pole contribution, giving
a large contribution in the ERBL region, is not taken into account.

At negative $x$, the only contribution is from the antiquark of the meson
substates. 
In the case of the helicity-dependent GPDs, the non-vanishing contributions
from the diagonal terms with the same meson in the initial and final states
are from  the $\rho$ and $\omega$ mesons, while the 
 pion can enter only in the 
interference terms with the $\omega$ and $\rho.$ 
Since the probability to find the $\rho$ or the $\omega$ meson in the dressed
nucleon is much smaller than that for the pion~\cite{PB07a},
we find that 
the meson contribution to the helicity-dependent GPDs is much smaller than 
in the case of the unpolarized GPDs, and it is almost vanishing
at all values of $x$.

\begin{table}[b]
\caption{Contributions to the angular momentum sum rule~(\ref{eq:jisumrule})
 from the bare nucleon (second column), the meson cloud at positive $x$ (third column), the meson cloud at negative $x$ (fourth column), and the total result (last column).}
\label{tab2}
\begin{center}
\begin{tabular}{ccccc}
\hline \\
{} & bare nucleon & $BM$ ($x>0$) & $BM$ ($x<0$) & TOT \\ \\ \hline \\
$A^u_{2,0}(0)$   & 0.61  & 0.047   & 0.0055   & 0.66 \\
$A^d_{2,0}(0)$   & 0.30  & 0.032   & 0.0086   & 0.34 \\
$B^u_{2,0}(0)$ & 0.43  & 0.027   & -0.0026  & 0.45 \\
$B^d_{2,0}(0)$ & -0.43 & -0.040  & 0.015    & -0.45  \\
$\Sigma^u$             &  0.90 & 0.062   & -0.00087 & 0.96 \\
$\Sigma^d$             & -0.23 & -0.014  & 0.00018  & -0.24 \\
$L^u$                  &  0.07  & 0.006  & 0.002    & 0.078\\
$L^d$                  &  0.05  & 0.003 &  0.012 & 0.065\\
\\ \hline \\
\end{tabular}
\end{center}
\end{table}
The different GPD moments entering the angular momentum sum
 rule~(\ref{eq:jisumrule}) are presented in Table~\ref{tab2}, with the separate
 contributions from the bare nucleon and the higher Fock-state components at $x>0$ and $x<0$. In all cases the contribution from the $BM$ components of the
 dressed nucleon 
are within a few percent of the dominant bare-nucleon contribution.
We note that $\Sigma^u$ is much larger than  $\Sigma^d,$ while the
angular momentum contribution is small and of the same amount
for both flavour quarks, despite the fact that there are less down quarks in the
 proton.
Consistently with the fact that in our model we do not have gluons and 
therefore the nucleon momentum is saturated by quarks and antiquarks, we find
$\sum_q B^q_{2,0}(0)=0$.

These results are all at the hadronic scale of the model, and evolution should be applied before comparing with other model calculations and experiments referring at higher scales. 


\section{Applications to observables}
\label{sect:appl}


\subsection{Electroweak form factors}
\label{sect:applone}

Electroweak form factors of the nucleon have played a privileged role in the investigation of the nucleon structure and a variety of models have been devised to account for data (for recent reviews, see~\cite{Gao03,Hyde04,ADZ06,PPV06,Kees06}). In the present context they are considered as first Mellin moments of GPDs. As such they have been used as constraints in building phenomenological GPDs as discussed in sect.  4.4. Conversely, model GPDs can be tested in a comparison with the observed form factors as done in a recent investigation based on the meson-cloud model~\cite{PB07a}. 

According to the analysis of Ref.~\cite{FW03} a pronounced bump structure in the neutron electric charge form factor $G_E^n(Q^2)$ (and a dip in  the other nucleon form factors) around $Q^2=0.2-0.3$ GeV$^2$ can be appreciated and  interpreted as a signature of a very long-range contribution of the pion cloud surrounding the bare-nucleon core and extending out to 2 fm~\footnote{
Throughout this section the notation $Q^2=-t$ is used for the four-momentum transfer.
}. 
This behaviour has been reproduced in the Lorentz covariant quark model of Ref.~\cite{FGLP06}. It must be said, however, that from dispersion relation analysis~\cite{Hammer04} the pion cloud should peak much more inside the nucleon, at $\sim 0.3$ fm, and the desired bump-dip structure of Ref.~\cite{FW03} can only be achieved at the cost of low-mass poles close to the $\omega$ mass in the isoscalar channel and to the three-pion threshold in the isovector channel~\cite{BHM07}. 

\begin{figure}[t]
\begin{center}
\epsfig{file=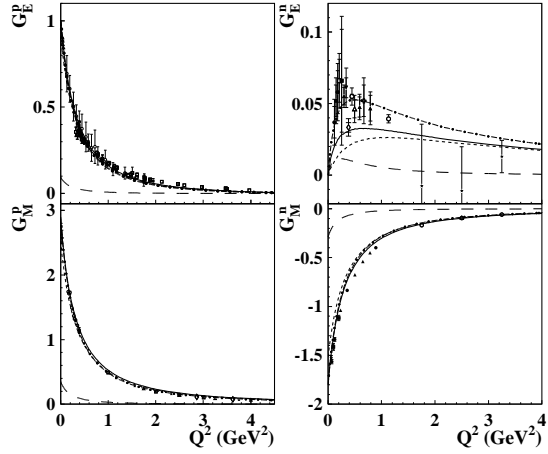,  width=8cm}
\end{center}
\caption{ The four nucleon electromagnetic form factors compared with the world data considered in the analysis of Ref.~\cite{FW03} and the recent JLab data~\cite{Punjabi05} using $G_E^p=(\mu^p G_E^p/G_M^p)/(1+Q^2/0.71{ \rm GeV}^2)^2$ (open squares). Long-dashed curves for the contribution of the meson cloud; dashed curves for the valence-quark contribution with SU(6) instant-form nucleon wave function; solid curves for the sum of the two contributions; dashed-dotted curves for the total result 
with 1\% mixed-symmetry  $S'$-state in the bare nucleon wave function (taken from Ref.~\cite{PB07a}). 
}
\label{form_factors}
\end{figure}

\begin{figure}[t]
\begin{center}
\epsfig{file=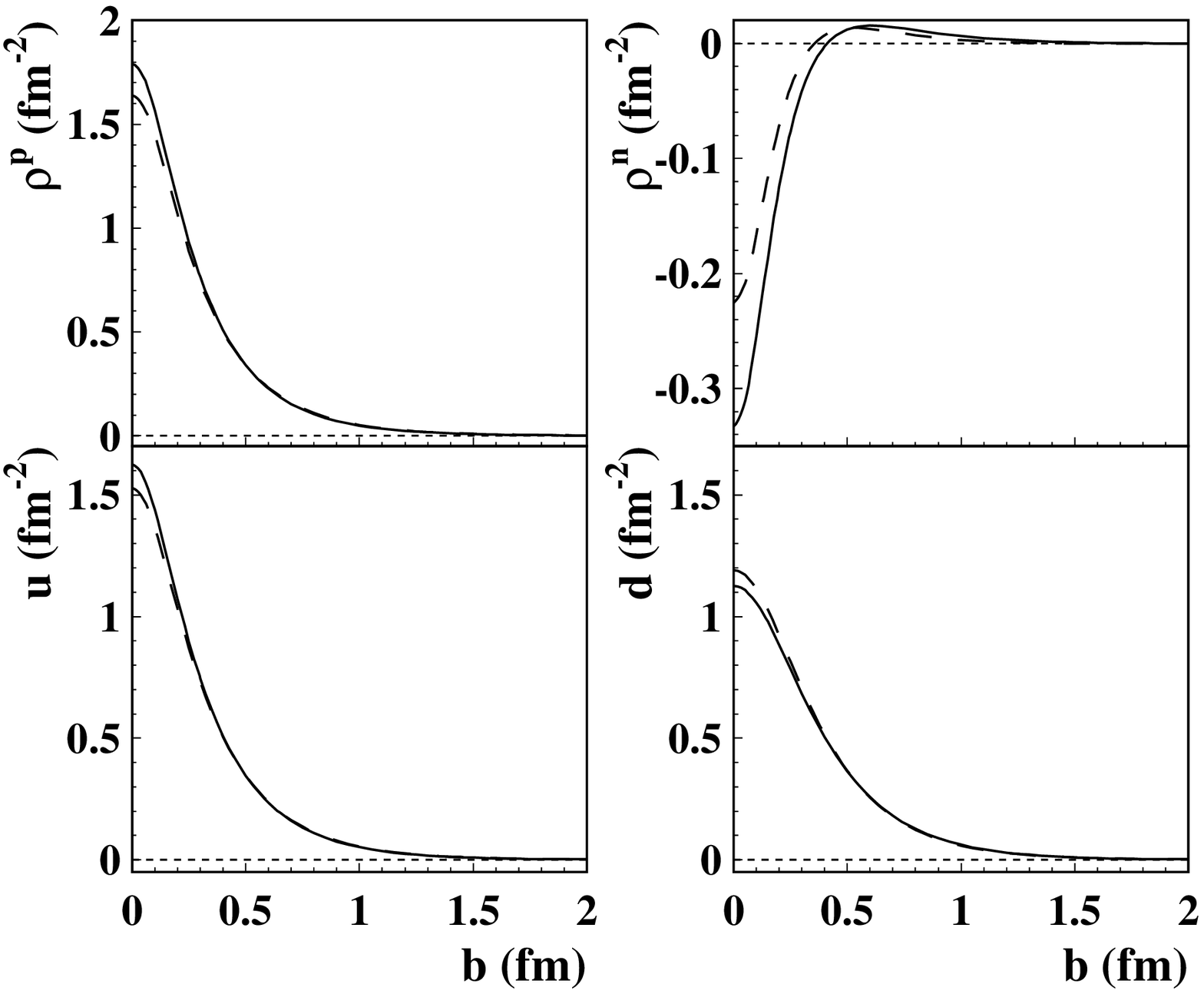,  width=9cm}
\end{center}
\vspace{-0.4cm}
\caption{\small Upper panels: The proton (left) and neutron (right) charge densities as a function of the impact parameter $b$. Lower panels: The up (left) and down (right) quark transverse charge densities in the proton. Solid curves for a SU(6)-symmetric instant-form wave function, dashed curves with mixed-symmetry components (taken from Ref.~\cite{PB07a}). }
\label{fig:fig8}

\begin{center}
\epsfig{file=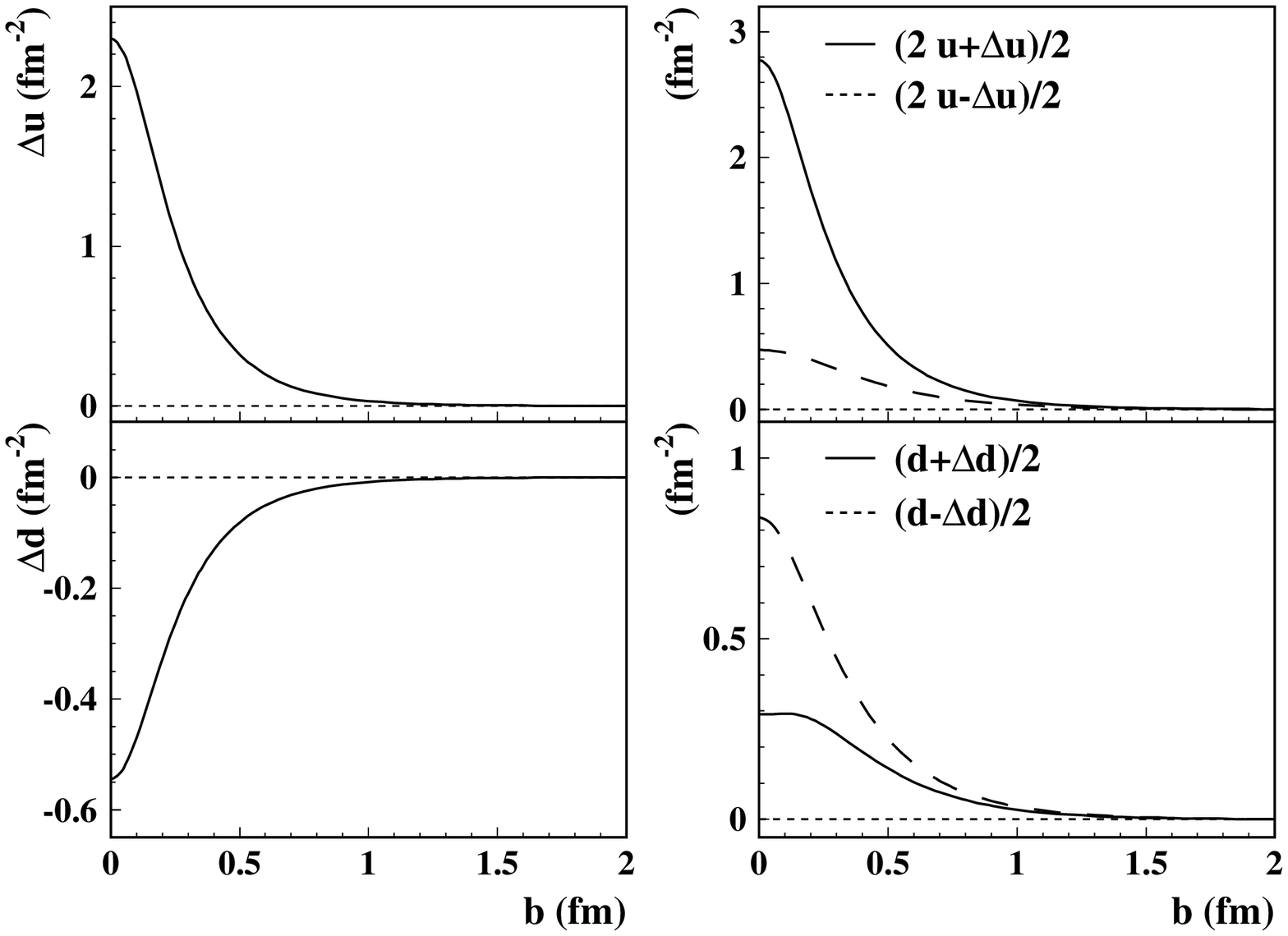,  width=9cm}
\end{center}
\vspace{-0.4cm}
\caption{\small Transverse distribution of up and down quarks in a longitudinally polarized proton as a function of the impact parameter $b$. Left panels: the axial contributions $\Delta u$ and $\Delta d$ for up and down quarks, respectively. Right panels: total contribution for quarks polarized in the longitudinal direction, either parallel (solid lines) or antiparallel (dashed lines) to the proton helicity (taken from~\cite{PB07a}).}
\label{fig:fig10}
\end{figure}

As a matter of fact, while giving an overall good description of all the electroweak form factors, the meson-cloud model is unable to reproduce such a bump/dip structure (Fig.~\ref{form_factors}).
In any case the contribution of the meson cloud is smooth and only significant below $Q^2=0.5$ GeV$^2$. For the neutron electric form factor it is essential to also include a small percentage ($1-2\%$) of mixed-symmetry $S'$-wave momentum component in the three-quark wave function, in agreement with earlier findings~\cite{mixed-s} and only slightly affecting the other form factors. 

Charge and magnetization densities are encoded in the electromagnetic form factors, but their radial distribution cannot be extracted without ambiguities~\cite{Kelly02}. Neglecting relativistic corrections, in the Breit frame they are given by the Fourier-Bessel transform of the nucleon electromagnetic Sachs form factors. In the meson-cloud model  the meson cloud manifests itself as a slight extension of the radial distribution up to $\sim 2 $ fm. In the neutron  case the resulting charge distribution shows a positive core surrounded by a negative surface charge pushed outwards by the meson cloud  and peaking at $\sim 0.8$ fm, in agreement with the analysis of Ref.~\cite{Kelly02} and the expectation based on the picture of a hadron's periphery caused by the pion cloud~\cite{Lvov}.

In contrast, the density $q(b)$ of partons of type $q$ in the transverse (impact parameter) plane with respect to the direction of the three-momentum transfer can be unambiguously determined by taking the two-dimensional Fourier transform of the Dirac form factor $F^q_1$:
\be
\label{eq:rhodib}
q(b) = \frac{1}{2\pi}\int_0^\infty dQ\,Q \,J_0(Qb) F^q_1(Q^2),
\ee
where $J_0 $ is a cylindrical Bessel function~\cite{Miller07}. Assuming that only up and down quarks are in the nucleon and invoking isospin symmetry, from the relations
\be
\label{eq:chargedensity}
\rho^p(b) = \frac{4}{3} u(b) - \frac{1}{3} d(b), \quad
\rho^n(b) = -\frac{2}{3} u(b) + \frac{2}{3} d(b)
\ee
one derives the proton and neutron transverse charge distributions. In Eq.~(\ref{eq:chargedensity}), $u(b)$ and $d(b)$ are the up and down quarks distributions in the proton, which are the same as for down  and up quarks in the neutron, respectively (with normalization $\int d^2 \tvec{b}\, u(b) = \int d^2\tvec{b}\, d(b) =1$). The corresponding charge densities for the proton and the neutron are plotted in the upper panels of Fig.~\ref{fig:fig8}. As in the phenomenological analyses of Refs.~\cite{Miller07,carlson} the densities are concentrated at low values of $b$ with a positive peak for the proton and a negative peak for the neutron. While the negative tail of the neutron distribution has the well known interpretation in terms of the pion cloud, the negative charge density near the origin appears to be mysterious. An intuitive  understanding of this result has been recently given in Ref.~\cite{Burkardt}, suggesting that the up quark in the neutron has a  larger p-wave component than down quarks, being therefore suppressed at the origin as shown in the lower panels of Fig.~\ref{fig:fig8}.

By Fourier transforming the quark contributions to the Dirac and axial form factors one has access to the probability $\rho^q(b,\lambda,\lambda_N)$ to find a quark with transverse position $b$ and light-cone helicity $\lambda$ ($=\pm 1$) in the nucleon with longitudinal polarization $\lambda_N$ ($=\pm 1$), i.e. 
\bea
\rho^q(b,\lambda,\lambda_N) &=& \frac{1}{2}\int d^2\tvec{q}_\perp \left[ F^q_1(Q^2=\tvec{q}_\perp^2) +\lambda\lambda_N g^q_A(Q^2=\tvec{q}_\perp^2) \right]\,e^{i\svec{q}_\perp\cdot\svec{b}} \\
&\equiv& \frac{1}{2}\left[q(b) +  \lambda\lambda_N \Delta q(b)\right],
\nonumber
\eea
where $q(b)$ was already defined in Eq.~(\ref{eq:rhodib}) and $\Delta q(b)$ is the Fourier transform of $g^q_A(Q^2)$, normalized as $\int d^2\tvec{b} \, \Delta u(b) = 0.96$ and $\int d^2\tvec{b} \, \Delta d(b) = -0.24$ (see Table~\ref{tab2}). Assuming a positive proton helicity ($\lambda_N=1$) the resulting probability is shown in the right panels of Fig.~\ref{fig:fig10}. The axial contributions $\Delta u (b)$ and $\Delta d (b)$ for up and down quarks (left panels), respectively, have opposite sign. When suitably combined with the corresponding transverse distributions $u(b)$ and $d(b)$ the positive helicity up quarks in the proton are found to be preferentially aligned with the proton helicity, while the opposite occurs for down quarks.


\subsection{Form factors of the energy-momentum tensor}

The nucleon form factors of the energy-momentum tensor were subject to modest interest in the literature for a long time. After some pioneering analysis of the mass structure of the nucleon in terms of quark and gluon contributions to matrix elements of the energy-momentum tensor~\cite{Ji95}, only very recently, in connection with the advent of GPDs accessible in hard exclusive processes, their importance has been appreciated. Generalized form factors of the quark part of the energy-momentum tensor have been calculated in lattice simulations in Refs~\cite{Mathur00,Gadiyak02} and more recently by the LHPC~\cite{LHPC03,LHPC04,LHPC05a,Hagler07,LHPC05} and QCDSF~\cite{QCDSF04,QCDSF05,QCDSF06c,QCDSF04a} collaborations. Quite recently they have also been studied in the $\chi$QSM~\cite{GGOPSSU07,GGOSSU07,WN07} and in the Skyrme model~\cite{CGOS07}.

In lattice simulations a comprehensive study of the lowest moments of the generalized form factors is possible at present for pion masses as low as 350 MeV and volumes as large as (3.5 fm)$^3$~\cite{Hagler07}, providing results at the scale $\mu^2=4$ GeV$^2$ in the $\overline{\rm MS}$ renormalization scheme.

The $t$ dependence of the different Mellin moments can be fitted to a dipole form, with a clear flattening of the $t$ slope with increasing order of moment~\cite{LHPC04,QCDSF05a}. The absolute values in the isovector and isosinglet channels are in qualitative agreement with the predictions from large $N_c$ counting rules for the unpolarized generalized form factors, i.e.
\bea
& & \vert A_{2,0}^{u+d}\vert \sim N_c^2 \gg \vert A_{2,0}^{u-d}\vert \sim N_c, \nonumber\\
& & \vert B_{2,0}^{u-d}\vert \sim N_c^3 \gg \vert B_{2,0}^{u+d}\vert \sim N_c^2, \nonumber\\
& & \vert C_{2,0}^{u+d}\vert \sim N_c^2 \gg \vert C_{2,0}^{u-d}\vert \sim N_c,
\eea
whereas in the polarized case counting rules are only partially satisfied~\cite{Hagler07}.

\begin{figure}[t]
\begin{center}
\epsfig{file=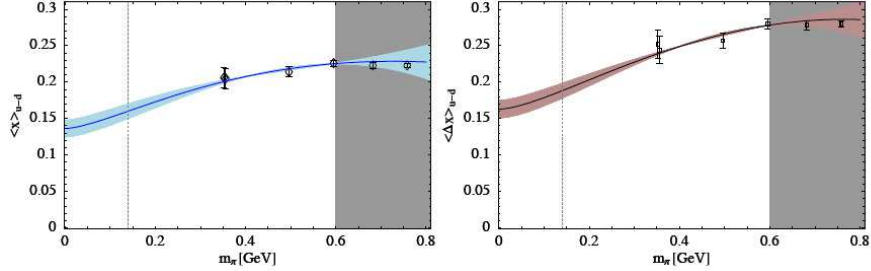,  width=12cm}
\end{center}
\caption{\small  Fit of the isovector moments $A^{u-d}_{2,0}(0)\equiv\langle x\rangle_{u-d}$ and $\tilde A^{u-d}_{2,0}(0)\equiv\langle\Delta x\rangle_{u-d}$ in the ${\cal O}(p^2)$ covariant BChPT~\cite{Dorati} to the LHPC lattice data of Ref.~\cite{Hagler07} (solid line). The bands shown indicate estimate of higher order possible corrections. The lattice results refer to a scale $\mu^2=4$ GeV$^2$ (taken from~\cite{Dorati}).}
\label{fig:fig9}
\end{figure}

Extrapolating to the physical pion mass requires a combination of full QCD lattice calculations and chiral perturbation theory (ChPT). Significant progress  has been made in several approaches to ChPT like heavy-baryon ChPT (HBChPT)~\cite{chiral,DMS,ACK06}, covariant ChPT in the baryon sector (BChPT)~\cite{Dorati}, self-consistently improved one-loop BChPT~\cite{Beane07}, and BChPT with finite regulators~\cite{Detmold01,Wang07}. As an example, in Fig.~\ref{fig:fig9} the isovector moments $A^{u-d}_{2,0}(0)$ and $\tilde A^{u-d}_{2,0}(0)$ are shown as a function of the pion mass obtained in lattice calculations~\cite{Hagler07} and in the covariant BChPT at ${\cal O}(p^2)$~\cite{Dorati}. The chiral curvature in both observables naturally bends down to the phenomenological value for lighter quark masses, leading to a very satisfactory extrapolation curve.

The generalized form factors $A^q_{2,0}(t=0)$ and $B^q_{2,0}(t=0)$ enable us to compute the total angular momentum contribution to the nucleon spin $J^q$ according to Ji's sum rule (\ref{eq:jisumrule}). In turn, the quark spin is given by $\oneh\Delta\Sigma^q=\oneh\tilde A^q_{1,0}(t=0)=\oneh\int dx\,\tilde H^q(x,0,0)$, so that the orbital angular momentum can be derived from lattice simulations versus the pion mass $m_\pi$. Two remarkable features are found. The first is that the magnitude of the orbital angular momentum contributions of the up and down quarks are separately sizable, $L^u\approx -L^d\approx 0.30$, yet they cancel nearly completely at all pion masses, $L^{u+d}\approx 0$~\cite{Hagler07,brommel}. The  second is the close cancellation between the orbital and spin contributions of the down quarks for all pion masses, $J^d\approx 0$~\cite{Hagler07,brommel}. Therefore these results indicate that the total angular momentum of quarks in the nucleon is of the same size as the quark spin contribution, while the orbital angular momentum is consistent with zero.

In the Breit frame characterized by $\Delta^0=0$ one defines the static energy-momentum tensor for quarks (and analogously for gluons)~\cite{GGOPSSU07},
\be\label{Def:static-EMT}
    {\Theta}^{\mu\nu}_q({\tvec r},{\tvec s}) =
    \frac{1}{2E}\int\frac{d^3\tvec \Delta}{(2\pi)^3}\;\exp(i\tvec \Delta\cdot{\tvec r})\;
    \bra{p^\prime,S^\prime}{\mit\Theta}^{\mu\nu}_q(0)\ket{p,S},
\ee
with the initial and final polarization vectors of the nucleon $S$ and $S^\prime$ defined such that they are equal to $(0,{\tvec s})$ in the respective rest-frame, where the unit vector ${\tvec s}$ denotes
the quantization axis for the nucleon spin.

The components of $\Theta_{q,0 k}({\tvec r},{\tvec s})$ and $\varepsilon^{i j k} r_j \Theta_{q,0k}({\tvec r},{\tvec s})$ correspond respectively to the distribution of quark momentum and quark angular momentum inside the nucleon. The components of $(\Theta_{q,ik}-\frac 13\delta_{ik}\Theta_{q,ll})({\tvec r},{\tvec s})$ characterize the spatial distribution of `shear forces' experienced by quarks inside the nucleon. The respective form factors are related to $\Theta^{\mu\nu}_q({\tvec r},{\tvec s})$ by
\begin{eqnletter}
\qquad{}  & &  J_q(t)+\frac{2t}{3}\, {J_q}^\prime(t)
    = \int d^3{\tvec r}\, e^{-i{\svec r}\cdot\svec\Delta}\,
    \varepsilon^{ijk}\,s_i\,r_j\,\Theta_{q,0k}({\tvec r},{\tvec s}) ,
    \label{Eq:ff.r1}\\
  & &  d_1^q(t)+\frac{4t}{3}\, {d_1^q}^\prime(t)
    +\frac{4t^2}{15}\, {d_1^q}^{\prime\prime}(t)
    = -\frac{M_N}{2} \int d^3{\tvec r}\,e^{-i{\svec r}\cdot\svec\Delta}\,
    \Theta_{q,ij}({\tvec r})\,\left(r^i r^j-\frac{{\tvec r}^2}3\,\delta^{ij}\right),
 \label{Eq:ff.r2}\\
  & &  M_2(t)-\frac{t}{4M_N^2}\left(M_2(t)-2 J(t)+\frac 45\, d_1(t) \right)
    = \frac{1}{M_N}\int d^3{\tvec r}\, e^{-i{\svec r}\cdot\svec\Delta}
    \, \Theta_{00}({\tvec r},{\tvec s}),
    \label{Eq:ff.r3}
\end{eqnletter}
where the prime denotes derivative with respect to $t$. For a spin-$\oneh$ particle only the $\Theta^{0\mu}$-components are sensitive to the polarization vector. Eq.~(\ref{Eq:ff.r3}) holds for the sum $\Theta_{00}\equiv \sum_q \Theta_{q,00}+\Theta_{g,00}$ with $M_2(t)\equiv \sum_q M_2^q(t)+M_2^g(t)$ and $J(t)$ and $d_1(t)$ defined analogously, but not for the separate quark and gluon contributions, since otherwise the form factor $\bar c(t)$ would not cancel out.

The form factors $M_2^{q,g}(t)$, $J_{q,g}(t)$ and $d_1^{q,g}(t)$ depend on the renormalization scale $\mu$. Their quark+gluon sums, however, are scale independent form factors. For $m_\pi\ne 0$ within the framework of the $\chi$QSM, $M_2(t)$, $J(t)$ and $d_1(t)$ can well be approximated by dipole fits~\cite{GGOPSSU07}. $J(t)$ exhibits a similar $t$ dependence as the electric Sachs form factor $G_E(t)$. However, $M_2(t)$ falls off with increasing $\vert t\vert$ slower than $G_E(t)$, while $d_1(t)$ shows a faster fall off. These results indicate that factorizing the $t$ dependence of GPDs is quite a rough approximation in the $\chi$QSM.

At $t=0$ $M_2(t)$, $J(t)$ and $d_1(t)$ satisfy the constraints,
\begin{eqnletter}
    M_2(0)&=&
    \frac{1}{M_N}\,\int d^3{\tvec r}\;\Theta_{00}({\tvec r},{\tvec s})=1
    ,\label{eq:m2.r1}\\
    J(0)&=&
    \int d^3{\tvec r}\;\varepsilon^{ijk}\,s_i\,r_j\,
    \Theta_{0k}({\tvec r},{\tvec s})=\frac12
    ,\label{eq:m2.r2}\\
    d_1(0)&=&
    -\frac{M_N}{2}\, \int d^3{\tvec r}\;\Theta_{ij}({\tvec r})\,
    \left(r^i r^j-\frac{{\tvec r}^2}3\,\delta^{ij}\right)\equiv d_1.
    \label{eq:m2.r3}
\end{eqnletter}
Eqs.~(\ref{eq:m2.r1}) and (\ref{eq:m2.r2}) simply mean that in the rest frame the total energy of the nucleon is equal to its mass and the spin of the nucleon is $\oneh$. The value of $d_1$ is not known a priori and must be determined experimentally. Remarkably, $d_1$ determines the behaviour of the $D$-term (and thus the unpolarized GPDs) in the asymptotic limit of renormalization scale $\mu\to\infty$~\cite{GPVdH}.

The form factor $d_1(t)$ is connected with the distribution of pressure and shear forces experienced by the partons in the nucleon~\cite{Polyakov03}, as can be verified by recalling that $\Theta_{ij}({\tvec r})$ is the static stress tensor which (for spin 0 and $\oneh$ particles) can be decomposed as
\be
\label{Eq:T_ij-pressure-and-shear}
    \Theta_{ij}({\tvec r})
    = s(r)\left(\frac{r_ir_j}{r^2}-\frac 13\,\delta_{ij}\right)
        + p(r)\,\delta_{ij} . 
\ee
Due to the conservation of the total energy-momentum tensor the functions $p(r)$ and $s(r)$ are related to each other by the differential equation
\be\label{Eq:p(r)+s(r)}
    \frac23\;\frac{\partial s(r)}{\partial r\;}+
    \frac{2s(r)}{r} + \frac{\partial p(r)}{\partial r\;} = 0\;.
\ee
Hereby $p(r)$ describes the radial distribution of the `pressure' inside the hadron, while $s(r)$ is related to the distribution of the `shear forces'~\cite{Polyakov03}. 

Another important property which can be directly derived from the conservation of the energy-momentum tensor is the so-called stability condition. Integrating $\int d^3{\tvec r}\,r^k(\nabla_i\Theta^{ij})\equiv 0$ by parts one finds that the pressure $p(r)$ must satisfy the relation
\be\label{Eq:stability}
    \int\limits_0^\infty \! d r\;r^2p(r)=0 \;.
\ee

\begin{figure}
\begin{center}
\includegraphics[width=13cm]{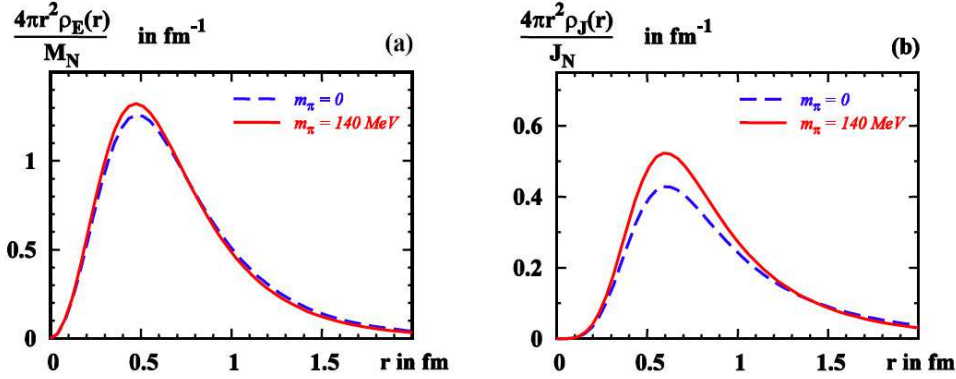}
\end{center}
\caption{\small (a) The normalized energy density $4\pi r^2\rho_E(r)/M_N$ from the $\chi$QSM as a function of $r$ in the chiral limit of $m_\pi=0$ and for $m_\pi=140\,{\rm MeV}$. (b) The same for the normalized angular momentum density $4\pi r^2\rho_J(r)/J_N$ (taken from~\cite{GGOPSSU07}).}
\label{fig:enspin-density}
\end{figure}

The energy density $4\pi r^2\rho_E(r)/M_N= 4\pi r^2\Theta_{00}(r)/M_N$ normalized according to (\ref{eq:m2.r1}) is shown in Fig.~\ref{fig:enspin-density}a as a function of $r$ in the chiral limit of a vanishing pion mass and for the physical situation with a pion mass of $140\,{\rm MeV}$. In this latter case the nucleon mass in the model is about $1250\,{\rm MeV}$. This overestimate of the physical nucleon mass of about 300 MeV is typical for the soliton approach and its origin is well understood~\cite{Pobylitsa92}. In the center of the nucleon the energy density is $\rho_E(0)=1.70$ GeV/fm$^3$ or $3.0\times 10^{15}$ g cm$^{-3}$, corresponding roughly to 13 times the equilibrium density of nuclear matter. As the pion mass decreases, the energy density is spread more widely. According to the role of the pion field in the $\chi$QSM, where one can associate the contribution of the discrete level to the quark core and the contribution of the negative continuum states to a pion cloud, this means that the range of the pion cloud increases and the nucleon becomes larger. Actually, the mean square radius $\langle r_E^2\rangle$ increases from 0.67 fm$^2$ in the case of the physical pion to 0.79 fm$^2$ in the chiral limit. With increasing pion mass up to 1.2 GeV this trend is confirmed with the nucleon becoming smaller and smaller~\cite{GGOSSU07}.

\begin{figure}
\begin{center}
\begin{tabular}{lll}
    \includegraphics[height=5cm]{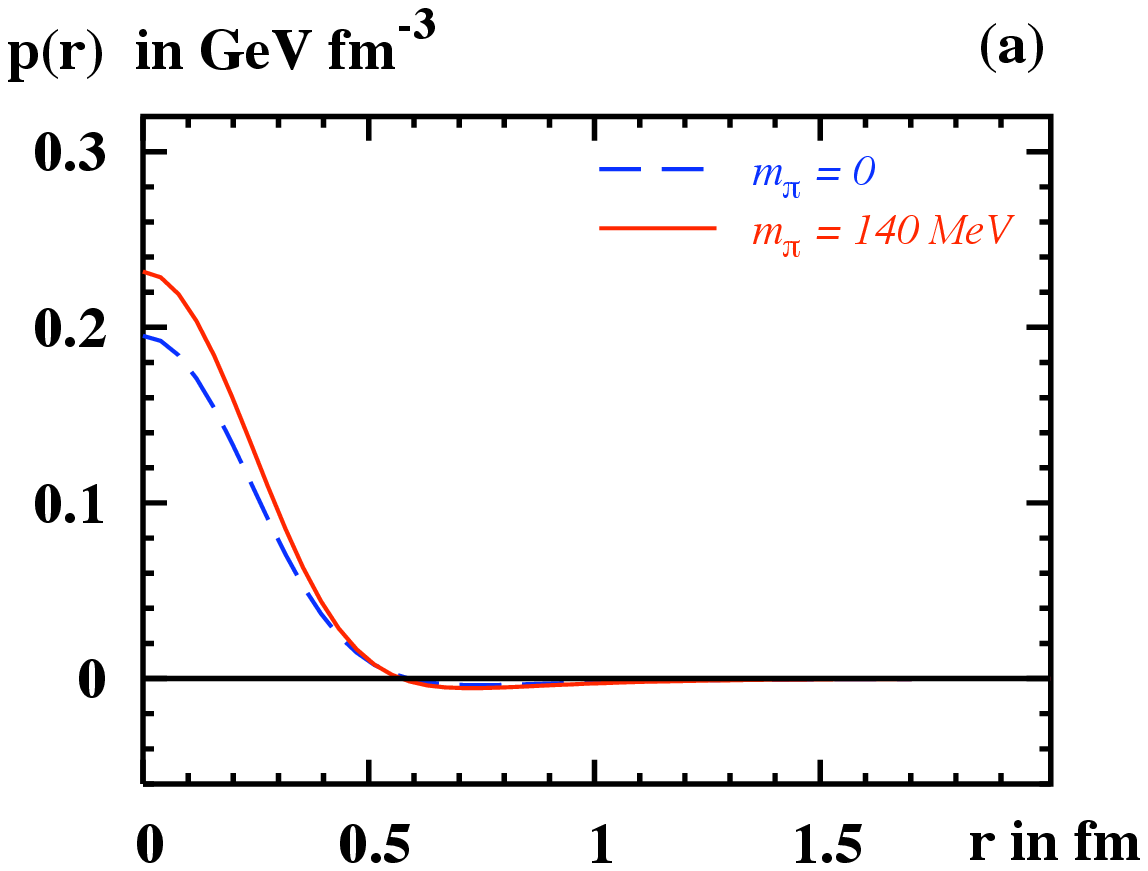}&
    \includegraphics[height=5cm]{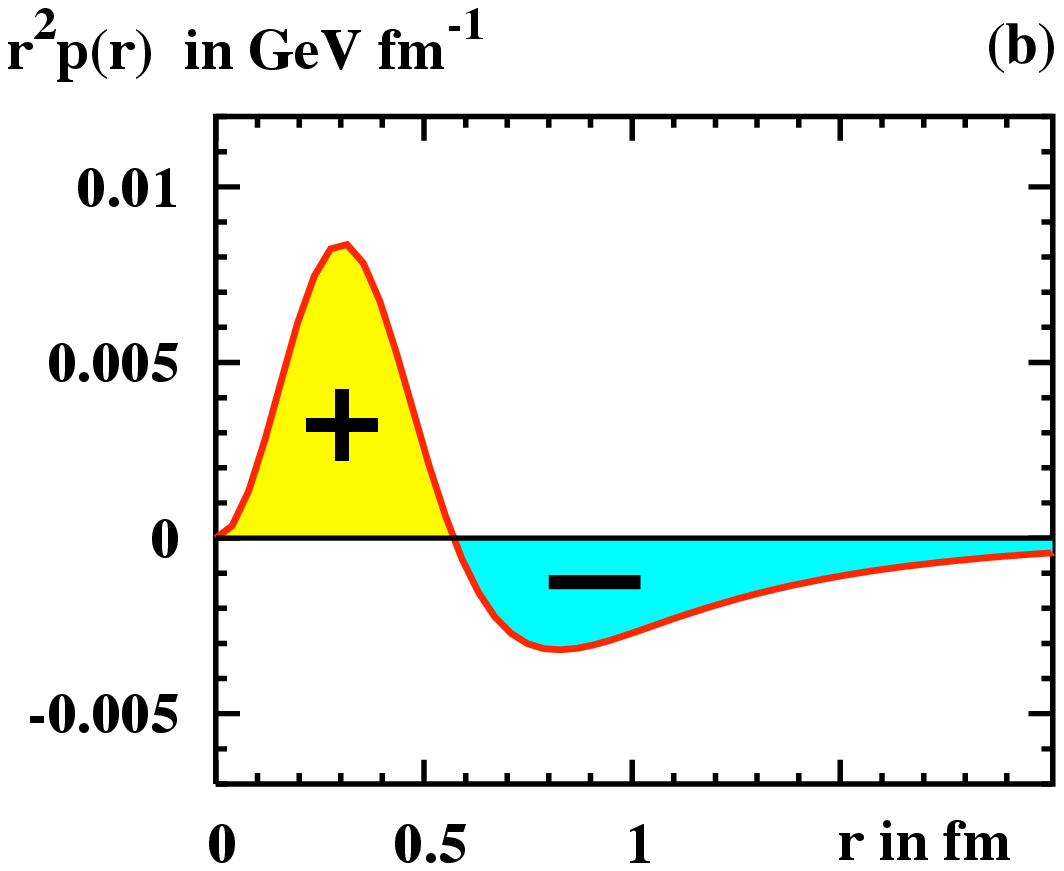} 
\end{tabular}
\end{center}
    \caption{\small (a) The pressure $p(r)$ from the $\chi$QSM as function of $r$ for $m_\pi=0$ and $140\,{\rm MeV}$. (b) $r^2 p(r)$ as function of $r$ at the physical value of $m_\pi$ (taken from~\cite{GGOPSSU07}).}
\label{fig:pressure}
\end{figure}

\begin{figure}
\begin{center}
  \includegraphics[height=5.3cm]{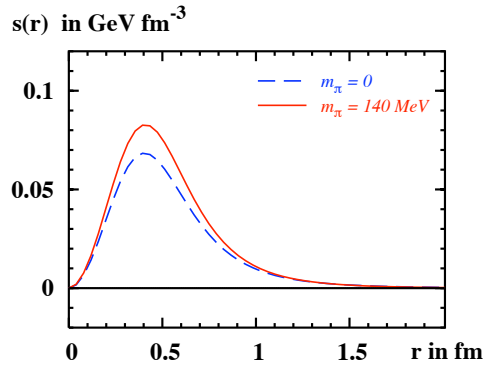}
 \caption{\small The function $s(r)$ describing the shear forces in the nucleon within the $\chi$QSM as a function of $r$ (taken from~\cite{GGOPSSU07}).}
\label{fig:shear}
\end{center}
\end{figure}

The angular momentum density $\rho_J(r)$ is related to the $\Theta_{0k}$ components of the static energy-momentum tensor as $\rho_J = \epsilon^{ijk} s_i x_j \Theta_{0k}$. The radial distribution of angular momentum $4\pi r^2\rho_J(r)/J_N$, normalized according to (\ref{eq:m2.r2}) with the nucleon spin $J_N=\oneh$, is shown in Fig.~\ref{fig:enspin-density}b as a function of $r$ for $m_\pi=0$ and $140\,{\rm MeV}$. For any $m_\pi$ at small $r$ one finds $\rho_J(r) \propto r^2$. The mean square radius $\langle r_J^2\rangle$ decreases with increasing $m_\pi$~\cite{GGOSSU07} in agreement with the idea of a shrinking pion cloud. For a physical pion one finds $\langle r_J^2\rangle= 1.32$ fm$^2$. At large $r$ in the chiral limit $\rho_J(r) \propto 1/r^4$ such that $\langle r_J^2\rangle$ diverges.

Fig.~\ref{fig:pressure}a shows the pressure $p(r)$ as function of $r$. In the physical situation $p(r)$ takes its global maximum at $r=0$ with $p(0) = 0.23\,{\rm GeV}/{\rm fm}^3 = 3.7\cdot10^{34}\,{\rm Pa}$. This is ${\cal O}(10\!-\!100)$ higher than the pressure inside a neutron star. Then $p(r)$ decreases monotonically (becoming zero at $r_0 = 0.57\,{\rm fm}$) till reaching its global minimum at $r_{p,\,\rm min}= 0.72\,{\rm fm}$, after which it increases monotonically remaining, however, always negative. The positive sign of the pressure for $r<r_0$ corresponds to the repulsion among quarks imposed by Pauli principle, while the negative sign in the region $r > r_0$ means attraction in agreement with the idea of a pion cloud responsible for binding the quarks to form the nucleon. The subtle balance between repulsion and attraction, ultimately producing a stable soliton, can be better appreciated from Fig.~\ref{fig:pressure}b where $r^2p(r)$ is shown as a function of $r$. The shaded regions have the same surface areas but opposite sign and cancel each other within numerical accuracy, thus satisfying the stability condition~(\ref{Eq:stability}).

Fig.~\ref{fig:shear} shows the distribution of the shear forces $s(r)$ obtained by solving the differential equation (\ref{Eq:p(r)+s(r)}). The  distribution of shear forces is always positive. For $m_\pi=140\,{\rm MeV}$ it reaches a global maximum at $r=0.40\,{\rm fm}$. The position of the maximum is weakly dependent on $m_\pi$. At small $r$, $s(r)\propto r^2$.


\subsection{Spin densities}

When $\xi=0$ and $x>0$, according to Refs.~\cite{Burkardt03,diehlhagler05} one defines three-dimensional densities 
\be
\rho(x,{\tvec b}, \lambda,\lambda_N) =  \oneh \left[{\mathcal H}(x,{b}^2) 
  + b^j\varepsilon^{ji} S^i  \frac{1}{M_N}\, 
       {\mathcal E}'(x,{b}^2)
  + \lambda \lambda_N \tilde{\mathcal H}(x,{b}^2) \,\right] ,
 \label{eq:long}
 \ee
\bea
 \label{eq:transv}
 \rho(x,{\tvec b},{\tvec s},{\tvec S}) 
&= &{}\dsp \oneh\left[ {\mathcal H}(x,{b}^2)  + s^iS^i\left( {\mathcal H}_T(x,{b}^2)  -\frac{1}{4M_N^2} \Delta_b \tilde {\mathcal H}_T(x,{b}^2) \right) \right.
\\
& &\quad{}\dsp + \frac{b^j\varepsilon^{ji}}{M_N}\left(
S^i{\mathcal E}'(x,{b}^2)  + s^i\left[ {\mathcal E}'_T(x,{b}^2)  + 2 \tilde{\mathcal H}'_T(x,{b}^2) \right]\right)
\nonumber\\
& &\quad\left.{}\dsp+ s^i(2b^ib^j - b^2\delta_{ij}) S^j\frac{1}{M_N^2} \tilde {\mathcal H}''_T(x,{b}^2) \right] .
\nonumber
 \eea
The distributions ${\mathcal H}$, ${\mathcal E}$, $\tilde {\mathcal H}$, ${\mathcal H}_T$, etc. are the Fourier transform of the corresponding GPDs as discussed in sect. 3.2 (see, e.g., Eq.~(\ref{impact-gpds})). In Eqs.~(\ref{eq:long}) and (\ref{eq:transv}) the shorthand notations
\be
f' = \frac{\partial}{\partial b^2}\, f ,
\qquad
f''= \Big( \frac{\partial}{\partial b^2} \Big)^2 f,
\qquad
\Delta_b f
= \frac{\partial}{\partial b^i}\, \frac{\partial}{\partial b^i}\, f
= 4\, \frac{\partial}{\partial b^2}
    \Big( b^2 \frac{\partial}{\partial b^2} \Big) f 
\label{eq:deriv}
\ee
have been used, and the two-dimensional antisymmetric tensor $\varepsilon^{ij}$ has been introduced with $\varepsilon^{12} = -\varepsilon^{21} = 1$ and $\varepsilon^{11} = \varepsilon^{22} = 0$. Roman indices are to be summed over.

Since $\Delta^+=0$, $\rho(x,{\tvec b}, \lambda,\lambda_N)$ and  $\rho(x,{\tvec b},{\tvec s},{\tvec S})$ represent the probability  to find a quark with longitudinal momentum fraction $x$ and transverse position $\tvec b$ either with light-cone helicity $\lambda$ ($=\pm 1$) in the nucleon with longitudinal polarization $\lambda_N$ ($=\pm 1$) or with transverse spin $\tvec s$ in the nucleon with transverse spin $\tvec S$. 

In Eq.~(\ref{eq:long}) the first term with ${\mathcal H}$ describes the density of unpolarized quarks in the unpolarized proton. The term with ${\mathcal E}'$ introduces a sideways shift in such a density when the proton is transversely polarized, and the term with $\tilde {\mathcal H}$ reflects the difference in the density of quarks with helicity equal or opposite to the proton helicity. 

In the three lines of Eq.~(\ref{eq:transv}) one may distinguish the three contributions corresponding to monopole, dipole and quadrupole structures. The unpolarized quark density $\oneh {\mathcal H}$ in the monopole structure is modified by the chiral-odd terms with ${\mathcal H}_T$ and $\Delta_b \tilde {\mathcal H}_T$ when both the quark and the proton are transversely polarized. Responsible for the dipole structure is either the same chiral-even contribution with ${\mathcal E}'$ from the transversely polarized proton appearing in the longitudinal spin distribution~(\ref{eq:long}) or the chiral-odd contribution with ${\mathcal E}'_T+2\tilde {\mathcal H}'_T$ from the transversely polarized quarks or both. The quadrupole term with $\tilde {\mathcal H}''_T$ is present only when both quark and proton are transversely polarized.

Lattice calculations accessing the lowest two $x$-moments of the transverse spin densities of quarks in the nucleon have recently been presented~\cite{QCDSF06a}, and impact parameter dependent parton distributions in phenomenological models of hadron LCWFs have been studied in Ref.~\cite{DMR07}.

\begin{figure}
\begin{center}
\epsfig{file=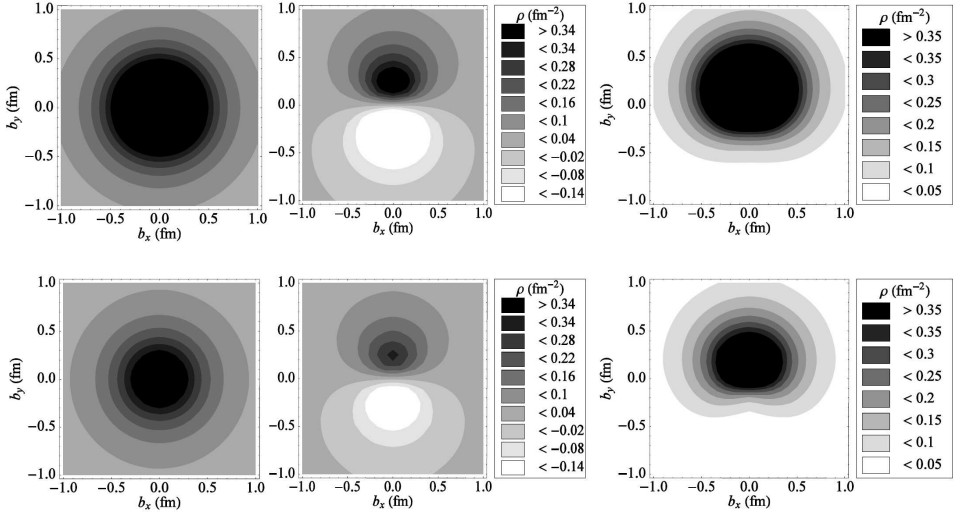,width=13 cm}
\end{center}
\caption{\small The monopole contribution $\frac{1}{2}{\mathcal H}$ (left) for unpolarized quarks, the dipole contribution $-\frac{1}{2} s_x b_y ({\mathcal E}'_T+2\tilde {\mathcal H}'_T)/M_N$ (middle) for (transversely) $\hat x$-polarized quarks, and the sum of both (right) in an unpolarized proton. The upper (lower) row gives the results for up (down) quarks (taken from~\cite{PB07}).}
\label{fig:spindens1}
\end{figure}

In Ref.~\cite{PB07} the first $x$-moments of the spin distributions
\be
\rho({\tvec b}, \lambda,\lambda_N) =\int dx\, \rho(x,{\tvec b}, \lambda,\lambda_N), \quad
 \rho({\tvec b},{\tvec s},{\tvec S}) = \int dx\,  \rho(x,{\tvec b},{\tvec s},{\tvec S})
 \label{eq:moments}
\ee
have been studied as functions of the transverse position and different quark and proton polarizations taking advantage of the overlap representation of LCWFs that was originally proposed in Refs.~\cite{DFKJ01,BDH01} and successfully applied to GPDs~\cite{BPT03,BPT04,PPB05,PPB06a,PPB06b}. 

\begin{figure}
\begin{center}
\epsfig{file= 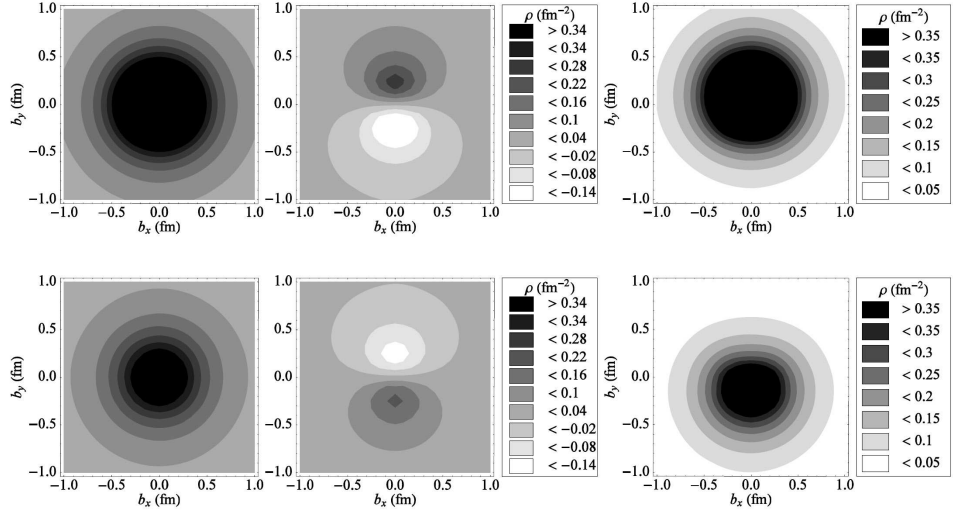,width=13 cm}
\end{center}
\caption{\small With unpolarized quarks the monopole contribution $\frac{1}{2}{\mathcal H}$ (left) for an unpolarized proton, the dipole contribution $-\frac{1}{2} S_x b_y {\mathcal E}'/M_N$ for a (transversely) $\hat x$-polarized proton, and the sum of both (right). The upper (lower) row gives the results for up (down) quarks (taken from~\cite{PB07}).}
\label{fig:spindens2}
\end{figure}

\begin{figure}
\begin{center}
\epsfig{file=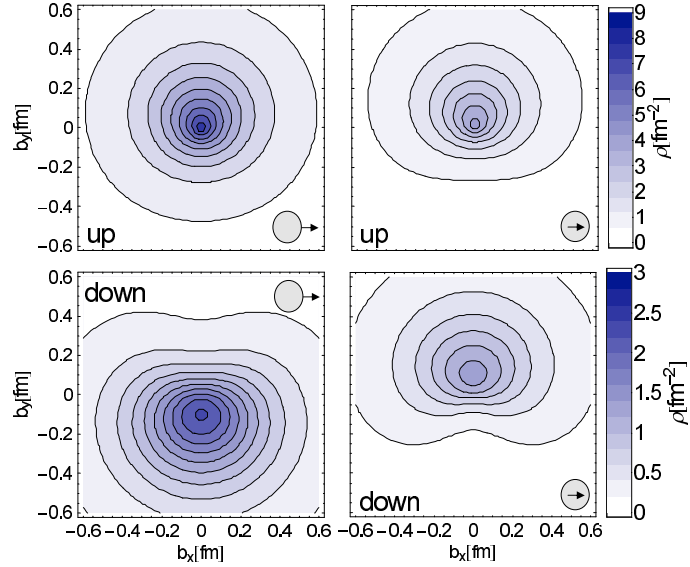,width=9 cm}
\end{center}
\caption{\small Densities of unpolarized quarks in a transversely polarized nucleon (left) and transversely polarized quarks in an unpolarized nucleon (right) for up (upper plots) and down (lower plots) quarks. The nucleon and quark spins are oriented in the transverse plane as indicated, where the inner arrow represents the quark and the outer arrow the nucleon spin (taken from~\cite{QCDSF06a}).}
\label{fig:spindenslat}
\end{figure}

With such a model only valence quarks are considered. Therefore, the integrals in Eq.~(\ref{eq:moments}) are restricted to $0\le x\le 1$.  In any case, for the unpolarized density this is not a limitation as the sea quark contribution drops out of the first moment of the GPD $H$. For the spin-dependent densities this restriction is not dramatic because valence quarks are known to dominate at large and intermediate $x$ ($x\ge 0.2$).

In Figs.~\ref{fig:spindens1} and \ref{fig:spindens2} the distorting effect of the dipole terms due to the transverse spin distributions on the monopole terms corresponding to  spin densities for unpolarized quarks in an unpolarized target is shown. If one multiplies the up and down monopole terms by the quark charge $e_q$ and sums over quark flavours, one obtains the nucleon parton charge density in transverse space, as discussed in sect.~\ref{sect:applone}.

For transversely polarized quarks in an unpolarized proton the dipole contribution introduces a large distortion perpendicular to both the quark spin and the momentum of the proton (Fig.~\ref{fig:spindens1}). Evidently, quarks in this situation also have a transverse component of orbital angular momentum. This effect has been related~\cite{Burkardt05b,Burkardt07} to a nonvanishing Boer-Mulders function~\cite{BM98} $h_1^\perp$ which describes the correlation between intrinsic transverse momentum and transverse spin of quarks. Such a distortion reflects the large value of  the anomalous tensor magnetic moment $\kappa_T$ for both flavours, i.e. $\kappa^u_T=3.98$ and $\kappa^d_T=2.60$, to be compared with the values $\kappa^u_T\approx 3.0$ and $\kappa^d_T\approx 1.9$ of Ref.~\cite{QCDSF06a} due to a positive combination ${\mathcal E}_T+2\tilde {\mathcal H}_T$. Since $\kappa_T\sim - h_1^\perp$, the results of Ref.~\cite{PB07} confirm the conjecture of Refs.~\cite{Burkardt05b,Burkardt07} that $h_1^\perp$ is large and negative both for up and down quarks.

As also noticed in Refs.~\cite{Burkardt00a,QCDSF06a} the large anomalous magnetic moments $\kappa^{u,d}$ are responsible for the dipole distortion produced in the case of unpolarized quarks in transversely polarized nucleons (Fig.~\ref{fig:spindens2}). With the model of Ref.~\cite{PB07}, $\kappa^u=1.86$ and $\kappa^d=-1.57$, to be compared with the values $\kappa^u=1.673$ and $\kappa^d=-2.033$ derived from data. This effect can serve as a dynamical explanation of a nonvanishing Sivers function~\cite{Siversa} $f_{1T}^\perp$ which measures the correlation between the intrinsic quark transverse momentum and the transverse nucleon spin. Such results, with the opposite shift of up and down quark spin distributions imply an opposite sign of $f_{1T}^\perp$ for up and down quarks~\cite{Burkardt02,Burkardt04a} as confirmed by the recent observation of the HERMES collaboration~\cite{Hermes05a}.

\begin{figure}
\begin{center}
\epsfig{file= 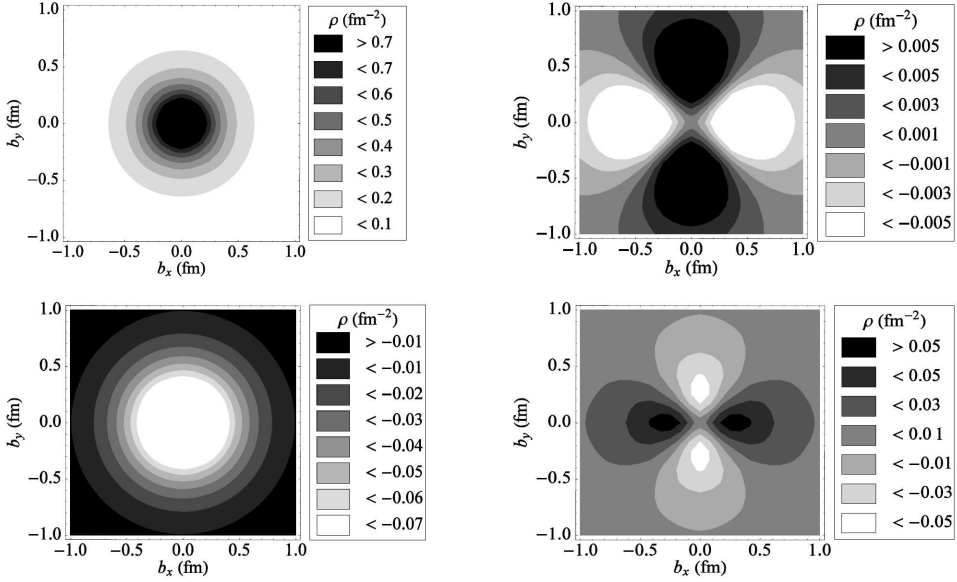,width=13 cm}
\end{center}
\caption{\small The monopole contribution $\oneh s_xS_x({\mathcal H}_T-\Delta_b\tilde {\mathcal H}_T/4M_N^2)$ (left) and the quadrupole contribution $\oneh s_xS_x(b_x^2-b_y^2)\tilde {\mathcal H}''_T/M_N^2$ (right) for $\hat x$-polarized quarks in a nucleon also polarized along $\hat x$. The upper (lower) row gives the results for up (down) quarks (taken from~\cite{PB07}).}
\label{fig:spindens3}
\end{figure}

\begin{figure}
\begin{center}
\epsfig{file= 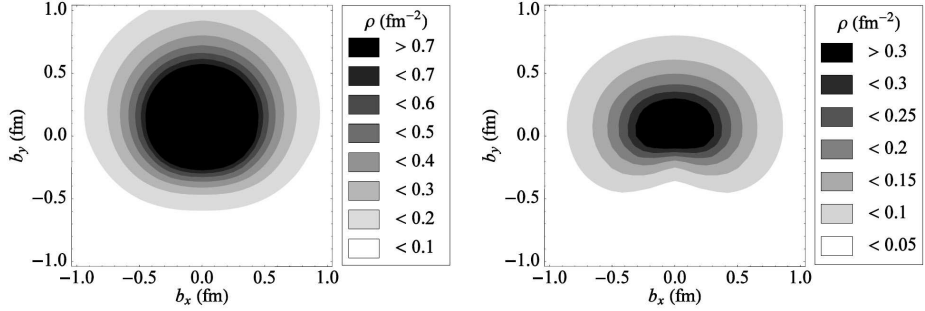,width=13 cm}
\end{center}
\caption{\small The total spin distribution as a sum of monopole, dipole and quadrupole terms,  for $\hat x$-polarized quarks in a proton also polarized along $\hat x$; left (right) panel for up (down) quarks (taken from~\cite{PB07}).}
\label{fig:spindens4}
\end{figure}

\begin{figure}
\begin{center}
\epsfig{file= 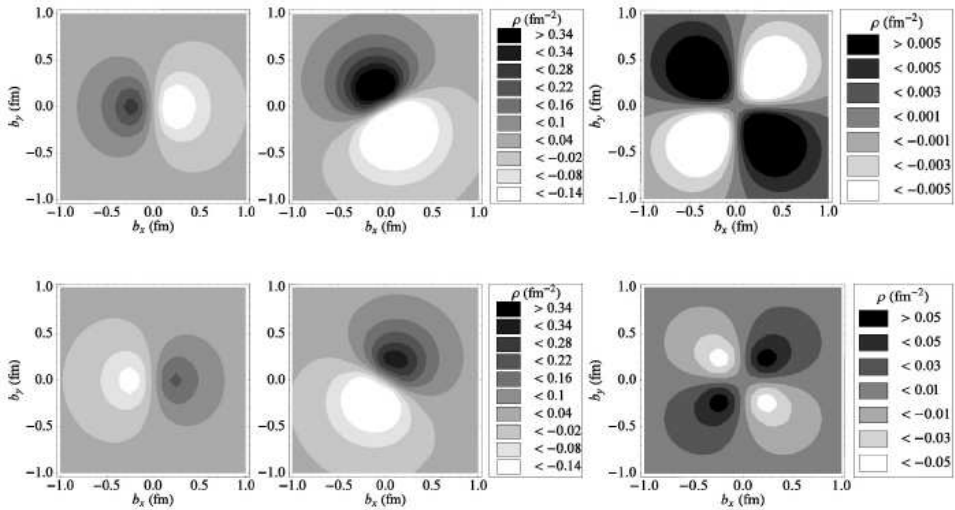,width=13 cm}
\end{center}
\caption{\small  The dipole contribution $\oneh S_yb_x{\mathcal E}'$ (left), the total dipole contribution $\oneh[S_yb_x{\mathcal E}'-s_xb_y({\mathcal E}'_T+2\tilde {\mathcal H}'_T)/M_N]$  (middle) and the quadrupole contribution $s_xS_yb_{x}b_{y}\tilde {\mathcal H}''_T/M_N^2$ (right) for $\hat x$-polarized quarks in a nucleon transversely polarized in the $\hat y$-direction. The upper (lower) row gives the results for up (down) quarks (taken from~\cite{PB07}).}
\label{fig:spindens5}
\end{figure}

\begin{figure}
\begin{center}
\epsfig{file= 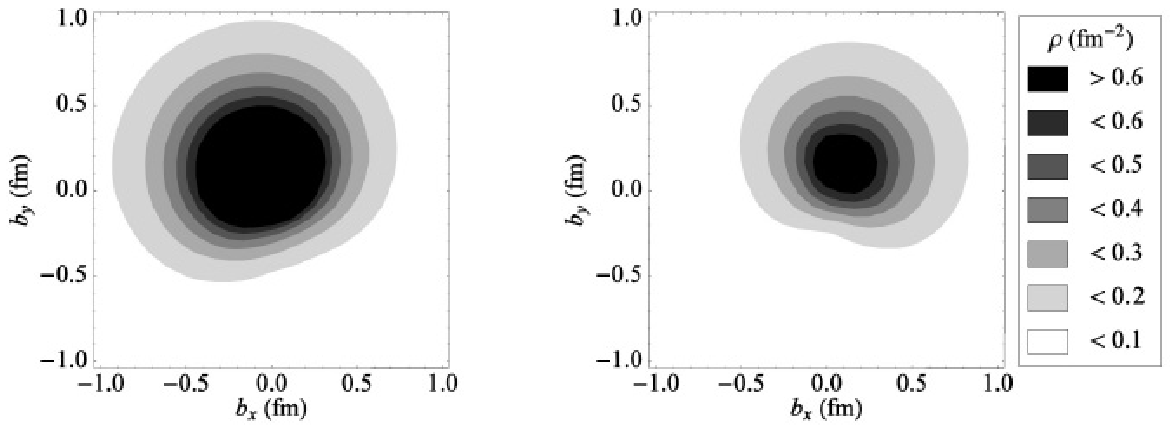,width=13 cm}
\end{center}
\caption{\small The total spin distribution, as a sum of monopole, dipole and quadrupole terms, for $\hat x$-polarized quarks in a proton transversely polarized in the $\hat y$-direction; left (right) panel for up (down) quarks (taken from~\cite{PB07}).}
\label{fig:spindens6}
\end{figure}

The results in Figs.~\ref{fig:spindens1} and \ref{fig:spindens2} are in qualitative agreement with those obtained in lattice calculations~\cite{QCDSF06a}, where strongly distorted spin densities for transversely polarized quarks in an unpolarized nucleon have been found (Fig.~\ref{fig:spindenslat}). One observes that the sideways distortion for down quarks is about twice as strong as for up quarks, even if the anomalous magnetic moment $\kappa^q$ and the anomalous tensor magnetic moment $\kappa_T^q$ have about the same magnitude. This is because the monopole distribution for up quarks is twice as large as for down quarks, and therefore adding the dipole contribution results in a larger distortion for down quarks than for up quarks.

In Fig.~\ref{fig:spindens3} quarks and proton all transversely polarized along $\hat x$ are considered. Quite remarkably, opposite signs of the up and down quadrupole and spin dependent monopole terms are found as a consequence of the opposite sign for up and down quarks  of the $x$ dependence of ${\mathcal H}_T$ and  $\tilde {\mathcal H}_T$ predicted by the model~\cite{PPB05}. The quadrupole distribution for up quark is more spread than for down quark, but it is much smaller, resulting in an average distortion equal to $-0.04$ to be compared with the value 0.07 for down quark.

The total spin distribution for quarks and proton transversely polarized along $\hat x$ is shown in Fig.~\ref{fig:spindens4} as the result of summing for each flavour the two monopole contributions in the left panels of Figs.~\ref{fig:spindens1} and \ref{fig:spindens3}, the two dipole contributions on the middle panels of Figs.~\ref{fig:spindens1} and \ref{fig:spindens2} and the quadrupole contribution of the right panel in Fig.~\ref{fig:spindens3}. For up quarks this distribution is almost axially symmetrical around a position slightly shifted in the $\hat y$-direction (Fig.~\ref{fig:spindens4}). This is a consequence of the dominating role of the monopole terms, the compensating action of the two dipole terms and the small quadrupole contribution. In contrast the down quark spin distribution has a much lower size and shows a strong and symmetric deformation about the $\hat y$-axis stretching along the same $\hat x$-direction of the quark and proton polarization. 

\begin{figure}
\begin{center}
\epsfig{file= 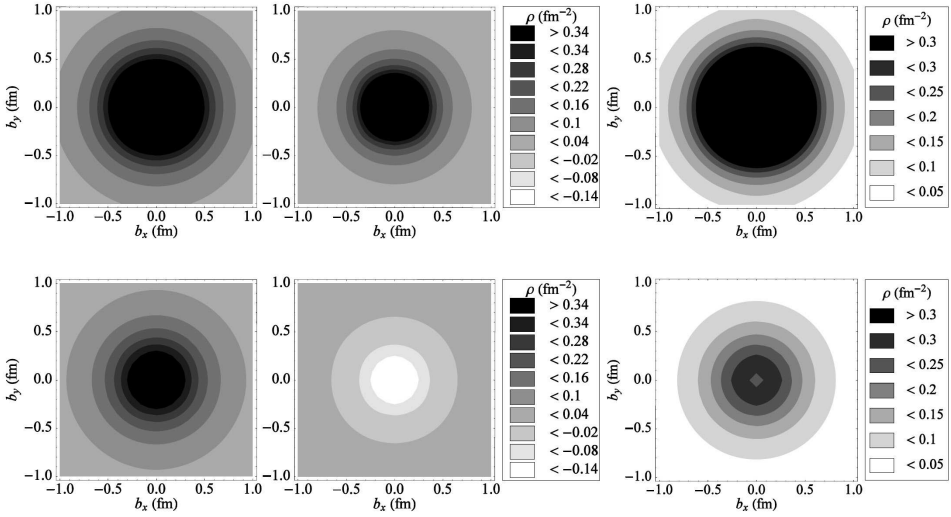,width=13 cm}
\end{center}
\caption{\small The monopole contribution $\frac{1}{2}{\mathcal H}$ (left) and $\frac{1}{2}\tilde {\mathcal H}$ (right), and their sum corresponding to the spin distribution for quark polarized in the longitudinal direction, parallel to the proton helicity. The upper (lower) row gives the results for up (down) quarks (taken from~\cite{PB07}).}
\label{fig:spindens7}
\end{figure}

The resulting transverse shift of both up and down distributions  in Fig.~\ref{fig:spindens4} is suggesting the presence of an effective transverse quark orbital angular momentum introduced in the LCWFs by the Melosh rotation required 
to transform the canonical spin to the light-front spin. Due to the shift in the positive $\hat y$-direction transverse quark spin and transverse quark orbital angular momentum seem to be aligned along the same $\hat x$-direction of the proton polarization.

In Figs.~\ref{fig:spindens5} and \ref{fig:spindens6} results are given for $\hat x$-polarized quarks in a proton polarized along $\hat y$. The distortion due to the dipole contribution $\oneh S_yb_x{\mathcal E}'$ in Fig.~\ref{fig:spindens5} is rotated with respect to the case shown in Fig.~\ref{fig:spindens1}, but the origin of opposite shift for up and down quarks is always the opposite sign of the anomalous magnetic moments $\kappa^{u,d}$. The total dipole distortion in Fig.~\ref{fig:spindens5} is obtained by considering also the second dipole term $-\oneh s_xb_y({\mathcal E}'_T+2\tilde {\mathcal H}'_T)/M_N$ displayed in Fig.~\ref{fig:spindens1}. The result is quite sizable, while the quadrupole term is rather small. Therefore, the total resulting distortion of the spin density (Fig.~\ref{fig:spindens6}) is  due to the dipole terms with a small contribution from the quadrupole terms. Correspondingly, the quark orbital angular momentum has positive $\hat x$ and $\hat y$ components for up quarks, and positive $\hat x$ and negative $\hat y$ components for down quarks. Here as well as in Fig.~\ref{fig:spindens4} the quark orbital angular momentum is entirely generated by the Melosh rotations.

Finally, the case of quark polarization parallel to the proton helicity is considered in Fig.~\ref{fig:spindens7}. Here only monopole terms occur (see Eq.~(\ref{eq:long})) and their role was first discussed in Ref.~\cite{Burkardt03}. The opposite sign of $\oneh \tilde {\mathcal H}$ for up and down quarks is responsible for quite a different radial distribution of the axially symmetric spin density. Since in the forward limit the GPD $\tilde {\mathcal H}$ reduces to the helicity distribution $\Delta q(x)$, $\tilde H(x,\xi=0,t=0)=\Delta q(x)$, this difference ultimately reflects the opposite behaviour of the helicity distributions and the opposite sign of the axial-vector coupling constants $g_A^{u,d}$ of up and down quarks (see also Ref.~\cite{BPT04}).


\section{Status of experimental investigation}
\label{sect:exp}

\begin{figure}
\begin{center}
    \epsfig{file=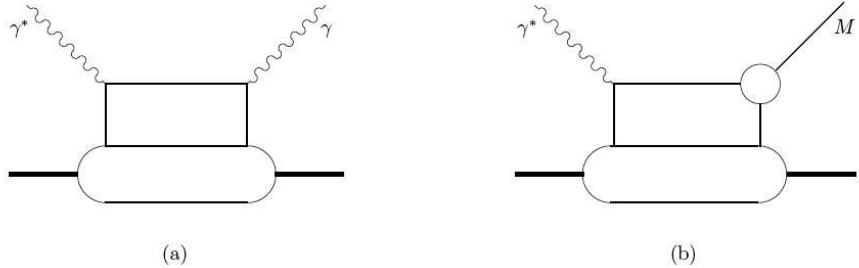, width=12cm}
\end{center}
\caption{Handbag diagrams: (a) for DVCS and (b) for exclusive meson production.}
\label{fig:handbag}
\end{figure}

Experimentally, the GPDs can be accessed in exclusive measurements such as hard exclusive meson leptoproduction $e N\rightarrow e' M N'$~\cite{FKS96,CFS97,VGG98}, electroproduction of the photon 
$eN\to e'N'\gamma$ which is sensitive to the DVCS amplitude~\cite{Ji97,Ji97a}, photoproduction of a lepton pair $\gamma N\to\ell\bar\ell' N'$ or time-like Compton scattering~\cite{Berger02} and electroproduction of a lepton pair 
$eN\to e'N'\ell\bar\ell'$ or double deeply virtual Compton scattering~\cite{Guidal03,MB03}. At leading order the DVCS and meson production amplitudes can be described in terms of the so-called handbag diagram (Fig.~\ref{fig:handbag}), where the lower blob represents the involved GPDs. In the asymptotic limit of $Q^2$ the corresponding differential cross sections scale as $Q^{-4}$ for DVCS and $Q^{-6}$ for meson production. These $Q^2$ dependences are strong experimental signatures that the appropriate regime is reached in order to rely on the validity of the handbag description and the interpretation of data in terms of GPDs.

The GPDs depend on three variables: $x$,  $\xi$ and $t$. However, only two of them are accessible experimentally in lepton scattering, i.e. $\xi$, fully defined by detecting the scattered lepton, and $t$, fully defined by detecting either the recoil proton or the outgoing photon or meson. Due to the loop in the handbag diagram (Fig.~\ref{fig:handbag}), $x$ is integrated over and the GPDs enter the integral with a weighting function given by the propagator of the quark between the incoming virtual photon and the outgoing photon or meson. This means that the amplitude of the process is proportional to the convolution
\be
\int_{-1}^1 dx \, \frac{F(x,\xi,t)}{\xi-x+i\epsilon} ={\cal P}\int_{-1}^1 dx \, \frac{F(x,\xi,t)}{\xi-x} -i\pi F(\xi,\xi,t),
\ee
where $F$ is any of the GPDs. The imaginary part of the amplitude is thus given directly by GPDs at the special point $x=\xi$, whereas the real part is sensitive to all $x$. However, also in the latter case the most relevant region of $x$ is determined by $\xi$. Only when an observable is proportional to the imaginary part of the amplitude, like in the case for instance of the beam-spin asymmetry in DVCS, one actually measures directly the GPDs at some specific value $x=\xi$. It is therefore a non-trivial task to extract GPDs from data by a deconvolution procedure, so that one has to rely on models and fitting procedures in a global study of several observables~\cite{Guidal02}. In addition, the effects of other mechanisms, such as Regge exchange contributions particularly important at large $Q^2$ and small $t$~\cite{Laget,Lagetbis,Londergan}, have to be considered. 

Hard exclusive meson production is harder to describe quantitatively than other processes, but it provides important complementary information such as gluon distributions that enter the Compton amplitude only at NLO order in the strong coupling constant $\alpha_s$. On the other hand, Compton scattering, especially double DVCS which provides a more direct means to measure quark GPDs and their dependence on the different kinematical variables, is much more demanding from an experimental point of view.

Exploratory and first dedicated experiments have been performed in recent years at the HERA collider (27 GeV electrons/positrons against 800-900 GeV protons) with the H1 and ZEUS collaborations, at DESY with the fixed-target HERMES detector and the 27 GeV electron/positron beam, at JLab in Hall A and B with the 6 GeV electron beam. Other measurements are in program, as e.g. at CERN with the COMPASS experiment working with the high-energy muon beam. In the following a brief review of the available data is presented.


\subsection{Hard exclusive meson production}

\begin{figure}
\begin{center}
    \epsfig{file=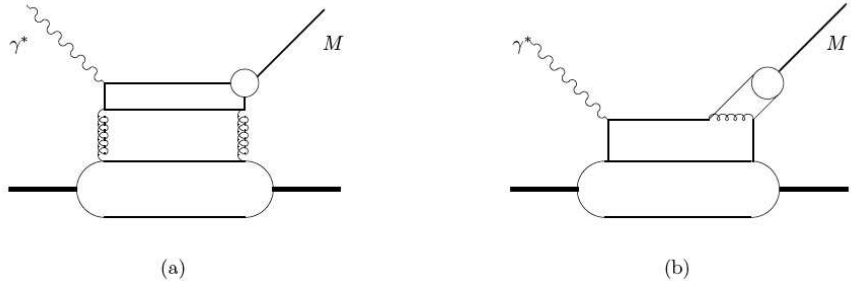, width=12cm}
\end{center}
\caption{ Gluon (a) and quark (b) GPDs contributing to the amplitude for meson production.}
\label{fig:gluonquark}
\end{figure}

Exclusive leptoproduction of light vector mesons, $\gamma^*p\to Vp$ ($V=\rho^0,\omega,\phi$), at low virtuality $Q^2$ of the virtual photon and for $\gamma^*p$ c.m. energies $W\simge 10$ GeV exhibits features typical of soft diffraction processes: a weak dependence on $W$ and a differential cross section that falls exponentially with $-t$ at low $|t|$ values (see, e.g., \cite{cassel,NMC94a,Zeus95,E665} and references therein). These features are consistent with the expectations of the vector-meson dominance model and the colour-dipole model according to which the photon fluctuates into a vector meson which then scatters elastically from the proton. In contrast, vector-meson production at high $Q^2$ (hard diffractive leptoproduction) was proposed as an important source of information of the gluon dynamics~\cite{BFGMS94,Radyushkin96b}. The production mechanism was assumed to proceed in three steps:  the virtual photon fluctuates into a $q\bar q$ state; the $q\bar q$ pair scatters  on the proton target via exchange of a gluon pair; and the scattered $q\bar q$ pair turns into a vector meson (Fig.~\ref{fig:gluonquark}a). It was also observed that longitudinally polarized photons with sufficiently large $Q^2$ produce small-size $q\bar q$ pairs configurations~\cite{FKS96}. This made possible to apply perturbative QCD to calculate the cross section when the vector meson is produced in the longitudinally polarized state by longitudinally polarized photons at small $x$. In this case a factorization theorem for the process amplitude can be established~\cite{FKS96}. The factorization theorem was proved in the collinear approximation for the full range of $x$ and the general case of hard exclusive electroproduction of any meson~\cite{CFS97}. Contributions from transverse photons are suppressed by at least one power of $1/Q$ with respect to the longitudinal ones~\cite{CFS97,DGP99,Collins00}. The analysis of the cross section for meson production with target polarization can be found in Refs.~\cite{DS05,Diehl07}.

In the scaling limit, $Q^2\to\infty$, at fixed Bjorken variable $x_B=Q^2/2(p\cdot q)$ and fixed invariant momentum transfer $t$ to the proton, the amplitude of vector-meson production is given by a convolution of the soft parts of the process, consisting of both gluon and quark GPDs~\cite{Radyushkin96b,CFS97} and the nonperturbative $q\bar q$ distribution amplitude for meson formation, with the perturbatively calculable hard-scattering kernels (Fig.~\ref{fig:gluonquark}). The kernels are known to NLO, i.e. to order $\alpha_s^2$~\cite{ISK04}. The description is restricted to sufficiently large $Q^2$ but can be used for both small and large $x_B$, thus providing a common framework for analyzing both collider and fixed-target data~\cite{Zeus07}. In the collinear factorization framework NLO corrections are not negligible in the $x_B$ range relevant for experiments at HERMES, JLab and COMPASS~\cite{diehlkugler07}. 

Hard exclusive meson productions are rather complex to analyze as they contain nonperturbative information on both the target and the produced meson. Nevertheless they offer the possibility to disentangle different GPDs. For mesons with natural parity $P=(-)^J$, such as vector mesons,  both quark and gluon GPDs in general contribute at leading order in $\alpha_s$, whereas for mesons with unnatural parity $P=(-)^{J+1}$, such as pseudoscalar mesons, only quark contributions appear. The quantum numbers of the produced meson select different flavour combinations of quark GPDs. At leading twist exclusive vector-meson production  is only sensitive to unpolarized GPDs ($H^q$ and $E^q$), whereas polarized GPDs 
($\tilde H^q$ and $\tilde E^q$) are involved in pseudoscalar meson production without the need for a polarized target or beam~\cite{FPPS99}. The gluon GPDs $H^g$ and $E^g$ can be accessed in neutral vector-meson production~\cite{VGG98,VGG99,Mankiewicz98b,Lehmann00,Lehmann01}.

Chirally odd GPDs cannot be accessed in exclusive electroproduction of a single vector meson~\cite{DGP99,Collins00,Hoodbhoy02}. They can be accessed in diffractive electroproduction of two vector mesons~\cite{IBST02}.

First results on hard diffractive $\rho^0$ production were obtained in the collider experiments H1~\cite{H100} and ZEUS~\cite{Zeus99,Zeus00a,Zeus07} by looking at the $W$, $Q^2$, and $t$ dependence of the cross section in terms of the invariant mass distribution of the two decay pions. Evidence for a dominant longitudinal contribution, increasing with $Q^2$, was found~\cite{Zeus99}. At large $Q^2$ the cross section also develops a stronger $W$ dependence than that expected from the behaviour of elastic and total hadron-hadron cross sections~\cite{H100}. Therefore, no comparison with models based on GPDs was possible. Also the recent high-precision data of the ZEUS collaboration~\cite{Zeus07} have been compared to various theoretical predictions, none of which are able to reproduce all the features of the data.

With fixed hydrogen target~\cite{Hermes00a,Clas05} the $\rho^0$ meson decay into $\pi^+\pi^-$ was used to identify the reaction of interest. The longitudinal part of the cross section was extracted from data~\cite{Hermes00a} or separated under the assumption of $s$-channel helicity conservation, i.e. that transitions with the same helicity for photon and meson are larger than those changing the helicity~\cite{Clas05}. A fair description of data is obtained with the phenomenological GPDs parametrization of Refs.~\cite{VGG98,VGG99} with the dominating quark contribution, while the gluon contribution starts to contribute significantly only at $Q^2=4$ GeV$^2$ and $W>10$ GeV.

Hard exclusive electroproduction of $\pi^+\pi^-$ pairs has also been measured~\cite{Hermes04a}. In this case both two-gluon and quark-antiquark exchange mechanisms contribute. The resonant $\pi^+\pi^-$ pair production via longitudinal $\rho^0$ (isospin $I=1,$ total angular momentum $J=1,3,\dots$ and C-parity $C=-1$) or $f$-meson ($I=0$, $J=0,2,\dots$, $C=+1$) decay is in competition with the two-gluon channel that gives rise to pion pairs with the quantum numbers of the $\rho$-meson  family only. Therefore a sizable admixture of isoscalar and isovector pion pairs is obtained.

The cross section of hard exclusive electroproduction of $\pi^+$ mesons was measured by the HERMES collaboration as a function of 
$t$ and $Q^2$~\cite{Hermes02a,Cynthia05,Hermes07a}. A model calculation based on GPDs with power corrections~\cite{VGG99} is in fair agreement with the data at low values of $\vert t\vert$. However, in the experiment separation of the longitudinal and transverse photon contributions to the cross section was not achieved. The $t$ and $Q^2$ dependence of the cross section was nicely described  by a model calculation based on the Regge formalism~\cite{Laget}. This good agreement is mainly due to the dominant pion-pole contribution which enters also in the GPD $\tilde E$.

The exclusive $\omega$ electroproduction off the proton was studied at JLab~\cite{Clas05a} in a large kinematical domain above the nucleon resonance region. The $t$-channel $\pi^0$ exchange, which is mostly due to transverse photons, seems to dominate the reaction even for $Q^2$ as large as 5 GeV$^2$. Thus the GPDs formalism is hardly applicable in this case.


\subsection{DVCS}

After some seminal work~\cite{Ji97,Ji97a,Radyushkin96a} a factorization theorem was proved valid also for DVCS to all orders in perturbation theory~\cite{Radyushkin97,Collins99a}. Compared to exclusive meson production DVCS is simpler to describe because the composite meson in the final state is replaced by an elementary particle, the photon, and thus there is no meson wave function in the factorization formula. Perturbative contributions to DVCS have been worked out to NLO accuracy~\cite{MB00,Mankiewicz98,JiOsborne}. A complete analysis in the twist-three approximation of the theory of DVCS was developed in~\cite{BMK02}, where an exhaustive set of analytical results was given for the cross section for all possible hadron and lepton polarizations involved. A detailed discussion of lepton scattering on longitudinally or transversely polarized protons is given in~\cite{DS05}.

\begin{figure}
\begin{center}
    \epsfig{file=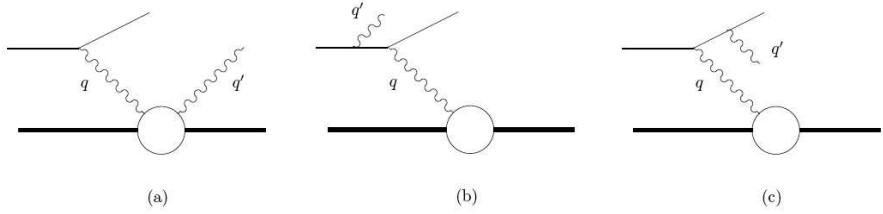, width=12cm}
\end{center}
\caption{(a) Compton scattering; (b) and (c) Bethe-Heitler contributions.}
\label{fig:dvcs-bh}
\end{figure}

The reaction $ep\to e'p'\gamma$ receives contributions from the DVCS process and the purely electromagnetic Bethe-Heitler (BH) process where the photon is emitted from the initial or the final electron (Fig.~\ref{fig:dvcs-bh}). While the BH cross section in most of the accessible kinematic regions is much larger than the DVCS cross section, the DVCS contribution can be measured through interference of DVCS and BH amplitudes. According to theory~\cite{DGPR97,BMK02}, the leading-order and leading-twist DVCS-BH interference  is proportional to the sum of two terms, one helicity independent ($d\sigma$) and one helicity dependent ($d\Sigma$), i.e.
\be
d\sigma + d\Sigma = \pm 
\left[
\cos\phi\frac{1}{\sqrt{\epsilon(1-\epsilon)}}\Re \tilde{\cal M}^{1,1} - P_l\sin\phi \sqrt{\frac{1+\epsilon}{\epsilon}} \Im\tilde{\cal M}^{1,1} 
\right],
\ee
where $+(-)$ denotes a negatively (positively) charged lepton with polarization $P_l$, $\phi$ is the azimuthal angle of the produced photon with respect to the lepton scattering plane, and $\epsilon$ is the polarization parameter of the virtual photon. The amplitude $\tilde{\cal M}^{1,1}$ is given by a linear combination of the Dirac and Pauli form factors $F_1$ and $F_2$ together with Compton form factors $\cal H$, $\tilde{\cal H}$ and $\cal E$ that are convolutions of the GPDs $H$, $\tilde H$ and $E$, respectively. $d\Sigma$ and $d\sigma$  provide complementary information: $d\Sigma$ measures the imaginary part of $\tilde{\cal M}^{1,1}$ providing direct access to GPDs at $x=\xi$, whereas $d\sigma$ determines the real part of the DVCS-BH interference and depends on the integral of GPDs over the full range of $x$. The beam-spin asymmetry, as the ratio of the difference to the sum of cross sections with  a polarized lepton beam, is associated with the imaginary part of the DVCS-BH interference with a $\sin\phi$ modulation. The beam-charge asymmetry, as the ratio of the difference to the sum of cross sections with opposite charge of the incoming lepton beam, accesses the real part with a $\cos\phi$ modulation.

First DVCS data on the proton have been reported by measuring the cross section in collider experiments at high energy by the H1~\cite{H101,H105} and ZEUS collaborations~\cite{Saull,Zeus03a} and by looking at the beam-spin asymmetry in fixed-target experiments at lower energy with polarized lepton beams and unpolarized targets by the HERMES collaboration~\cite{Hermes01a} and at JLab~\cite{Clas01}. More precise DVCS data are now available on beam-spin asymmetry~\cite{Munoz06} as well as on the longitudinal target-spin asymmetry~\cite{Clas06} from JLab and on transverse target-spin asymmetry from HERMES~\cite{Hermes06}. The lepton beam-charge asymmetry has also been measured at HERMES~\cite{Hermes02,Hermes07} and H1~\cite{Schoeffel07a}, and will be a dedicated experiment in a future program at COMPASS with the high-energy muon beam at CERN~\cite{Compass04}. Quite recently the neutron contribution to the DVCS off a deuterium target has been extracted from the helicity-dependent cross section measured at JLab in Hall A~\cite{Mazouz07}.

The consistent H1~\cite{H101,H105} and ZEUS~\cite{Saull,Zeus03a} cross section results for $Q^2$ up to 100 GeV$^2$ and $W$ up to 140 GeV are compatible with NLO calculations using GPDs parametrizations~\cite{FMcD} as well as with a description in terms of colour dipole models~\cite{Donnachie01,Favart04} which have been successful in describing both the inclusive and diffractive DIS cross sections at high energy.

First results on the beam-spin asymmetry with the positron beam at HERMES~\cite{Hermes01a} and the electron beam at JLab~\cite{Clas01} confirmed a $\sin\phi$ modulation in agreement with calculations based on GPDs parametrizations predicting a dominant contribution of the GPD $H$~\cite{VGG98,KPVdH01}.

The full potential of the beam-spin asymmetry has been first explored at JLab~\cite{Munoz06} by separating $d\Sigma$ and $d\sigma$. The absence of $Q^2$ dependence in the range $1.5\le Q^2\le 2.3$ GeV$^2$ of the $\sin\phi$ term of $d\Sigma$ supports the  twist-two dominance of the DVCS amplitude driven by the GPD $H$. The $d\Sigma$ data are in qualitative agreement with the predictions from the model of Refs.~\cite{VGG99,GPVdH,Guidal04}, but $d\sigma$ is significantly underestimated.

First measurements of the beam-charge asymmetry have been reported by the HERMES collaboration using the electron and positron beams at HERA~\cite{Hermes02,Hermes07}. A $\cos\phi$ dependence has been observed in the range $0.03\le -t\le 0.12$ and $x_B\simeq 0.1$. The data are in agreement with the dual parametrization of GPDs~\cite{GP06} and the dominance of the GPD $H$ with a large negative $D$-terms~\cite{GPVdH,KPVdH01}. However, the data allow sufficient freedom in modeling the unknown small-$x$ behaviour of the double distribution part of GPDs, so that it is also possible to describe them without the $D$-term~\cite{BMK02}.

The longitudinal target-spin asymmetry $A_{UL}$ in $e\vec p\to e'p'\gamma$, as the ratio of the difference to the sum of cross sections with target polarization antiparallel and parallel to the beam direction, has been first measured at JLab~\cite{Clas06}. A dominating $\sin\phi$ modulation with a large contribution from GPD $\tilde H$ is observed, consistent with predictions based on  the GPD formalism.

The transverse target-spin asymmetry $A_{UT}$ associated with DVCS on the proton is measurable as the ratio of the difference to the sum of cross sections using an unpolarized lepton beam and a transversely polarized target proton at angles $\phi_S$ and $\phi_S+\pi$ with respect to the lepton scattering plane~\cite{Hermes06}. The interest of such an experiment lies in the sensitivity of $A_{UT}$ to the GPD $E$ entering the Compton form factor driving a $\cos\phi$ modulation. In turn, $E$ can be modeled using the total angular momentum $J^q$ carried by quarks in the nucleon~\cite{GPVdH}. Since the contributions of the up and down quarks are proportional to the corresponding squared charge, the down quark contribution is suppressed and $A_{UT}$ can give constraints on $J^u$~\cite{Ellinghaus06}. Model-dependent constraints can indeed be derived from experiment~\cite{Hermes06} in good agreement with lattice simulations~\cite{QCDSF04,Hagler07}. Using the model of Ref.~\cite{GPVdH} the result is $J^u+J^d/2.9=0.42\pm 0.21\pm 0.06$. The quite recent analysis of the beam-spin asymmetry on the deuteron~\cite{Mazouz07} provides a correlated constraint on $J^u$ and $J^d$ from a fit to the neutron data ($J^d+J^u/0.5=0.18\pm 0.14$) intersecting the constraint from the proton in close vicinity of the lattice results. However, one should remind that the lattice results were obtained by neglecting disconnected diagram contributions which only drop out of the isovector combination $J^u-J^d$ but can be relevant for the isoscalar combination $J^u+J^d$.

At the present stage of experimentation it is clear that DVCS is a quite selective process to study GPDs. However it is also clear that, in absence of any model-independent deconvolution procedure, one has to rely on some global fitting of data~\cite{Kumericki}. Therefore, a full experimental program aiming at the extraction of the individual GPDs requires to study several observables in different channels with dedicated experiments.


\section{Conclusions}
\label{sect:conclude}

The study of the internal structure of the nucleon through the newly developed concept of GPDs
has seen substantial progresses in the last decade. The physical content of the GPDs has fully been disclosed through the theoretical study of these functions both in momentum space, where they
allow to study the momentum correlations of partons in the nucleon, and in 
 impact parameter space, where they enable to map the spatial densities of the nucleon. At the same time, the rich spin structure of the GPDs provides ways to study aspects of the nucleon spin which are otherwise difficult to obtain, like for example the role of the orbital angular momentum of nucleon's constituents in building up the total spin.

The challenging task of the theory is to develop models able to capture the dynamics responsible for these many non trivial features of hadron structure. Efforts have been made in several directions, some of them with more emphasis on the phenomenological implications, others within the framework of effective models aimed to give complementary descriptions of the nucleon dynamics.

Among the phenomenological approaches, we discussed different strategies which incorporate the constraints of the parton distributions in the forward limit and of the electroweak form factors for the first moments in $x$ of the GPDs. As a very promising tool to handle the phenomenology of the GPDs we also considered recent fitting procedures which use additional inputs from lattice QCD simulations for the GPD moments. 

As far as effective models are concerned, we reviewed the basic ideas and results obtained in the chiral quark-soliton model, light-front constituent-quark and meson-cloud models, with the aim to unravel the different role of the valence and sea quarks in the nucleon's structure. These hadronic models refer to low energy parametrizations of the GPDs and need to be supplemented by QCD evolution before the comparison with experimental results at larger scales. 

{\it Ab initio} calculations of the GPDs can be obtained through lattice QCD simulations. Lattice QCD methods have made steady progress during the last years and  first results became recently available for moments of parton distributions, form factors and GPDs.  Although the actual simulations refer to
unrealistic large quark masses and only to the flavour non-singlet quark contribution, they have provided important insights into the transverse structure of unpolarized nucleons, the lowest moments of polarized and tensor GPDs, and transverse spin densities of quarks in the nucleon. Significant investments are being made to improve the accuracy of the lattice methods and to extend the calculations also to the gluon and the singlet-quark GPDs, holding the promise to provide a complementary tool to the experimental investigation of the GPDs.

Experimental information about GPDs come from measurements of a variety of hard exclusive processes in $eN$ scattering, which rests on the possibility to apply factorization theorems to unambiguously separate the short-distance information specific to the probe from the long-distance information contained in the GPDs, representing universal, process-independent quantities.

Deeply virtual Compton scattering and meson production in $eN$ scattering have been measured in several experiments at fixed-target facilities (HERMES at DESY, Jefferson Lab in Hall A and B) and the HERA collider. The most problematic task in such measurements is to disentangle the dependence of GPDs on the three variables $x,$ $\xi$ and $t$. This is becoming possible with the refinement of procedures which combine theoretical studies, phenomenological parametrizations, non-perturbative calculations such as lattice QCD, and increasingly more accurate experimental data. New measurements both for DVCS and meson production are in program at CERN with the COMPASS experiment working with the high-energy muon beam. The H1 and ZEUS collaborations will continue to analyze data on DVCS and
 vector-meson production at high $Q^2$ and small $x.$ The JLab 12 GeV upgrade program will double the energy beam of the CEBAF continuous-beam electron scattering accelerator from 6 to 12 GeV. Thanks also to the upgrade in the detection equipment of the experimental halls~\cite{Jlab12}, high quality data are expected for exclusive processes in a larger $Q^2$, $x$ and $t$ kinematical range. The $Q^2$ dependence will provide stringent tests of the factorization 
theorem, while the possibility to scan the $t$ dependence of the observables as a function of 
$x$ will allow to study the transverse profile of the nucleon as a function of quark light-cone momentum fraction~\cite{Hyde06}. To separate the different spin components of the GPDs, measurements of a variety of polarization observables (beam and target spin) are also planned. Moreover, the flavour decomposition of the GPDs will come both from measurements of DVCS with proton and deuterium targets, and from meson production experiments.

Finally, the construction of a new Electron Ion Collider (EIC) is currently under discussion for a location in the USA~\cite{EIC}. The capabilities provided by this machine would complement the information from
fixed-target experiments, offering the opportunity to map the gluon and quark GPDs at very small $x.$ 





\end{document}